\documentclass[twocolumn]{aastex7}
\usepackage{booktabs}
\usepackage{graphicx}
\usepackage{float}
\usepackage{amsmath}
\usepackage{CJK}

\newcommand{\code}[1]{\texttt{#1}}

\begin{document}
\begin{CJK*}{UTF8}{gbsn}

    \title{PEPSI Investigation, Retrieval, and Atlas of Numerous Giant Atmospheres (PIRANGA). II. Phase-Resolved Cross-Correlation Transmission Spectroscopy of KELT-20b}
    
    \author[0009-0001-1459-3738]{Calder Lenhart}
    \affiliation{Department of Astronomy, The Ohio State University, 4055 McPherson Laboratory, 140 West 18th Avenue, Columbus, OH 43210, USA}
    \email[show]{lenhart.106@osu.edu}
    
    \author[0000-0002-5099-8185]{Marshall C. Johnson}
    \affiliation{Department of Astronomy, The Ohio State University, 4055 McPherson Laboratory, 140 West 18th Avenue, Columbus, OH 43210, USA}
    \email{johnson.7240@osu.edu}
    
    \author[0000-0002-4361-8885]{Ji Wang (王吉)}
    \affiliation{Department of Astronomy, The Ohio State University, 4055 McPherson Laboratory, 140 West 18th Avenue, Columbus, OH 43210, USA}
    \email{wang.12220@osu.edu}
    
    \author[0000-0002-8823-8237]{Anusha Pai Asnodkar}
    \affiliation{Department of Astronomy, The Ohio State University, 4055 McPherson Laboratory, 140 West 18th Avenue, Columbus, OH 43210, USA}
    \email{paiasnodkar.1@osu.edu}
    
    \author[0009-0008-8016-6591]{Sydney Petz}
    \affiliation{Department of Astronomy, The Ohio State University, 4055 McPherson Laboratory, 140 West 18th Avenue, Columbus, OH 43210, USA}
    \email{petz.16@osu.edu}
    
    \author[0000-0002-4531-6899]{Alison Duck}
    \affiliation{Department of Astronomy, The Ohio State University, 4055 McPherson Laboratory, 140 West 18th Avenue, Columbus, OH 43210, USA}
    \email{duck.18@osu.edu}
    
    \author[0000-0002-6192-6494]{Klaus G. Strassmeier}
    \affiliation{Leibniz-Institut f{\"u}r Astrophysik Potsdam, An der Sternwarte 16, D-14482 Potsdam, Germany}
    \email{kstrassmeier@aip.de}
    
    \author[0000-0002-0551-046X]{Ilya Ilyin}
    \affiliation{Leibniz-Institut f{\"u}r Astrophysik Potsdam, An der Sternwarte 16, D-14482 Potsdam, Germany}
    \email{ilyin@aip.de}

\begin{abstract}
KELT-20b is a well-studied ($T_{\text{eq}}=2262$ K) ultra hot Jupiter, but its multidimensional atmospheric structure remains unconstrained. We performed high-resolution cross-correlation transmission spectroscopy (HRCCTS) on a single transit time series of KELT-20b, observed with PEPSI on the LBT. Upon combining nineteen in-transit exposures, we detect Fe I $(11.9\sigma)$ and Fe II $(23.7\sigma)$ and tentatively detect Na I $(3.4\sigma)$ and Cr I $(3.3\sigma)$. The full-transit velocity offsets of the strongest absorbers are $\Delta V_{\text{Fe I}} = -1.0 \pm 0.7$ km s$^{-1}$ and $\Delta V_{\text{Fe II}}= 0.0\pm 0.5$ km s$^{-1}$, which are mostly inconsistent with previously published values for KELT-20b, although the previous measurements are mostly inconsistent with each other. By correcting for discrepant systemic velocity solutions of up to $1.7$ km s$^{-1}$ between studies, our Fe II offset becomes consistent with previous measurements ($\leq 1.7\sigma$), while Fe I remains significantly less blueshifted than in earlier studies ($ \geq  2.2-4.5\sigma$). We propose a set of detection criteria to improve future reproducibility in HRCCTS work. Phase-resolving the Fe I and Fe II absorption signatures into eight orbital phase bins reveals distinct dynamical regimes: Fe II exhibits a strong phase-dependent blueshift from ingress to egress along with significant limb asymmetry, while Fe I shows weaker signals and a more modest blueshift with phase. These patterns indicate day-to-night winds and suggest scale height differences are a significant driver of limb asymmetry in KELT-20b.
\end{abstract}


    \section{Introduction} \label{sec:intro}
    Ultra hot Jupiters (UHJs) are an exceptional subclass of exoplanets that are highly irradiated, inflated, and expected to be tidally locked. They have equilibrium temperatures $T_{\text{eq}}$ $\gtrsim 2000$ K \citep{Mansfield2021, Petz2025, Snellen2025}, making them particularly amenable to atmospheric characterization. KELT-20b's equilibrium temperature is $T_{\text{eq}}=2262$ K \citep{Lund2017}, making it a member of the UHJ regime.
    
    Transmission spectroscopy is the primary tool used to probe low-pressure ($\sim10^{-2}-10^{-6} \, \text{bar}$) UHJ upper atmospheres at the terminator. The advent of ground-based spectrographs with resolving power $\lambda/\Delta\lambda = \mathcal{R}\sim10^{5}$ on 8-10 m telescopes enabled high-resolution cross-correlation transmission spectroscopy (HRCCTS). \citet{Snellen2010} introduced the cross-correlation technique for exoplanet transmission spectroscopy. Instead of analyzing transmission spectra line-by-line, a generated template spectrum and the observed spectrum were cross-correlated, enhancing the atmospheric absorption signal for species with many lines in the bandpass. The cross-correlation functions (CCFs) of each usable in-transit exposure were stacked together to increase signal strength. Absorption lines, and thus CCFs Doppler shifted with respect to the planetary rest frame, quantified the atmospheric dynamics. Work by \citet{Ehrenreich2020} on WASP-76b presented CCFs of each exposure with respect to time, introducing phase-resolved HRCCTS and enabling observational probes of multidimensional wind structures \citep{MillerRicciKempton2012, Cauley2019, Hoeijmakers2020, Ehrenreich2020, Cauley2021, Rainer2021, Pino2022, Kesseli2022, Prinoth2023, Simonnin2024, Wardenier2024, Langeveld2025, Basinger2025, Seidel2025}.

    Phase-resolved HRCCTS analyses can bin a limited number of in-transit exposures together to retain time resolution, yet maintain a significant signal during transit. We can fit Gaussian profiles to each phase-resolved CCF and record their centroids, then trace their velocities during transit. Theoretical models have long presented simulated phase-resolved velocities \citep{Wardenier2021, Wardenier2023, Savel2022, Savel2023, Savel2024, Beltz2021, Beltz2022, Beltz2023}, so connecting the many models with observation allows mapping to a range of dynamical, radiative, and chemical mechanisms. KELT-20 is rapidly rotating, so stellar spectral lines are broadened to the point that only an upper limit is established on the planetary mass which in turn forces a large uncertainty in the modeled physics or, as in \citet{Chachan2025}, a $\log g$ assumption corroborated by retrieval frameworks.

    Phase-resolved Doppler-shifted cross-correlation signals of individual species templates and high-resolution transmission spectra are one of the most promising tools in exoplanetary characterization, granting an unprecedented interface between exoplanet atmosphere theory and observation. However, in its current state, the exoplanet HRCCTS subfield is replete with inconsistent results (see Section \ref{subsec:discrepancies}). To promote consistency between our work and future studies of KELT-20b in identifying and validating cross-correlation signals, we explicitly outline our detection criteria. Phase-resolved velocity offsets of UHJs have been analyzed for HAT-P-70b \citep{Langeveld2025}, KELT-9b \citep{Cauley2019, Pino2022}, KELT-20b \citep{Hoeijmakers2020, Rainer2021}, TOI-1518b \citep{Simonnin2024, Basinger2025}, WASP-33b \citep{Cauley2021}, WASP-76b \citep{Ehrenreich2020, Kesseli2022}, WASP-121b \citep{Borsa2020, Wardenier2024, Seidel2025}, and WASP-189b \citep{Prinoth2023}.

    This work seeks to reproduce previous detections of atomic species in the atmosphere of KELT-20b, report previously undetected species, and phase-resolve Doppler-shifted signals at an increased time resolution. To date, HRCCTS of KELT-20b has yielded detections of Fe I, Fe II, Na I, Ca II, and tentative detections of Mg I, Cr II, and FeH \citep{CasasayasBarris2019, Hoeijmakers2020, Nugroho2020, Rainer2021, Langeveld2022, Sicilia2022}. We intend to complement simulated wind velocity versus orbital phase plots, e.g., \citet{Savel2022, Savel2023, Wardenier2023}, with our observational data. We remove the Doppler shadow from the phase-resolved CCFs to maintain signal clarity throughout transit. We diagnose drivers of observed phase asymmetries by relating a set of phase-binned CCFs of each species to several model predictions.
 
    This paper is organized as follows: In Section \ref{sec:Observations}, we describe the observation. In Section \ref{sec:Methodology}, we describe model spectra generation and data processing of observed spectra, data cleaning steps, cross-correlation of the model and data, Doppler shadow removal, and phase-binning procedure. We also detail our detection criteria. In Section \ref{sec:Results}, we compile full-transit detections from all previous high-resolution transmission spectroscopy and HRCCTS papers on KELT-20b alongside our own. We also present phase-resolved velocity offsets, compare the absorption traces to theory, and search for behavior that is consistent with limb asymmetry drivers. We discuss the impact of discrepancies in methodology on results across the literature for KELT-20b and correct systemic velocity offsets and also discuss interesting nondetections and aliasing. In Section \ref{sec:Conclusions}, we review our results and motivate future observation and modeling of KELT-20b.
    
\section{Observation}\label{sec:Observations}
\begin{figure}
        \centering
        \includegraphics[width=0.5\textwidth]{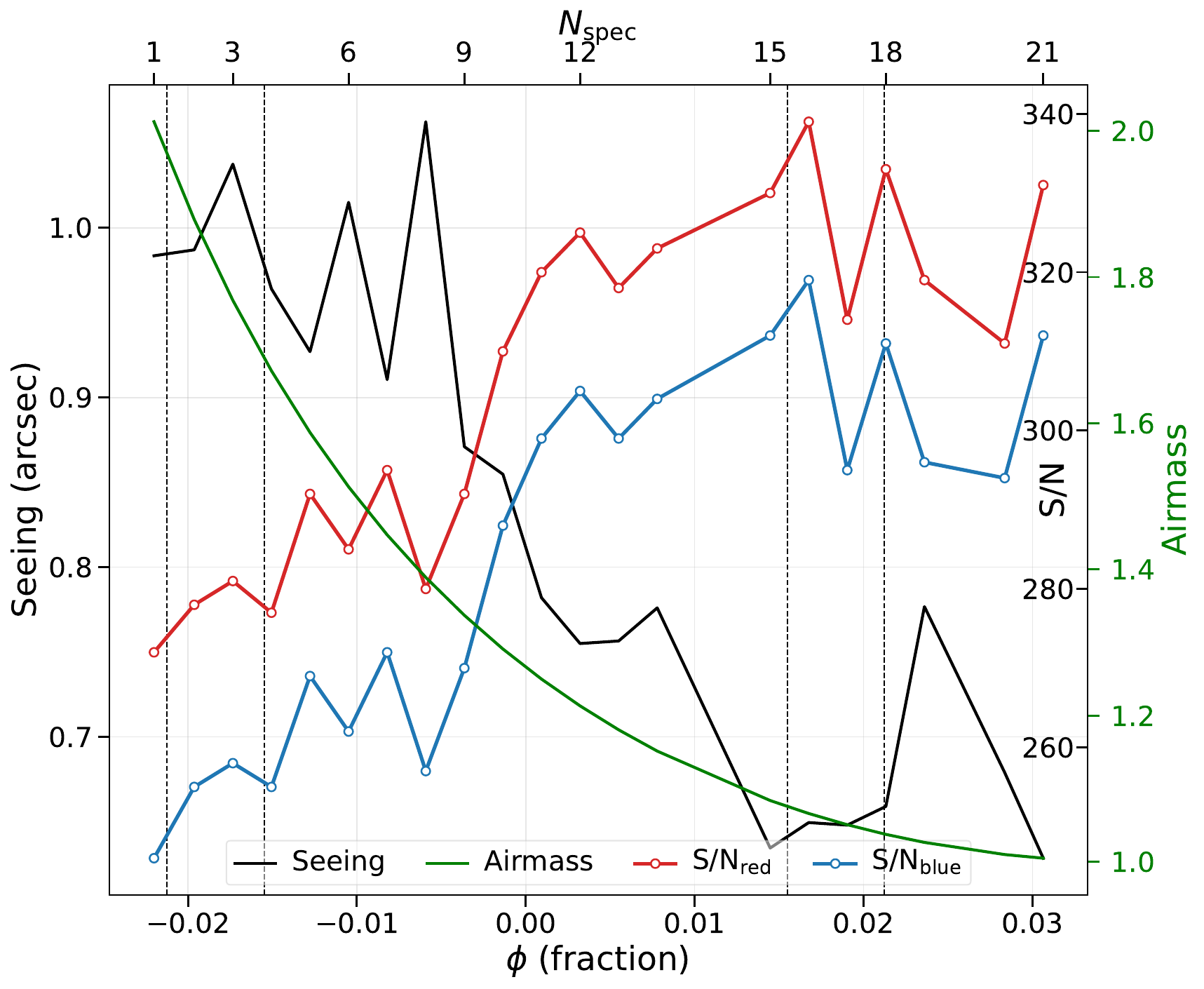}
        \caption{S/N, seeing, and airmass versus the corresponding orbital phase at the center of each exposure. The red and blue arms of PEPSI are color-coded. For the Seeing value, we use SEEINGSX, derived from measurements on the left side of the LBT. The vertical lines correspond to $T_1$, $T_2$, $T_3$, $T_4$, in order. There is a marked increase in S/N with phase, which impacts our interpretation of time-resolved CCFs, further discussed in Section \ref{subsec:Atmospheric Dynamics}.}
        \label{fig:observation_quality}
    \end{figure}

    We observed a single transit event with the Potsdam Echelle Polarimetric and Spectroscopic Instrument (PEPSI) \citep{Strassmeier2015} on the Large Binocular Telescope (LBT), with a resolving power ${\mathcal{R} = 130000}$. We utilized the PEPSI CD III+V bandpass, with a blue arm of 4800--5441\AA\ and a red arm of 6278--7419\AA. To maximize signal recovery without sacrificing line measurement precision due to Doppler smearing, we took $600~\text{s}$ exposures.\footnote{Doppler smearing occurs when the planetary velocity signal changes by more than one resolution element during an exposure \citep{Wardenier2024, BoldtChristmas2024}. At mid-transit, the maximum exposure time for PEPSI observation of KELT-20b that avoids smearing is $\Delta t \sim \frac{cP}{2\pi \mathcal{R} K_p} \sim 650~\text{s}$, where $c$ is the speed of light, $P$ is the period of the planet, $\mathcal{R}$ is the resolving power of the spectrograph, and $K_p$ is the planetary radial velocity semi-amplitude.} For comparison, previous observations of KELT-20b used exposure times of 192--600 s with HARPS and HARPS-N \citep{CasasayasBarris2019, Nugroho2020, Stangret2020, Rainer2021, Langeveld2022, Sicilia2022}, 200 s with FIES \citep{BelloArufe2022} and CARMENES \citep{Nugroho2020}, and approximately 200 s with EXPRES \citep{Hoeijmakers2020}. The blue arm spectrograph signal-to-noise ratio $\text{S/N}_{\text{blue}}$ ranged from $272-330$ during the observation, and $\text{S/N}_{\text{red}}$ ranged from $246-319$. $\text{S/N}_{\text{blue}}$ and $\text{S/N}_{\text{red}}$ are the mean of the 95\textsuperscript{th} quantile per-pixel S/Ns in the blue and red arms, respectively, during transit. We plotted each arm's $\text{S/N}$ with respect to exposure count in Figure \ref{fig:observation_quality}, noting phases of first contact $T_1$, second contact, $T_2$, third contact $T_3$, and fourth contact $T_4$. Both seeing and airmass decreased over the course of transit which likely accounts for the increase in $\text{S/N}$, as visible in Figure \ref{fig:observation_quality}. We note that there was only a single baseline exposure shortly before ingress. Since the planetary signal moved significantly over the course of transit, extraction of the planetary signal should not have been meaningfully impacted. We list the remaining key observation parameters in Table~\ref{tab:observation_details}. We reduced this observation using the Spectroscopic Data Systems (SDS) Pipeline, as detailed by \citet{Ilyin2000, Strassmeier2018}. Limited analysis of these data was done by \citet{PaiAsnodkar2022b} and \cite{Johnson2023}, but this work treats them comprehensively.
    
    \begin{deluxetable}{lc}
        \tabletypesize{\footnotesize}
        \tablecaption{Parameters of the single transit observation. $N_{\text{spec}}$ is the number of exposures, or spectra, collected in the observation, while $N_{\text{spec, in-transit}}$ are those taken between $T_{14}$. $T_{\text{exp}}$ is the time of each exposure.   \label{tab:observation_details}}
        \tablehead{
            \colhead{\textbf{Parameter}} & \colhead{\textbf{Value}}
        }
        \startdata
            Date (UT) & 2019 May 4\\
            Telescope & LBT \\
            Instrument & PEPSI \\
            $N_{\text{spec}}$ & $23$ \\
            $N_{\text{spec, in-transit}}$ & $19$ \\
            $T_{\text{exp}}$ (s) & $600$ \\
            Airmass Range & $[1.00,2.01]$ \\
            Phase Range ($^{\circ}$) & $[-8.28 , 12.24]$ \\
            $\text{S/N}_{\text{blue}}$ & $288$ \\
            $\text{S/N}_{\text{red}}$ & $308$ \\
        \enddata
    \end{deluxetable}

    \section{Methodology}\label{sec:Methodology}

        \subsection{Data preparation}\label{subsec:Data Preparation}
             The following pipeline is nearly identical to \citet{Johnson2023} and \citet{Petz2023}. Our systematics correction method employs both the~\code{molecfit} package~\citep{Smette2015, Kausch2015} and a modified version of the \code{PySysRem}~\footnote{\url{https://github.com/stephtdouglas/PySysRem}} package that accepts PEPSI data. As determined in~\citet{Johnson2023}, applying both methods yields greater significance than either alone. 
             
             ~\code{molecfit} fits a model of the telluric spectrum to the data, finds the best-fit model, then removes it. We apply ~\code{molecfit} to the red arm spectra only, as the blue arm lacks significant tellurics~\citep{Smette2015}. We split the red arm spectra into a region with $>1\%$ line depth in adjacent lines (telluric region) and a region with $\leq 1 \%$ line depth in adjacent lines (non-telluric region. We import the flux values for each spectrum, then regrid the fluxes to common wavelength values for each spectrum to correct for sub-pixel drift between exposures. Next, we flatten both arms of the time-series spectra by subtracting off the median flux from each, yielding residual spectra with stellar lines and time-invariant telluric lines removed. Finally, we apply SYSREM to the residual spectra. The SYSREM detrending algorithm~\citep{Tamuz2005} fits and removes systematic effects, with the number of systematics removed $N_{\mathrm{Sys}}$ being the relevant input parameter. For the red arm, we run SYSREM on the telluric and non-telluric regions separately.
             We empirically determine the $N_{\mathrm{Sys}}$ that yields the greatest recovered SNR in each species as calculated with Equation \ref{equ:SNR}. Optimal SYSREM iterations for the non-telluric regions of the red arm $\text{N}_{\mathrm{Sys},\text{Red, NT}}$, telluric regions of the red arm $\text{N}_{\mathrm{Sys},\text{Red, T}}$, and for the blue arm $\text{N}_{\mathrm{Sys},\text{Blue}}$ are given for each species in Table~\ref{tab:sysrem-and-vmrs}.

           \begin{deluxetable}{lcccc}
           \tabletypesize{\footnotesize}
            \caption{Optimal SYSREM iterations ($\text{N}_{\mathrm{Sys}}$) for systematics reduction for each species meeting the tentative detection. We separate the telluric and non-telluric-contaminated wavelength regions, finding the best $\text{N}_{\mathrm{Sys}}$ for each. The telluric regions are as follows: $6278 \, \text{\AA} < \lambda \leq 6328 \, \text{\AA}$, $6459 \, \text{\AA} < \lambda \leq 6527 \, \text{\AA}$, $6867 \, \text{\AA} < \lambda < 6867.5 \, \text{\AA}$, $6390 \, \text{\AA} \leq \lambda < 7168 \, \text{\AA}$, and $7312 \, \text{\AA} \leq \lambda \leq 7419 \, \text{\AA}$. $\text{N}_{\mathrm{Sys},\text{Red, NT}}$ refers to the number of systematics in non-telluric regions of the red arm and $\text{N}_{\mathrm{Sys},\text{Red, T}}$ refers to the number of systematics in telluric regions.}
            \label{tab:sysrem-and-vmrs}    
            \tablehead{
                \colhead{\textbf{Species}} & 
                \colhead{$\text{N}_{\mathrm{Sys},\text{Red, NT}}$} & 
                \colhead{$\text{N}_{\mathrm{Sys},\text{Red, T}}$} & 
                \colhead{$\text{N}_{\mathrm{Sys},\text{Blue}}$} & 
                \colhead{VMR}
            }
            \startdata
            Na I & 1 & 10 & 2 & $2.8 \times 10^{-6}$ \\
            Cr I & 5 & 5 & 5 & $7.1\times 10^{-7}$ \\
            Fe I & 5 & 5 & 6 & $5.0 \times 10^{-5}$ \\ 
            Fe II & 5 & 5 & 5 & $5.0 \times 10^{-5}$ \\
            \enddata
        \end{deluxetable}
     \subsection{Model spectra}\label{subsec:Model Spectra}
        We generate model transmission spectra in 0.01 \AA\ steps from 3850--7500 \AA{}, or a spectral resolution range of $385000$--$750000$, with the \code{petitRADTRANS} package~\citep{petitRADTRANS}; these templates can be viewed in Appendix~\ref{app:Detections}. We use physical and orbital parameters from one of the discovery papers~\citep{Lund2017} and ephemeris by Duck et al. (in prep). We also use pressure-temperature (P-T) profile parameters from \citet{Johnson2023}, and systemic velocity derived by \citet{Petz2023}; these are listed in Table~\ref{tab:parameters_summary}. We select a P-T profile given by Equation 29 of~\citet{Guillot2010} with best-fit parameters $\gamma = 30$ and $\kappa_{\text{IR}} = 0.04$ as obtained in a grid search in~\citet{Johnson2023}. While these parameters were derived using emission spectra, transmission spectra are less sensitive to detailed vertical temperature structures~\citep{Kempton2014}. The absorption lines of refractory species observable in the optical are expected to probe the dayside throughout the entire transit \citep{Wardenier2023, Pelletier2023}, since the dayside is more inflated than the nightside. Moreover, refractory species are expected to condense out on the nightside. Hence, using dayside P-T structure inferred from emission observations is reasonable. Since our analysis concerns the atmospheric dynamics instead of chemistry, we deem the same profile sufficient for this analysis.
        
            \begin{deluxetable*}{llcc}
                \tabletypesize{\footnotesize}
                \tablehead{
                    \colhead{\textbf{Parameter}} & \colhead{\textbf{Symbol (Unit)}} & \colhead{\textbf{Value}} & \colhead{\textbf{Source}}}
                \tablecaption{A summary of the stellar, planetary, and updated ephemeris used in our analysis. We reuse the systemic velocity found by \citet{Petz2023} since it utilized three PEPSI observations: two emission datasets and one transmission dataset, including ours. The updated transit timings are part of a larger effort---to be presented by Duck et al. (in prep)---to update ephemerides with \code{EXOFASTv2} \citep{Eastman2019} fits to \textit{TESS} photometry. Sources: (a)~\citet{Lund2017}; (b)~\citet{Johnson2023}; (c)~\citet{Petz2023}; (d) Duck et al. (in prep).}
                \startdata
                \multicolumn{4}{c}{\textbf{Stellar Parameters}} \\
                \hline
                Stellar radius & $R_{\ast}$ ($R_{\sun}$) & $1.565$ & (a) \\
                Stellar mass & $M_{\ast}$ ($M_{\sun}$) & $1.76$ & (a) \\
                Stellar effective temperature & $T_{\text{eff}}$ (K) & $8720$ & (a) \\
                Metallicity & $[$Fe/H$]$ & $-0.29$ & (a) \\
                Stellar surface gravity & $\log g_{\ast}$ & $4.290^{+0.017}_{-0.020}$ & (a) \\
                Stellar rotational velocity & $v\sin i_{\ast}$ (km s$^{-1}$) & $117.4\pm2.9$ & (a) \\
                \hline
                \multicolumn{4}{c}{\textbf{Planetary Parameters}} \\
                \hline
                Planet radius & $R_P$ ($R_\text{J}$) & $1.741$ & (a) \\
                Planet mass & $M_P$ ($M_{\text{J}}$) & $<3.382$ & (a) \\
                Equilibrium temperature & $T_{\text{eq}}$ (K) & $2262$ & (a) \\
                Planetary radial velocity semi-amplitude & $K_{p,\text{expected}}$ (km s$^{-1}$) & $169.0 \pm 6.1$ & (a) \\
                Projected equatorial rotational velocity & $v \sin i_P$ (km s$^{-1}$) & $2.5$ & (a) \\
                Orbital inclination & $i ~(^{\circ})$ & $86.12^{+0.28}_{-0.27}$ & (a) \\
                Semimajor axis & $a$ (AU) & $0.0542$ & (a) \\
                Eccentricity & $e$ & $0$ & (a)\\
                Infrared opacity & $\kappa_{\text{IR}}$ & $0.04$ & (b) \\
                Ratio of optical to infrared opacities & $\gamma$ & $30$ & (b) \\
                Reference pressure & $P_0$ (bar) & $1$ & (b) \\
                Atmospheric abundance of H & $X_{\text{H}_2}$ & $0.7496$ & (b) \\
                Atmospheric abundance of He & $X_{\text{He}}$ & $0.2504$ & (b) \\
                Volume mixing ratio of H$^-$ & VMR (H$^-$) & $1 \times 10^{-9}$ & (b) \\
                \hline
                \multicolumn{4}{c}{\textbf{Updated Ephemeris}} \\
                \hline
                Systemic velocity & $v_{\text{sys}}$ (km s$^{-1}$) & $-22.78\pm0.11$ & (c) \\
                Orbital period & $P$ (d) & $3.47410151 \pm 0.00000012$ & (d) \\
                Epoch of mid-transit & $T_0$ ($\text{BJD}_{\text{TDB}}$) & $2459757.811176 \pm 0.000019$ & (d) \\
                Transit duration & $T_{14}$ (d) & $0.147565^{+0.000091}_{-0.000092}$ & (d) \\
                Ingress/egress duration & $\tau$ (d) & $0.02007 \pm 0.00011$ & (d) \\
                \enddata
            \end{deluxetable*}\label{tab:parameters_summary}
        
            We calculate volume mixing ratios (VMRs) for each species using the \code{PyFastChem} equilibrium chemistry model~\citep{Stock2018, Stock2022, Kitzmann2023}, assuming solar abundances~\citep{Asplund2021} and the aforementioned P-T profile. To account for quenching, we assume that the chemical abundances below pressures of $1$ bar remain constant. We set the VMRs of ionized species equal to those of their neutral counterparts to simplify our analysis. We again note that our VMR limits are approximate since they are dependent on a simplified P-T profile previously constrained for an emission spectrum and a $1$ bar quench pressure.
            
            Atomic and molecular opacities were sourced from the~\code{petitRADTRANS} database\footnote{\url{https://petitradtrans.readthedocs.io/en/2.7.7/content/available_opacities.html}}, with missing entries supplemented from the DACE Opacity Database\footnote{\url{https://dace.unige.ch/opacityDatabase/}} and converted to \code{petitRADTRANS}-compatible format (P. Molli\`ere, private communication). \code{petitRADTRANS} sources its atomic opacities from the Kurucz line list, and we ensured the DACE opacities---Na I and Co I---came from the same line list\footnote{\url{http://kurucz.harvard.edu/}}. We used several DACE molecular opacity datasets, including the \citet{ATP} line list for AlO, \citet{VBATHY} line list for CaO, and \citet{Rivlin} line list for NaH.
        
        \subsection{Cross-correlation}\label{subsec:cross-correlation}
             With data preparation complete, we cross-correlate model and observed spectra in the stellar rest frame, with the same procedure as described in~\citet{Johnson2023}. The cross-correlation procedure is adapted from the BANZAI-NRES package~\citep{McCully2022}. Per-spectrum SNR is defined as
            \begin{equation} \label{equ:SNR}
                R_j(\phi) =\text{P}_{90}\left(\frac{f_j(\phi)}{\sigma_{f_j}(\phi)}\right)
            \end{equation}
            where $R_j$ is the $\text{P}_{90}$ (90\textsuperscript{th} percentile) of the ratio of residual flux $f_j(\phi)$ to residual flux error, as calculated by the PEPSI pipeline, for the j\textsuperscript{th} spectrum and $\phi(t)$ is the planet orbital phase. The cross-correlation function is defined as
            \begin{equation} \label{equ:CCF}
                C(\phi,v)= \sum_{i=1}^{N_{\text{pixels}}} \frac{f_i(\phi) T_i(v)}{\sigma_{f_i}^2(\phi)}
            \end{equation} 
            where the index $i$ is the pixel, $f_i(\phi)$ is residual flux at phase $\phi$, $T_i(v)$ is the \code{petitRADTRANS} template spectrum Doppler-shifted to a radial velocity $v$, and $\sigma^2_{f_i}(\phi)$ is the variance of the residual flux. We compute CCF weights for the j\textsuperscript{th} spectrum with
            \begin{equation}\label{equ:CCF_weight}
                    w_j(\phi) = WR_j^2(\phi)
            \end{equation}
            where $W$ is the sum of the equivalent widths of all of the lines in the template spectrum of the species. The radial velocity of the planet as a function of time is
            \begin{equation} \label{equ:RV_p}
            \mathrm{RV}(\phi) = K_p\sin \left[ 2\pi\phi(t) \right]+ v_{\text{sys}} + \Delta V(\phi)
            \end{equation}
            where $\Delta V$ is the velocity offset of the lines or CCF being observed. We select a velocity range of $\left[-400,400\right]$ km s$^{-1}$ in $1$ km s$^{-1}$ steps. We stack each exposure's CCF with the weighted sum, then subtract off the median value $M$ of the 2D map to correct for any offsets introduced by this process, yielding the full-transit CCF of a single observation. We normalize the full-transit CCF amplitude with the standard deviation of the CCF $\sigma_C$ at systemic velocity ${|v_{\text{sys}}| > 100 \, \text{km}\,\text{s}^{-1}}$, yielding the SNR. Next, we remove the Doppler shadow (described in more detail in Section \ref{subsec:DopplerShadowRemoval}). We then grid search for a SNR peak over planet radial velocity semi-amplitude $K_p \in [50, 350]$ $\text{km}\,\text{s}^{-1}$ in steps of $1 ~ \text{km}\,\text{s}^{-1}$, yielding a ${K_p-v_{\text{sys}}}$ SNR map. For a point $(K_p, v_{\text{sys}})$ in the SNR map, this gives
            \begin{equation}
                \text{SNR}(K_p, v_{\text{sys}}) = \frac{1}{\sigma_C} 
                \left[ \sum_{j=1}^{N_{\text{spec}}} w_j(\phi) C(\phi, v) - M \right]
            \end{equation}
    
           We use Equation \ref{equ:RV_p} to calculate the planetary velocity at each exposure, assuming $\Delta V=0$, then shift the full-transit CCFs into the planetary rest frame, i.e., we transform to $K_p-\Delta V$ space. Since $e$ is consistent with zero, the orbital contribution to $\Delta V$ is nominally zero \citep{Lund2017}. Assuming the $v_{\text{sys}}$ correction is accurate, a nonzero $\Delta V$ implies an atmospheric process is causing a line velocity offset; this value is also referred to as $v_{\text{wind}}$ in some works.
            
            We fit a Gaussian to the 1D CCF at $K_p$ corresponding to the peak of the SNR map, i.e., peak $K_p$ slice. We record the central radial velocity of the Gaussian fit in Table \ref{tab:detection-summary}. We adapt the methods of \citet{Kesseli2022} for calculating uncertainties of $\Delta V$ and $K_p$. That is, we divide the width of the fitted Gaussian by its SNR, then combine in quadrature with $v_{\text{sys}}$ to yield the $\Delta V$ uncertainty. We also propagate error on the phase-resolved absorption traces to the full-transit CCFs, as discussed in Section \ref{subsubsec:fit-to-absorption-trace}. We calculate uncertainty on $K_p$ by selecting upper and lower intervals that correspond to a $1\sigma$ decrease in SNR.
            
            After completing the above process for each PEPSI arm separately, we coadd each arm into a CCF we henceforth refer to as the combined arm. We generate a combined arm ${K_p-\Delta V}$ SNR map via grid search and fit a Gaussian to the peak $K_p$ slice. 
            
           \subsection{Doppler shadow removal}\label{subsec:DopplerShadowRemoval}
\begin{figure*}
                \centering
                \includegraphics[width=0.6\textwidth]{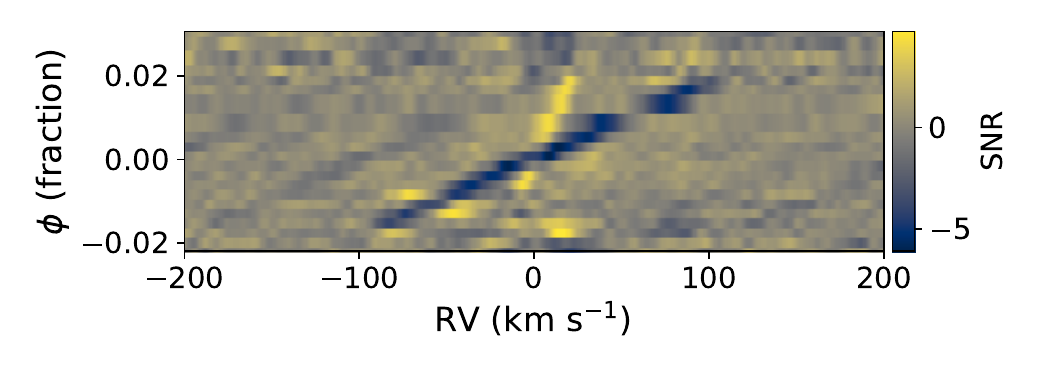}
                \includegraphics[width=0.6\textwidth]{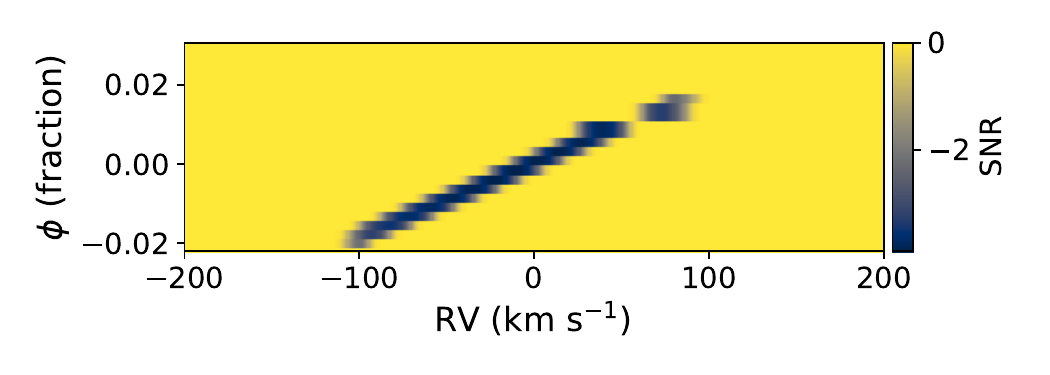}
                \includegraphics[width=0.6\textwidth]{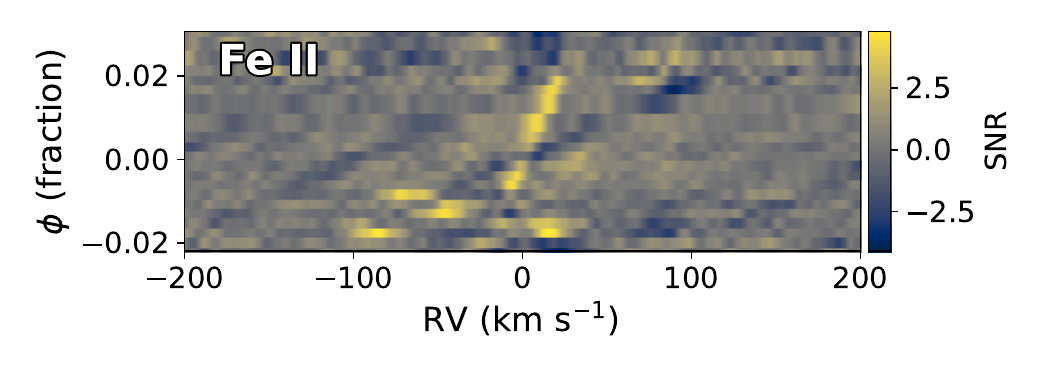}
                \caption{Removal of the Doppler shadow in the blue arm 2D CCF of Fe II. \textit{Top}: 2D CCF of the blue arm Fe II signal for all $N_{\text{spec}}$ exposures stacked. The Doppler Shadow (blue streak) intersects the atmospheric velocity signal (yellow streak near $\mathrm{RV}=0$, at the center of the Doppler shadow). \textit{Middle}: Doppler shadow CCF generated by our model and fitted to our CCF data via non-linear least squares optimization. \textit{Bottom}: Residual CCF after subtracting off the Doppler shadow.}
                \label{fig:doppler-shadow-removal}
            \end{figure*}

               As the planet crosses the stellar disk, the disk-integrated rotationally-broadened stellar absorption lines are deformed—a phenomenon known as the Rossiter-McLaughlin effect \citep[RME;][]{Rossiter1924, McLaughlin1924}. A spectrally resolved manifestation of the RME in the CCF is the Doppler shadow. To correct for the RME, we employ a Doppler shadow model, detailed in ~\citet{Johnson2016} and further discussed in~\citet{Johnson2014, Johnson2017}. In brief, we begin by dividing the stellar disk into surface elements defined by a $300\times300$ Cartesian grid, finding the fraction of each exposure time grid spaces are obscured or partially obscured by the planet. We scale the line profile contribution of each obscured surface element by this fraction, then convolve the line profiles with a Gaussian instrumental line profile. We assume solid-body rotation and a Gaussian line profile in each surface element, scaled by quadratic limb-darkening coefficients. We neglect macroturbulent broadening and gravity darkening~\citep{Johnson2016}. We subtract the out-of-transit line profile from the full line profile, revealing time-series CCF contribution of the transiting planet, known as the Doppler shadow. Using a normalized line profile, we fit to the Doppler shadow signature by limited non-linear least squares optimization with respect to spin-orbit alignment $\lambda$, impact parameter $b$, width of the Gaussian instrumental line profile, and a multiplicative scaling factor to transform the Doppler shadow into SNR space\footnote{We are referring to the CCF map computed by Equation \ref{equ:CCF}, not the function to compute SNR in ${K_p-\Delta V}$ space.}. Finally, we subtract the scaled Doppler shadow from the CCF. This process is demonstrated in Figure \ref{fig:doppler-shadow-removal}.

            \subsection{Detection criteria}\label{subsec:DetectionCriteria}
                We first examine \code{petitRADTRANS}-generated spectra, noting if the species has an abundance of lines in only one arm or both arms in the PEPSI CD III+V bandpass. We then inspect each arm 1D CCF, selecting whichever has a peak SNR $\geq3\sigma$. If each arm is $\geq3\sigma$, we instead select the combined CCF.
                
                We validate our detections by checking five criteria:
                \begin{enumerate}
                    \item If there is $\geq{3\sigma}$ SNR peak, it is a tentative detection, and if the SNR is $\geq5\sigma$, it is a detection. Otherwise, we do not detect the species.
                    \item The CCF has a physically reasonable shape in ${K_p-\Delta V}$ space. A physically reasonable 2D CCF has $|\Delta V| \leq 15~$ km s$^{-1}$ and $K_p$ consistent with $K_{p,\text{expected}}$, the planetary radial velocity semi-amplitude calculated from planetary parameters in Table \ref{tab:parameters_summary}.
                    \item There are no other CCF peaks on the map with an SNR within $1\sigma$ of the peak SNR that are within
                    \begin{enumerate}
                        \item $1\sigma$ of $K_p$
                        \item $25$ km s$^{-1}$ of $\Delta V$.
                    \end{enumerate}
                    See Appendix \ref{sec:2d-ccf-check-example} for examples.
                    \item If each arm of the spectrograph has a $\geq{3\sigma}$ SNR peak, Gaussian fits to their line profiles must have centroids within $\pm 5$ km s$^{-1}$ of each other. We show one example of detection and one example of nondetection in Appendix \ref{sec:template-spectra-and-1d-ccf-check-example} to further illustrate the criteria as they apply to template spectra and 1D CCFs. This condition could be generalized to other spectrographs that calculate CCFs spectra order-by-order by enforcing a consistent CCF velocity across the whole bandpass.
                    \item The detection is only valid if there is insignificant or no aliasing between the search species and strong lines from other species in the spectrum, as described in~\citet{Borsato2023, Petz2023}. In 1D CCFs, these artifacts manifest as peaks at both visually detectable unphysical $\Delta V$ values, and as peaks overlapping with the absorption signal. In most cases, species with many lines, such as Fe I, interfere with species with few lines, such as Ba II, causing false atmospheric signals at otherwise reasonable velocity offsets. To correct for this, we first compute an aliased CCF between the search species template and Fe I template, since Fe I is a strongly detected species with the most absorption lines in the PEPSI bandpass. We visually confirm that the aliased CCF peaks do not peak in the same location as the true CCF peak. Future studies can fit the aliased CCF to the true CCF with a multiplicative scaling factor plus a constant offset in $K_p-\Delta V$ space, then subtract off, yielding the residual CCF. If the residual CCF remains unaltered in the immediate area of the 2D CCF and in the 1D line profile and Gaussian fit, they can verify the detection. If the residual CCF is altered, but the signal structure in $K_p - \Delta V$ space is consistent with the original CCF, they can report the detection characteristics of the original 2D full-transit CCF. They would discard the detection if the residual CCF is significantly altered.
                    \end{enumerate}

        We completed this process for each species we searched for, including Li I, Be I, B I, N I, Na I, Mg I, Mg II, Si I, SiO, K I, K II, Ca I, Ca II, CaH, CaO, Sc I, Sc II, Ti I, Ti II, TiO, V I, Cr I, Cr II, Mn I, Fe I, Fe II, FeH, Co I, Ni I, Cu I, Zn I, Ga I, Ge I, Rb I, Sr I, Y I, Nb I, Mo I, Ru I, Rh I, Pd I, In I, Sn I, Cs I, Ba I, Ba II, Hf I, W I, Os I, Ir I, Tl I, Pb I.

        \subsection{Phase-resolved absorption signatures}\label{subsec:Line Velocities}
                We repeat the fitting procedure described in Section \ref{subsec:cross-correlation} but present each CCF in two distinct phase-binning schemes:
                \begin{enumerate}
                    \item We divide the 1D CCF into an ingress-inclusive first half of transit bin $(T_{1C})$  and an egress-inclusive second half of transit bin $(T_{C4})$ at the $K_{p, \text{expected}}$ slice. We plot these along with the full-transit CCF to get the broadest view of any asymmetries between the leading and trailing limbs of the planet. We use the same arm as the full-transit CCFs and fix $K_p=K_{p, \text{expected}}$. This is a common way to present phase-resolved CCFs, but it washes out the asymmetric signals from the limbs \citep{Savel2023}.
                    \item We confine each CCF function to an ingress bin ($T_{12}$), egress bin ($T_{34}$), three bins in an ingress-excluding first half of transit bin $(T_{2C})$, and three bins in an egress-excluding second half of transit bin $(T_{C3})$. We use the same arm for phase-resolved CCFs as the full-transit CCF, but we do not apply the stringent detection criteria given in Section \ref{subsec:DetectionCriteria} to each bin. Instead, we fix $K_p=K_{p, \text{expected}}$ and fit a Gaussian to the SNR peak at $|\Delta V| \leq 15$ km s$^{-1}$ in each bin. By dividing $T_{23}$ into six bins, we seek to show the behavior of high SNR species during transit with a higher time resolution, more in line with theoretical models. We plot the CCF amplitude and $\Delta V$ as a function of phase, observing the change in line velocity between bins. 
                \end{enumerate}
     The results of each method are presented in Section \ref{subsubsec:fit-to-absorption-trace}. We examined the impact of the Doppler shadow removal on the results of the second phase-binning scheme in Appendix \ref{app:doppler-shadow-impact}.
       
\section{Results and Discussion}\label{sec:Results}        
    \subsection{Detections}\label{subsec:Detections}
\begin{figure*} 
            \centering
            \gridline{
                \includegraphics[width=0.4\textwidth]{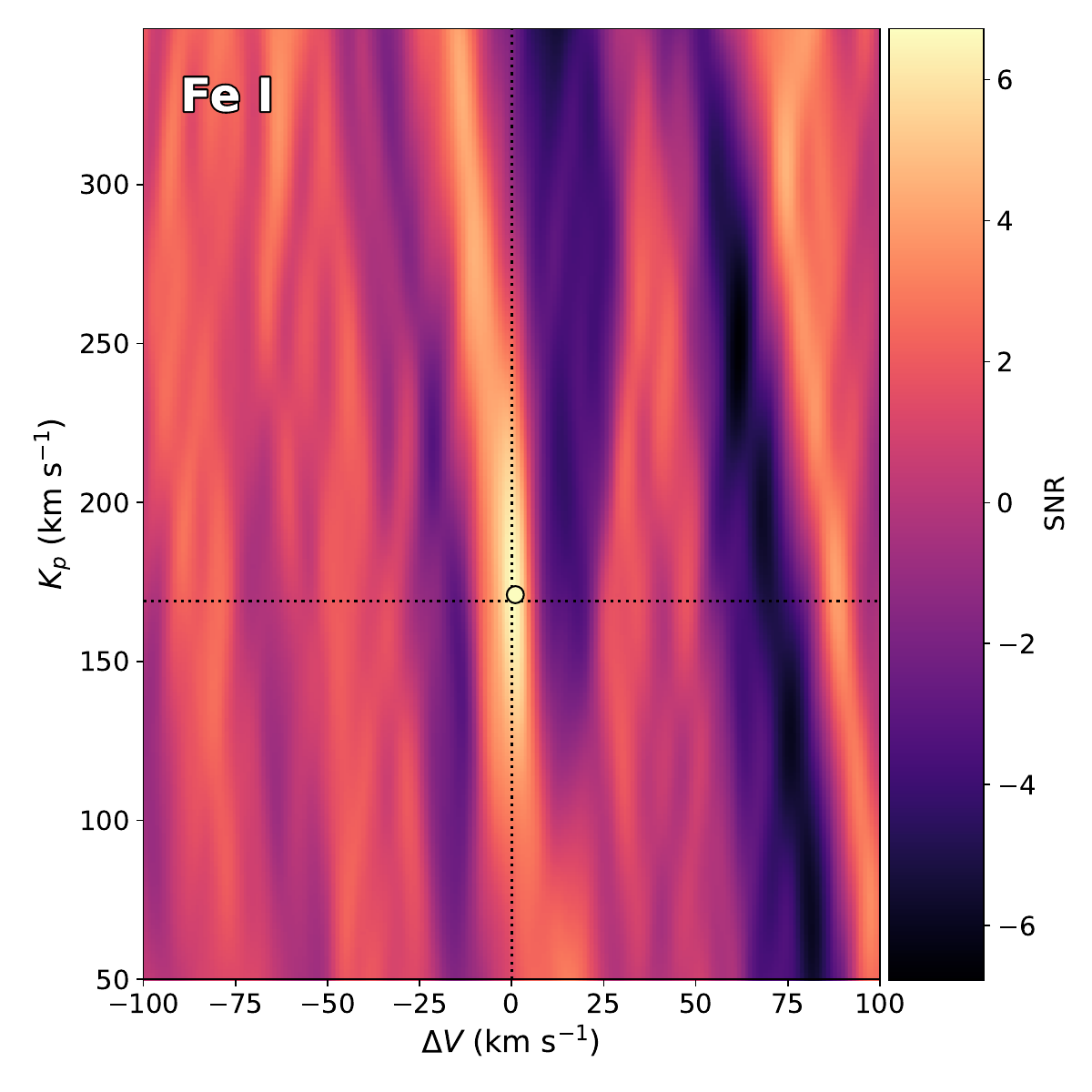}
                \includegraphics[width=0.4\textwidth]{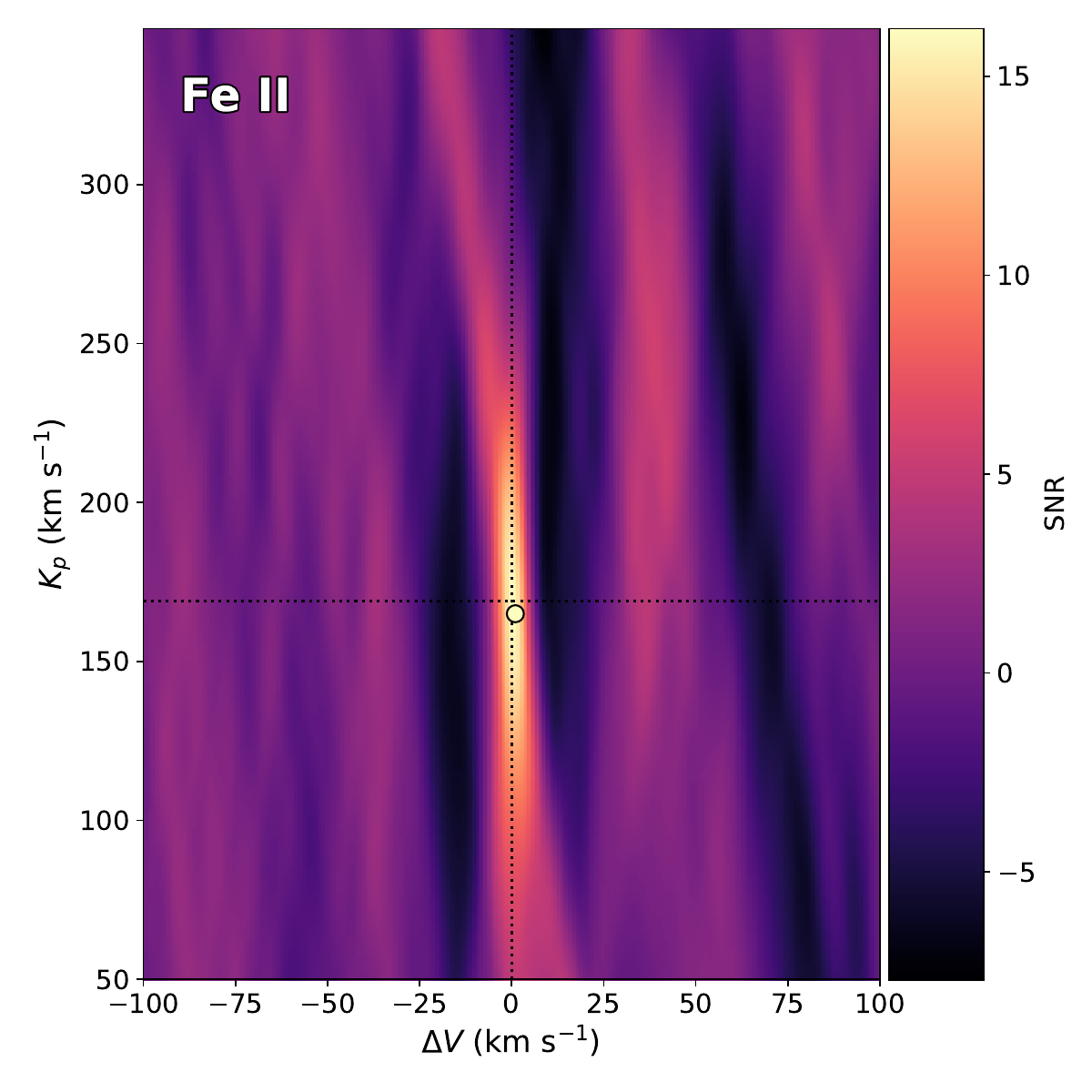}
                }
               \gridline{
                \includegraphics[width=0.4\textwidth]{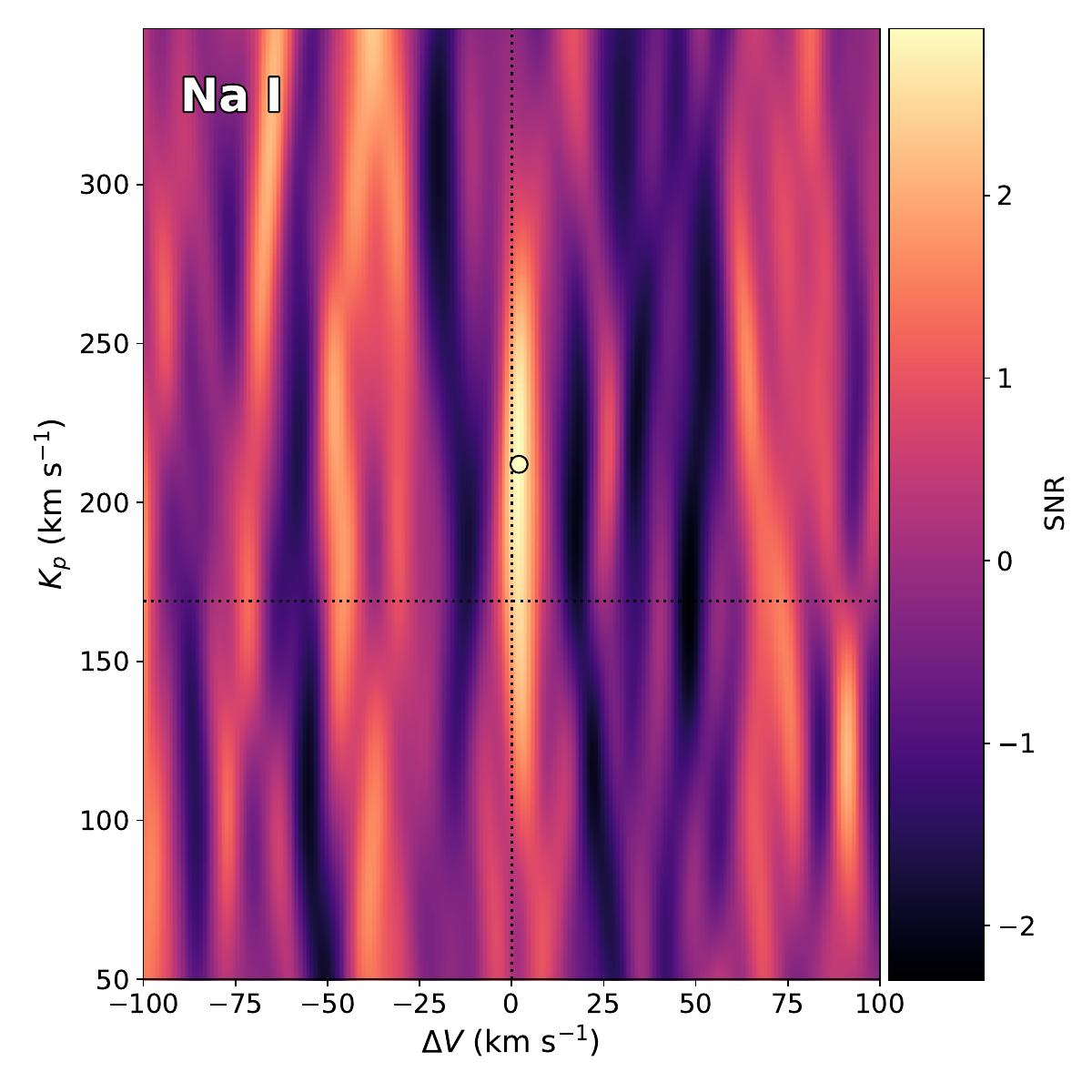}
                \includegraphics[width=0.4\textwidth]{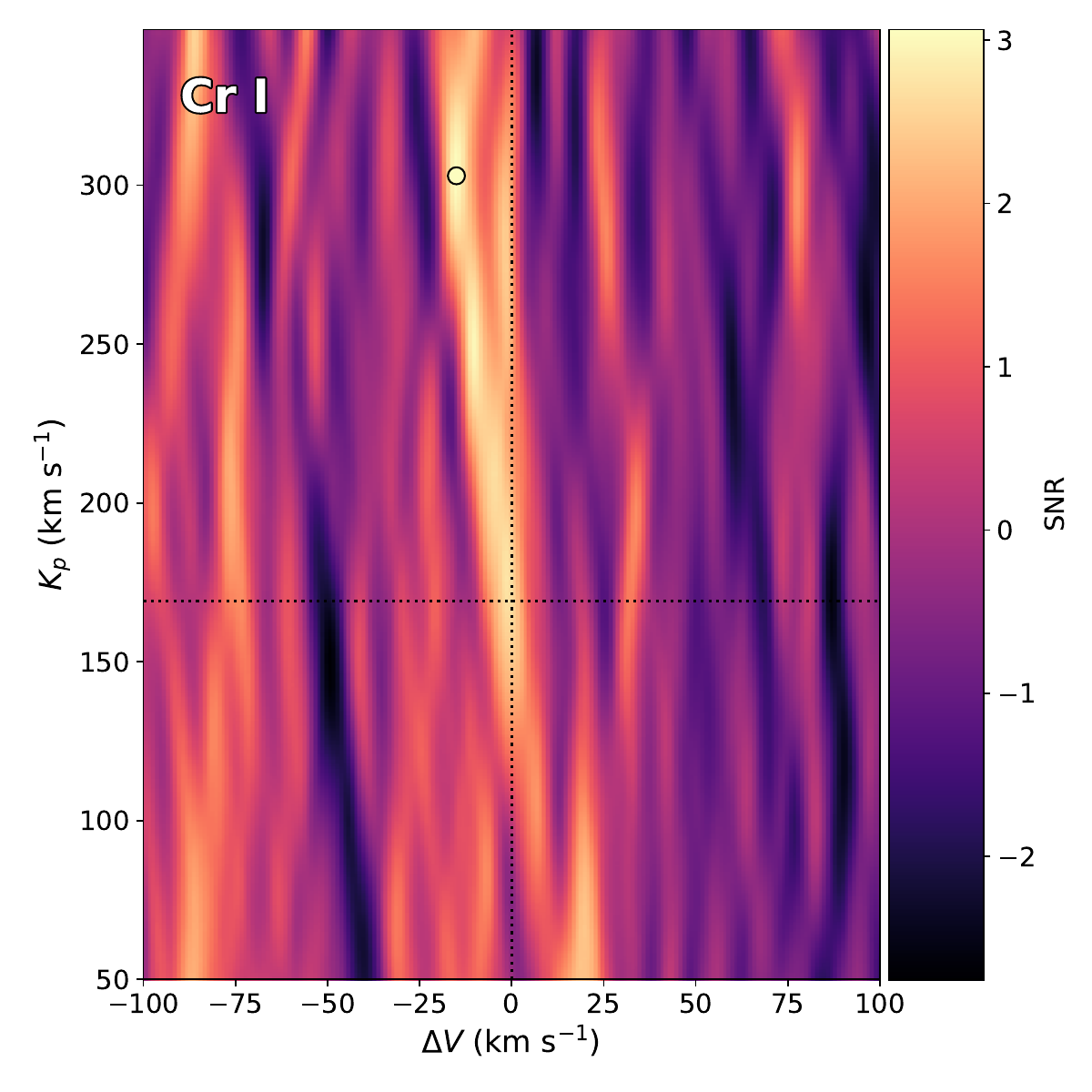}
                }
            \caption{Combined arm 2D full-transit CCFs of species exceeding the tentative detection threshold. \textit{Top left:} Fe I. \textit{Top right:} Fe II. \textit{Bottom left:} Na I. \textit{Bottom right:} Cr I. }
            \label{fig:2d-ccfs}
        \end{figure*}

        We confirm detections of Fe I and Fe II and tentative detections of Cr I and Na I (see Table \ref{tab:detection-summary} and Figure \ref{fig:2d-ccfs}). The raw CCFs, 1D CCFs, 2D CCFs, and aliases with the Fe I template are shown for each detection and tentative detection in Appendix \ref{app:Detections}.
        
      Along with a majority of other observations, we detect Fe I and Fe II with high confidence. We find a blueshifted Fe I velocity offset $\Delta V_{\text{Fe I}} = -1.0 \pm 0.7$ km s$^{-1}$. All studies that detected Fe I \citep{Stangret2020, Hoeijmakers2020, Nugroho2020, Rainer2021} reported a greater blueshift than ours, ranging from $-6.3$ to $-3.6$ km s$^{-1}$. 
      
      We find the velocity of Fe II to be consistent with zero, which agrees with the CARMENES data of \citep{Nugroho2020}. Each of \citet{CasasayasBarris2019, Hoeijmakers2020, BelloArufe2022, PaiAsnodkar2022b} find a blueshifted Fe II velocity offset. \citet{Nugroho2020} reports a blueshifted Fe II velocity offset with the HARPS-N spectrograph. The \citet{Nugroho2020} observations, taken on different nights, suggest variability between observations and/or spectrographs, although the cause---whether physical, instrumental, or otherwise---remains unclear.

        We tentatively detect redshifted $\Delta V_{\text{Na I}} = 2.5 \pm 1.3$ km s$^{-1}$, which is inconsistent with all other Na I detections that reported blueshifts and were discrepant by $\sim 2-5$ km s$^{-1}$ from our reported value \citep{CasasayasBarris2020, Hoeijmakers2020, Nugroho2020, Langeveld2022, Sicilia2022}.
        
        Cr I is a novel tentative detection that is significantly blueshifted $\Delta V_{\text{Cr I}} = - 6.3 \pm 1.9$ km s$^{-1}$.
        
       We do not detect ions Ca II and Cr II at $\geq 3\sigma$, which were detected in several other studies. Ca II's lines fall outside of the PEPSI bandpass, unlike CARMENES \citep{CasasayasBarris2019, Nugroho2020} and HARPS-N \citep{Nugroho2020}. We also do not recover molecular species FeH, which was tentatively detected in a multi-planetary survey by \citet{Kesseli2020}.
    
            \begin{deluxetable*}{ccccccccc}[]\label{tab:detection-summary}
            \tabletypesize{\footnotesize}
            \tablecaption{Results of high-resolution line-by-line and cross-correlation transmission spectroscopy species detections in the atmosphere of KELT-20b. Note that $\Delta V$ is the full-transit averaged 1D CCF Gaussian velocity offset, so phase-resolved data is not present in this table. The source letters refer to the following papers: (a) \citet{CasasayasBarris2019}, (b) \citet{Stangret2020}, (c) \citet{Hoeijmakers2020}, (d) \citet{Nugroho2020}, (e) \citet{Kesseli2020}, (f) \citet{Rainer2021}, (g) \citet{Langeveld2022}, (h) \citet{Sicilia2022}, (i) \citet{BelloArufe2022}, (j) \citet{PaiAsnodkar2022b}, (k) \citet{Lund2017}, (m) \citet{Talens2018}, (n) \citet{Petz2023}.}
            \tablehead{
                \colhead{\textbf{Source}} 
                & \colhead{\textbf{Spectrograph}}  & \colhead{\textbf{$N_{\text{obs}}$}} &
                \colhead{\textbf{$v_{\text{sys}}$ ($\text{km}\,\text{s}^{-1}$)}}&
                \colhead{\textbf{Species}} & \colhead{\textbf{SNR ($\sigma$)}} & \colhead{\textbf{$\Delta V$ ($\text{km}\,\text{s}^{-1}$)}} & \colhead{\textbf{FWHM ($\text{km}\,\text{s}^{-1}$})} & \colhead{\textbf{$K_p$ ($\text{km}\,\text{s}^{-1}$)}}
            }
            \startdata
                This work & PEPSI & 1 & $-22.78\pm0.11$\textsuperscript{(n)} & Na I & $3.4$  & $2.5 \pm 1.3$ & $8.7 \pm 0.5$  & $212^{+42}_{-59}$ \\
                & & & & Cr I& $3.3$   & $-6.3 \pm 1.9$  & $15.4 \pm 1.8$ & $252^{+2}_{-98}$ \\
                & & & & Fe I & $11.9$   & $-1.0 \pm 0.7 $  & $15.8 \pm 0.6$ & $172^{+39}_{-4}$ \\
                & & & & Fe II & $23.7$ & $ 0.0\pm 0.5$ & $9.8 \pm 0.3$ & $169^{+21}_{-1}$ \\
                \hline
                (a) & HARPS-N& 3 & $-21.07\pm 0.03$\textsuperscript{(m)} & Na I & --- & $-3.1 \pm 0.9$ & $9.2 \pm 2.0$ & $176.6 \pm 11.7$ \\
                & & & & Fe II & --- & $-2.8 \pm 0.8$ & $7.2 \pm 1.2$ & $174.4 \pm 14.0$ \\
                & & & & $H_{\alpha}$ & --- & $-3.0 \pm 1.2$ & $19.0 \pm 1.6$ & $165.6 \pm 16.7$ \\
                & & & & $H_{\beta}$ & --- & $-1.2 \pm 1.4$ & $19.4 \pm 2.5$ & $136.2 \pm 18.6$ \\
                & & & & $H_{\gamma}$ & --- & $-2.3 \pm 2.7$  & $16.6 \pm 4.2$ & $135.0 \pm 34.8$ \\
                & CARMENES & 1 &  & Na I & --- & $-3.2 \pm 1.7$ & $8.0 \pm 1.2$ & $182.5 \pm 14.3$ \\
                & & & & Ca II & --- & $-1.9 \pm 0.6$ & $9.2 \pm 1.0$ & $157.7 \pm 8.2$ \\
                & & & &  $H_{\alpha}$ & --- & $-4.5 \pm 0.5$ & $22.6 \pm 0.9$ & $166.2 \pm 7.4$ \\
                (b) & HARPS-N & 3 & $-21.07\pm 0.03$\textsuperscript{(m)} &  Fe I & $10.5 \pm 0.4$& $-6.3\pm 0.8$ & $15.1 \pm 0.6$ & $121^{+86}_{-29}$\\
                & & & & Fe II & $8.6 \pm 0.5$  & $-2.8 \pm 0.8$  & $8.5\pm0.6$ & $155^{+18}_{-18}$ \\
                (c) & EXPRES & 1 & $-23.30\pm 0.40$\textsuperscript{(k)} & Mg I & $3.33$ & $-8.40 \pm 1.40$ & $33.45 \pm 3.30$ & 175 \\
                & & & & Na I & $3.40$ & $-4.38 \pm 0.54$ & $10.07 \pm 1.26$ & $175$ \\
                & & & & Fe I & $3.45$ & $-4.81 \pm 0.72$ & $12.39 \pm 1.69$ & $175$ \\
                & & & & Fe II & $4.60$ & $-0.75 \pm 0.37$ & $8.54 \pm 0.87$ & $175$ \\
                & & & & Cr II & $3.69$ & $-3.40 \pm 0.42$ & $5.31 \pm 0.99$ & $175$ \\
                (d) & HARPS-N & 3 & $-22.06\pm 0.35$ & Na I & $7.72$ & $-1.4 \pm 0.7$ & ---& $180.0 \pm 11.8$ \\
                & & & & Fe I & $14.30$ & $-3.6 \pm 0.3$ & ---& $200.1 \pm 5.2$ \\
                & & & & Fe II & $14.61$ & $-1.4 \pm 0.2$ &--- & $165.0 \pm 3.5$ \\
                &  & & & Ca II & $7.53$ & $2.2 \pm 1.2$ & ---& $138.5 \pm 15.3$\\
                & CARMENES &  1 & $-22.02\pm 0.47$ & Na I  &$6.83$ & $-0.6 \pm 0.7$ & --- & $167.3 \pm 12.1$\\
                &          &    & & Fe I & $6.44$ & $-6.5 \pm 0.6$ & --- & $263.1 \pm 8.8$\\
                &          &    & & Fe II &$3.60$ & $-0.4 \pm 0.6$ & --- & $139.2 \pm 2.5$\\
                &          &    & & Ca II &$8.60$ & $0.1 \pm 0.6$ &  --- & $174.8 \pm 8.2$ \\
                (e) & CARMENES & 1 & $-23.3 \pm 0.3$\textsuperscript{(k)}& FeH & $3.02$ & $-0.5$ & --- & $158$ \\
                (f) & HARPS-N  & 5 & $-24.48 \pm 0.04$ & Fe I & --- & $-4.7^{+0.3}_{-0.7}$ & --- & $147^{+7}_{-6}$\\
                (g) & HARPS-N & 3 & --- & Na I & --- & $-3.1 \pm 1.9$ & --- & --- \\
                (h) & HARPS-N & 3 & --- & Na I D$_1$ & --- & $-3.8^{+0.7}_{-0.7}$ & $12.5^{+2.9}_{-2.3}$ & $192.5^{+12.4}_{-13.5}$\\
                & & & & Na I D$_2$ & --- & $-3.7^{+0.9}_{-0.9}$ & $14.8^{+2.8}_{-2.4}$ & $170.0^{+14.9}_{-13.5}$\\
                & & & & $H_{\alpha}$ & --- & $-5.2^{+1.8}_{-1.7}$ & $27.9^{+3.7}_{-3.3}$ &$156.3^{+28.5}_{-27.0}$ \\
                (i)  & FIES & 2 & $-23.2 \pm 0.4$ &  Fe II & $5.3$& $-20^{+6}_{-6}$ & --- & $120^{+78}_{-46}$\\
                (j) & (PEPSI, HARPS-N) & (1,3) & $-22.0978^{+0.4540}_{-0.4541}$ & Fe II & --- & $\sim-2.0^{+1.0}_{-0.7}$ & --- & $168.8^{+21.0}_{-12.4}$\\
            \enddata
            \end{deluxetable*}

        \subsection{Discrepancies between KELT-20b studies}\label{subsec:discrepancies}

    \subsubsection{Systemic velocity offset}
\begin{figure*}
                \centering
                \gridline{
                \includegraphics[width=0.48\linewidth]{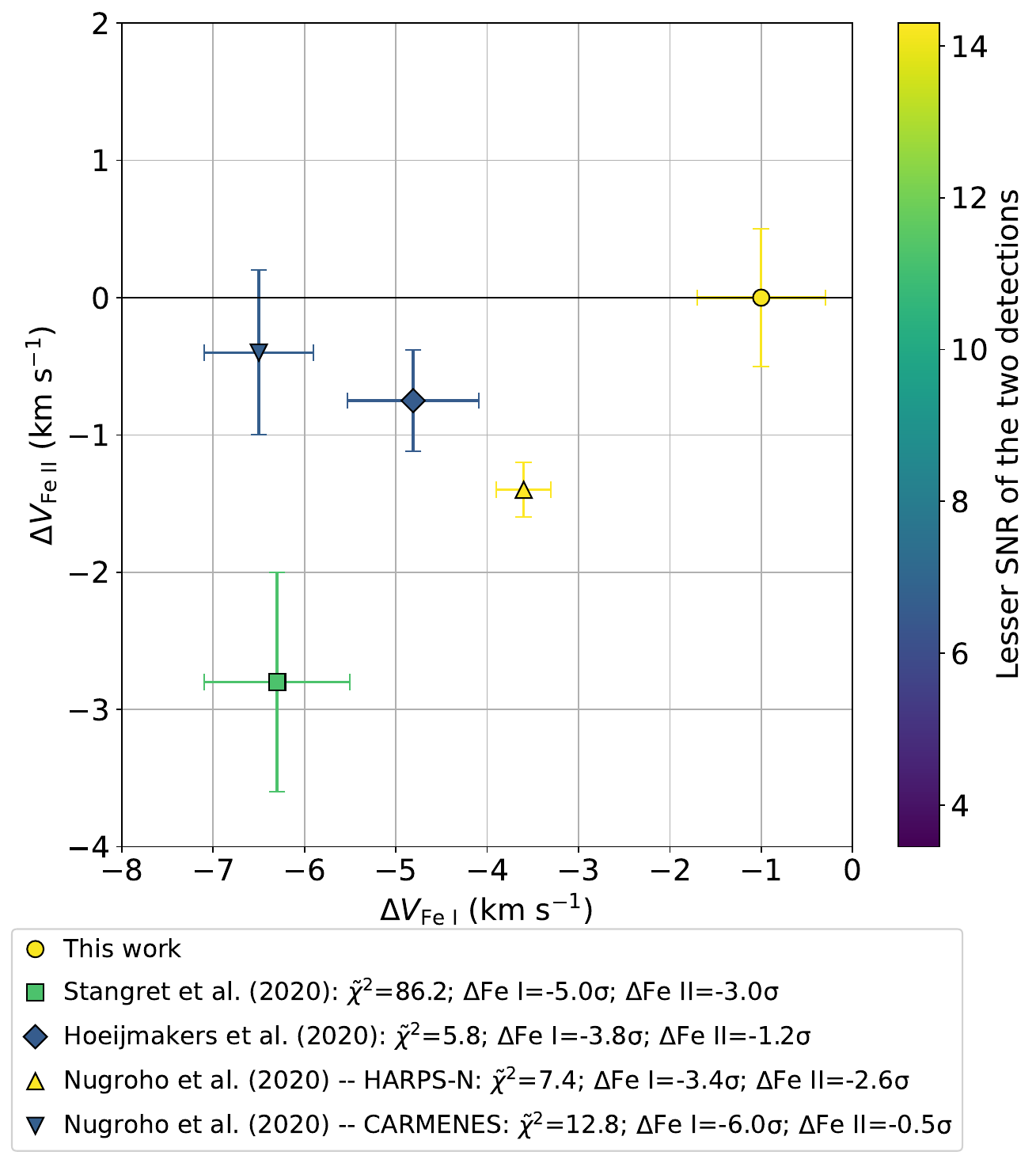}
                \includegraphics[width=0.48\linewidth]{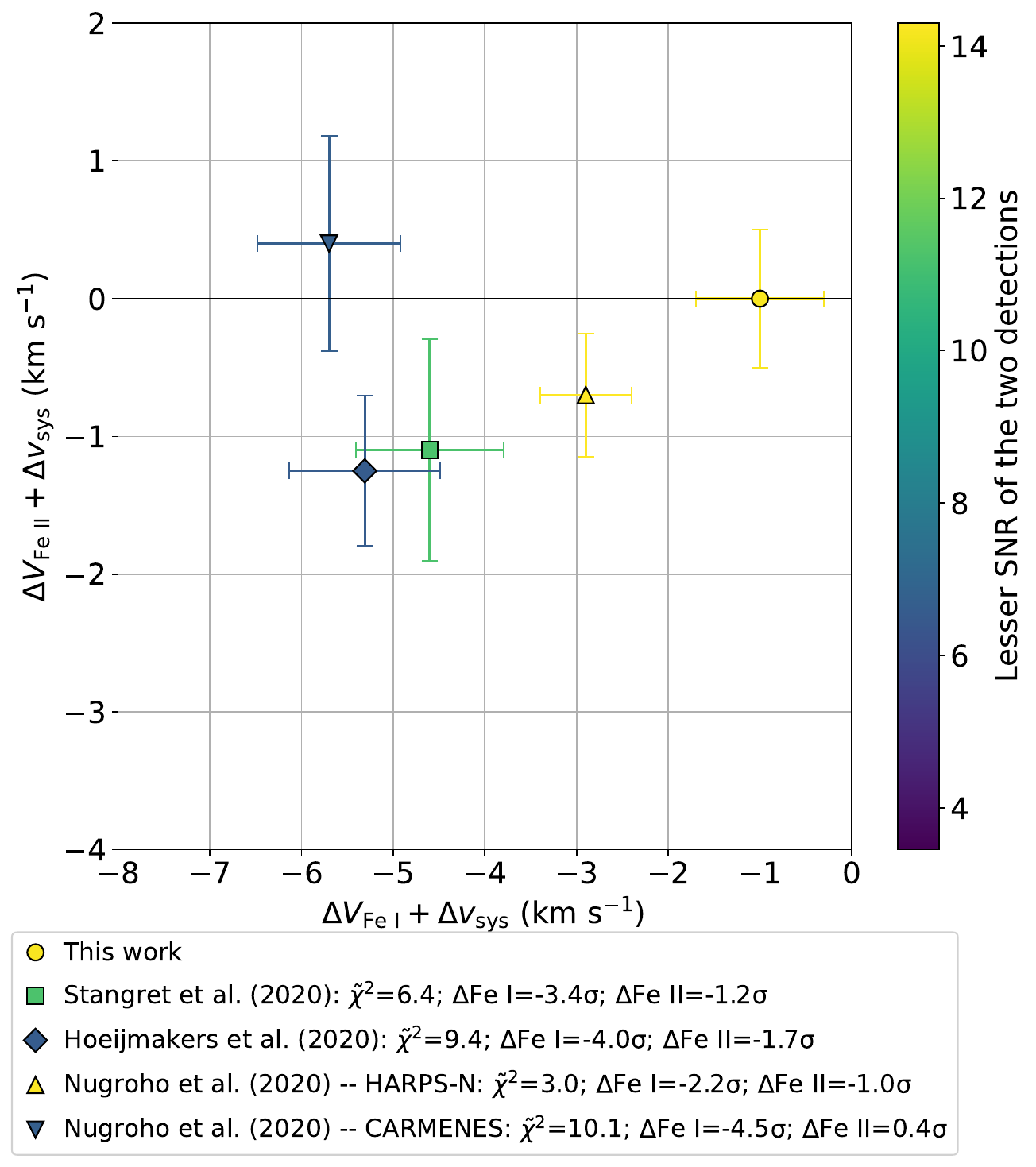}
                }
           \caption{\textit{Left:} Scatter plot of the Fe I and Fe II full-transit binned velocity offsets in studies that detect both species. We find the lowest wind speeds for each species, but the blueshift of Fe I relative to Fe II is somewhat preserved: Fe I is blueshifted relative to Fe II by $\sim-1$ km s$^{-1}$, which is inconsistent with other studies. Fe II is consistent with zero, in agreement with CARMENES observations by \citet{Nugroho2020} and roughly in agreement with \citet{Hoeijmakers2020}. Our findings are inconsistent with the remaining studies of KELT-20b which detected both Fe I and Fe II. \textit{Right:} The same plot but with systemic velocity discrepancies $\Delta v_{\text{sys}}$ corrected for. Our Fe II velocity offset is consistent with all \citet{Nugroho2020} observations. In the legend, $\Delta{\rm Fe\,I}$ and $\Delta{\rm Fe\,II}$ give, for each study, the difference between that study's Fe I/Fe II velocity offsets and ours, expressed in units of the combined $1\sigma$ uncertainty. The quoted $\tilde{\chi}^{2}$ is the reduced chi-squared obtained by summing the squared tensions in Fe I and Fe II and dividing by the number of species.}
                \label{fig:FeI-FeII-offset}
            \end{figure*}

    We define a self-consistent $v_{\mathrm{sys}}$ as one measured from the stellar lines in the same dataset, and a self-inconsistent $v_{\mathrm{sys}}$ as one adopted from a discovery paper or another work. Differences in self-consistent measurements or self-inconsistent assumptions of $v_{\mathrm{sys}}$ can significantly affect reported full-transit velocity offsets and make studies of the same system difficult to reconcile. \citet{CasasayasBarris2018, Stangret2020, Hoeijmakers2020, Kesseli2020} each assume their $v_{\mathrm{sys}}$ from one of the two discovery papers, \cite{Lund2017, Talens2018}, rather than deriving it from the stellar lines. $v_{\mathrm{sys, Talens}}$ and $v_{\mathrm{sys, Lund}}$ differ by $2.2$ km s$^{-1}$ ($5.6\sigma$). Due to KELT-20's rapid rotation (see Table \ref{tab:parameters_summary}) and correspondingly broad stellar lines, accurately measuring $v_{\mathrm{sys}}$ is difficult. Our $v_{\mathrm{sys}}$ is inconsistent with $v_{\mathrm{sys, Talens}}$ by $15.0\sigma$ and with $v_{\mathrm{sys, Rainer}}$ \citep{Rainer2021} by $7.4\sigma$. The \citet{Lund2017} discovery paper and the remaining measurements are each $\leq 2\sigma$ discrepant with our $v_{\mathrm{sys}}$ (see Table \ref{tab:detection-summary}). Fe I and Fe II are among the strongest detections in most analyses of KELT-20b \citep{Stangret2020, Hoeijmakers2020, Nugroho2020}. Their relative velocity offsets in Figure \ref{fig:FeI-FeII-offset} are largely inconsistent before correcting for $\Delta v_{\mathrm{sys}}$, but after correction, the scatter does decrease somewhat. Therefore, we propose that $\Delta v_{\mathrm{sys}}$ must be carefully accounted for in comparative studies. We note that $\Delta v_{\mathrm{sys}}$ does not, by itself, account for the full $2-5$ km s$^{-1}$ offset seen in some studies. However, it is the most significant methodological discrepancy we can confidently correct for. The remaining differences are likely due to a combination of other factors, including different data reduction pipelines, telluric correction methods, or potential stellar/planetary variability between epochs, which are beyond the scope of this paper to deconstruct.

            \subsubsection{Checking reproducibility of this dataset}
            We note a discrepancy between our Fe II full-transit CCF velocity offset and the reported SNR-weighted median Fe II velocity offset of four different datasets (including this one), computed by both line-by-line and cross-correlation spectroscopy of $\sim-2.0^{+1.0}_{-0.7}$ km s$^{-1}$ by \citet{PaiAsnodkar2022b}. They also report their full-transit CCF velocity offset of Fe II with only the PEPSI data (the same data used in this study) as $\sim-1.1\pm0.4$ km s$^{-1}$ which is inconsistent by $1.6\sigma$ with our reported $\Delta V_{\text{Fe II}}= 0.0\pm 0.5$~km s$^{-1}$. Note that \citet{PaiAsnodkar2022b} only used blue arm data, whereas we report the full-transit velocity offset of the telluric-corrected red and blue arms, combined. When isolating our blue arm data, fixing $K_p=177~\text{km s}^{-1}$\,\,\footnote{ $K_p$ of KELT-20b's cross-correlation signal as measured by PEPSI and used for the Fe II cross-correlation analysis in \citet{PaiAsnodkar2022b} but not listed in the paper (A. Pai Asnodkar, personal correspondence, 2025 February 19). The $K_p$ listed for \citet{PaiAsnodkar2022b} in Table \ref{tab:detection-summary} was obtained from line-by-line Fe II analysis and was the KELT-20b $K_p$ presented in that paper.}, and correcting for the systemic velocity difference $\Delta v_{\mathrm{sys}}= 0.7\pm0.5$ km s$^{-1}$, we reduce the discrepancy to $0.2\sigma$, meaning that our results are consistent with theirs.
            
    \subsection{Dynamics}\label{subsec:Atmospheric Dynamics}
        \subsubsection{Assessing phase-dependence}\label{subsubsec:fit-to-absorption-trace}
\begin{figure}
                \centering
                \includegraphics[width=\linewidth]{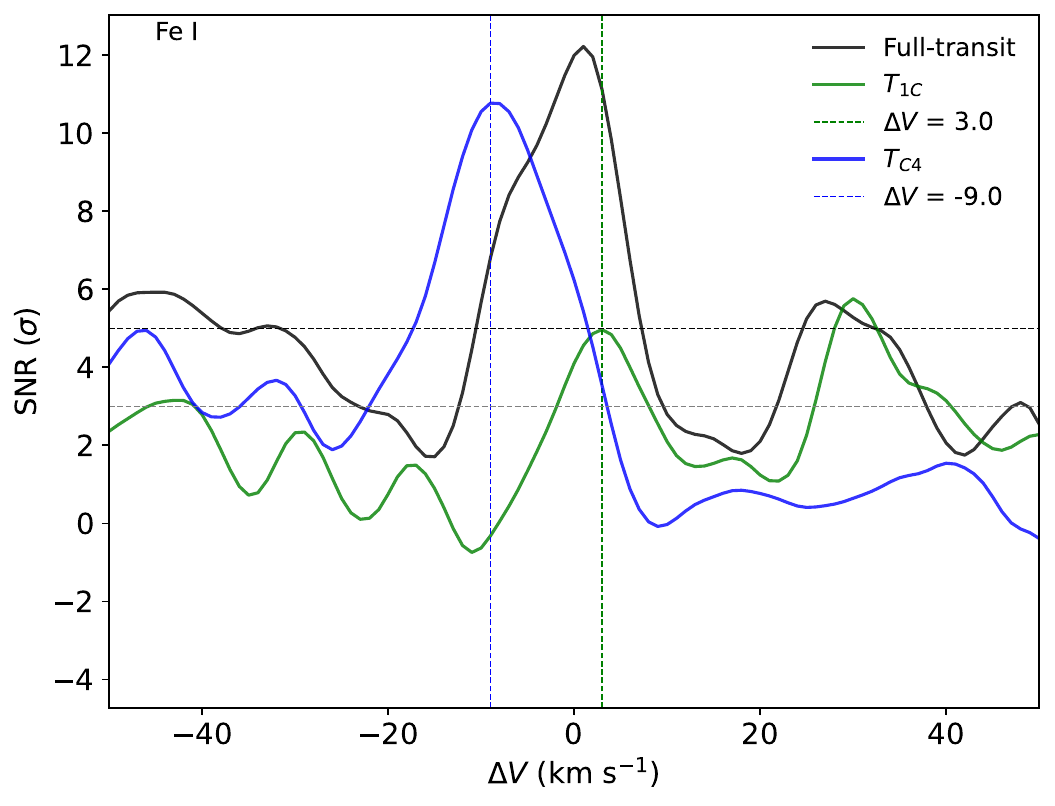}
               \includegraphics[width=\linewidth]{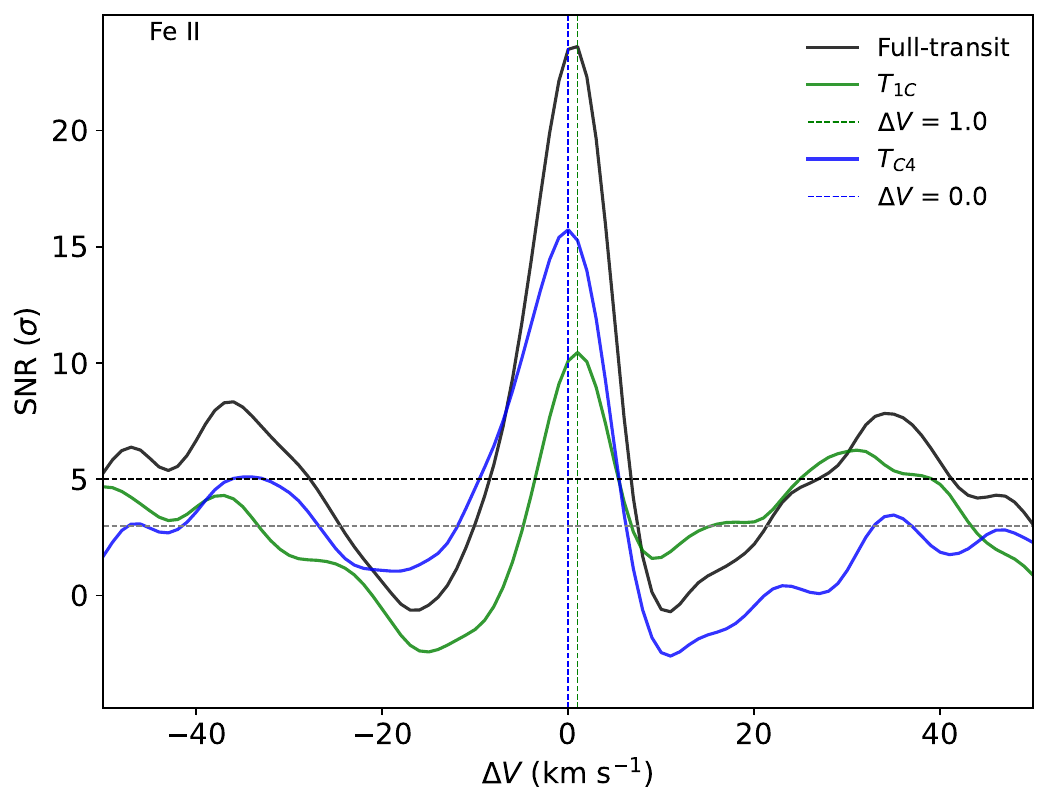}
                \caption{\textit{Top:} Coadded Fe I 1D CCFs from $T_{1C}$ in green, $T_{C4}$ in blue, and full-transit in black. The gray horizontal dashed line is the tentative detection threshold and the black horizontal dashed line is the detection threshold. The center of a Gaussian fit to the $T_{1C}$ and $T_{C4}$ 1D CCFs are displayed in the legend. \textit{Bottom:} The same plot for Fe II.}
                \label{fig:1d-ccfs-halves}
            \end{figure}

           To first check that the absorption traces vary with phase, we binned the first half $T_{1C}$ and second half $T_{C4}$ CCFs of Fe I and Fe II, selected the $K_{p,\text{expected}}$ slice, then plotted the 1D CCFs, as shown in Figure \ref{fig:1d-ccfs-halves}. Fe I exhibits clear blueshifting and confirms an asymmetrical absorption velocity prior to and post mid-transit. Fe II merely blueshifts by $-1~\text{km s}^{-1}$ and thus does not demonstrate phase-dependence in the half-transit binning scheme.
            
            Next, we compared a constant fit of the phase-binned CCFs to a linear fit. We empirically added a minimum error of $1/\sqrt{7}$~km~s$^{-1}$ in quadrature to each point to reduce the goodness of fit of the stronger Fe II fit $\chi^2_\nu \sim1$. We found that for Fe II, the linear fit was statistically favored over the constant fit\footnote{$\chi^2_{\nu,\text{linear}}=1.0$ and $\chi^2_{\nu,\text{constant}}=11.8$.}. We also propagated the $1/\sqrt{7}$~km~s$^{-1}$ error to the full-transit velocity offsets.
            
        \subsubsection{Features of Fe I and Fe II absorption traces}\label{subsubec:evidence-for-mechanisms}
\begin{figure*} 
                \centering
                    \includegraphics[width=0.48\textwidth]{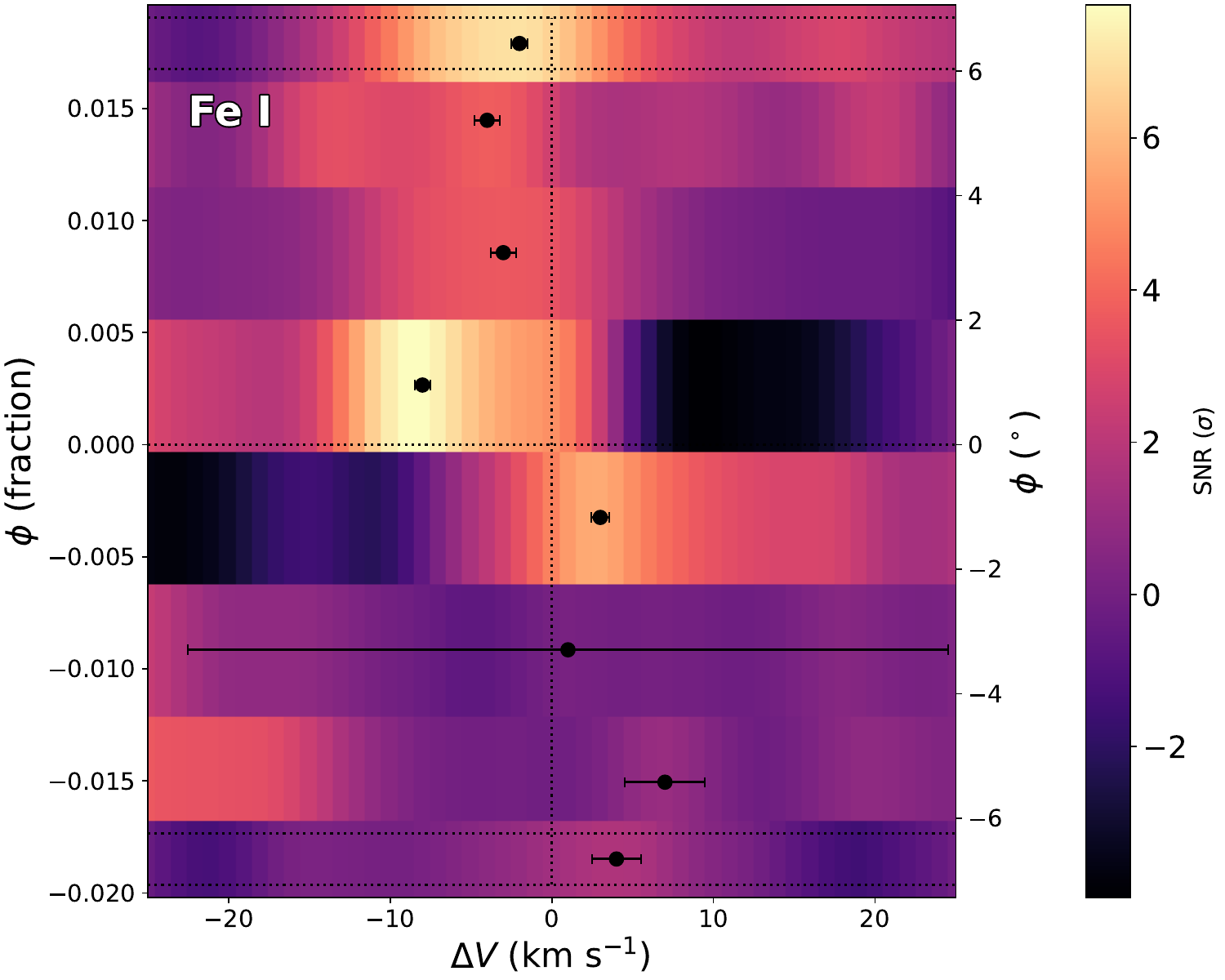}
                    \includegraphics[width=0.48\textwidth]{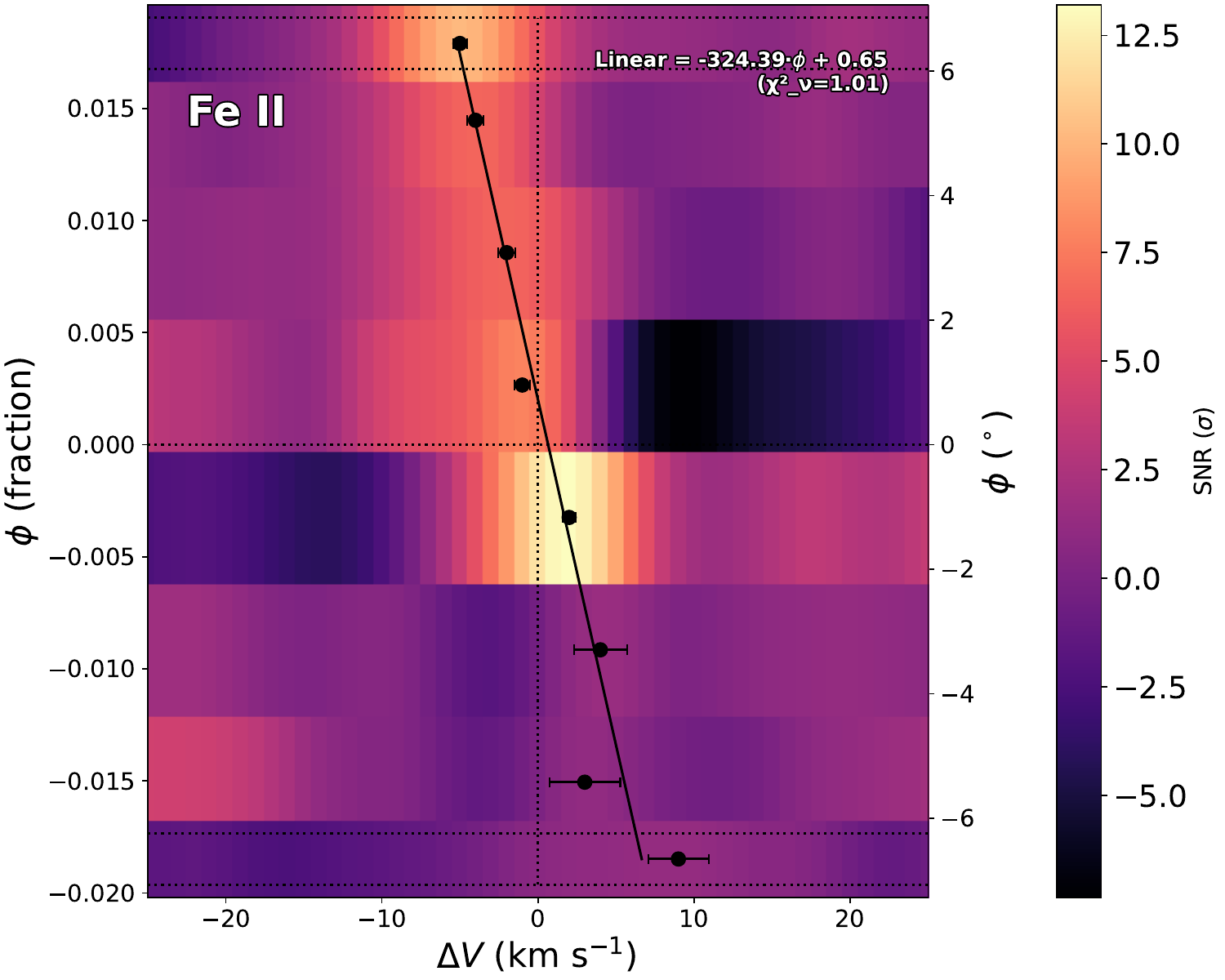}
                \caption{Phase-binned absorption traces of Fe I (left) and Fe II (right) with a linear fit. The parameters of the Fe II fit are shown in the top right. The horizontal dashed lines at the bottom delineate the ingress phase bin and those at the top delineate the egress phase bin.}
                \label{fig:phase-binned-1d-ccfs-Fe/Fe+}
            \end{figure*}

            Fe I and Fe II exhibit distinct phase-resolved absorption signatures, as can be seen in Figure \ref{fig:phase-binned-1d-ccfs-Fe/Fe+}. To quantify the difference between their velocity offsets, we calculated the reduced chi-squared statistic given by
            \begin{equation}
                \chi^2_\nu= \frac{1}{\nu}\sum_{n=1}^{N_{\text{bins}}} \frac{\left(\Delta V_{\text{Fe II},n} - \Delta V_{\text{Fe I},n}\right)^2}{\sigma_{\Delta V_{\text{Fe II},n}}^2 + \sigma_{\Delta V_{\text{Fe I},n}}^2}
            \end{equation}
            where $\nu=N_\text{bins}=8$. We obtained $\chi^2_\nu=20.1$, suggesting Fe I and Fe II are indeed probing distinct atmospheric regimes.
            
            Fe I's velocity is consistent with zero at ingress, redshifts somewhat up until $T_C$, then blueshifts. The largest blueshift is in the bin immediately following mid-transit, which then decreases and settles at a $-2$ km s$^{-1}$ blueshift by egress. Fe I's signal is weaker than Fe II's in each phase bin, but both signals grow stronger with phase. Fe I's signal is strongest at egress, where it exceeds $5\sigma$. 
            
            Fe II is significantly redshifted at ingress, then steadily blueshifts from mid-transit to egress where it reaches a blueshift of $-5$ km s$^{-1}$. Fe II's signal at the limbs is asymmetric in both signal strength and velocity. Egress SNR is $\sim7\sigma$ greater than ingress SNR, and the egress signal is blueshifted $-14$ km s$^{-1}$ relative to ingress. The Fe II signal is stronger than Fe I's at all phases, so it is considered the best phase-resolved trace of the atmospheric absorption signal available in this work.
            
        \subsubsection{Comparison with other phase-resolved studies of KELT-20b}
            Our results are largely inconsistent with other KELT-20b observations with phase-resolved results. The Fe and Fe II velocity offsets in Figure \ref{fig:1d-ccfs-halves} are asymmetric between first half and second half of transit, unlike those in \citet{Hoeijmakers2020}. However, each of our Fe I and Fe II phase binned signal strengths increase in the second half of transit, which is consistent with \citet{Hoeijmakers2020}'s findings. 
            
            \citet{Rainer2021}'s findings conflict with ours entirely. They observed five transits: three of five absorption traces do not exhibit a net Doppler shift, the other two redshift with phase. SNRs are consistent with each other in four of five transits, and one transit's signal decreases in the second half. \citet{Rainer2021} also combines the five transits together and does not find a velocity shift or signal strength asymmetry. These discrepancies again motivate direct comparisons to KELT-20b-specific GCMs (e.g., \citealt{Chachan2025})---ideally post-processed into phase-resolved, high-resolution transmission predictions---and a continued observational campaign with rigorous detection methodology.

       \subsubsection{Limb asymmetry and potential driving mechanisms}\label{subsubsec:limb-asymmetry}
       At egress, both Fe I and Fe II are blueshifted relative to ingress, which indicates day-to-nightside winds are present. We assume that as the hotter trailing limb rotates into view at egress, the transmission spectra probes a hotter region than at ingress. Thus, species with ions that have sufficient absorption within the PEPSI bandpass that are mostly unionized in the leading limb are expected to increase in signal strength as the trailing limb comes into view. Using the \citet{Guillot2010} P-T profile from \citet{Johnson2023} in \code{petitRADTRANS}, we collect the temperatures corresponding to $10^{-3}$ bar and $10^{-6}$ bar, as well as $T_{\text{eq}}$ from \citet{Lund2017}. Using FastChem, we calculate the ionization fraction to connect the relative signal strength of each species during transit to the range of temperature regions probed by the spectra. As mentioned in Section \ref{subsec:Model Spectra}, the P-T profile is unlikely to be particularly accurate, but we can rudimentarily constrain the regimes probed at ingress and egress.
        
\begin{figure}
                \centering
                \includegraphics[width=\linewidth]{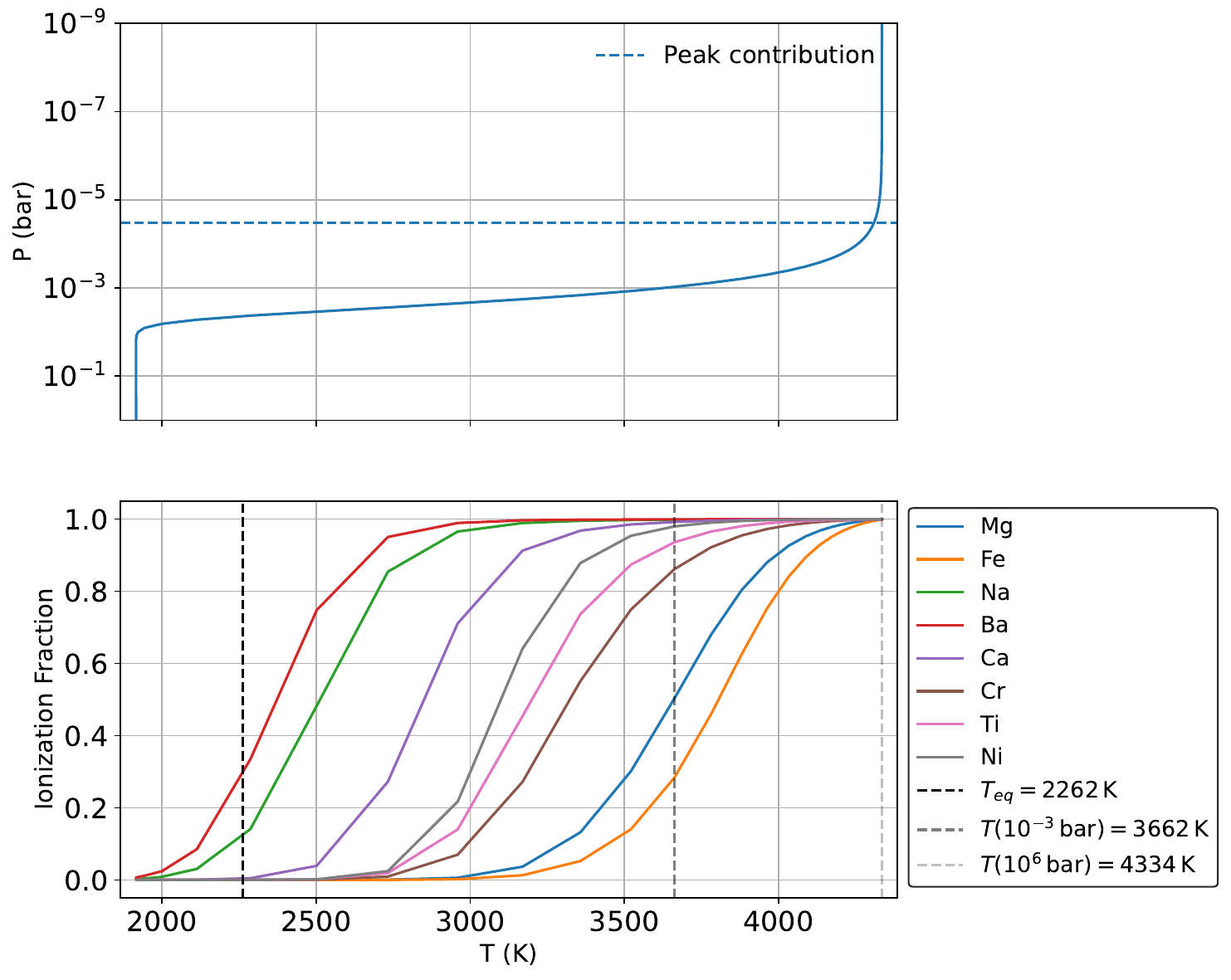}
                \caption{\textit{Top:} P-T profile used in this work. The horizontal dashed line shows the estimated pressure at which the transmission contribution function peaks. \textit{Bottom:} Ionization fraction for the atoms deemed detectable. We are using \citet{Asplund2021} extended solar abundances. Vertical lines correspond to $T_{\text{eq}}$, temperature in the P-T profile at $10^{-3}$ bar and temperature at $10^{-6}$ bar, in ascending order.}
                \label{fig:ionization-fraction}
            \end{figure}

             By $T(10^{-6}~\text{bar})=4300$ K, Fe is completely ionized, as seen in Figure \ref{fig:ionization-fraction}. However, Fe I's peak SNR is at egress, suggesting the temperature at trailing limb is lower than $T(10^{-6}~\text{bar})$. We again note that Fe II's signal is stronger at egress than ingress as well, yet it still exceeds an SNR of $3\sigma$ in the ingress bin, suggesting a lower limit of $T=3000$ K at the ingress limb. Note that we used a simplified P-T profile that assumes isothermal structure from the quench point until the thermal inversion, where it is linear in log space, then isothermal up to $T(10^{-6}~\text{bar})$ \citep{Johnson2023}. Nevertheless, this limb asymmetry suggests an interplay of asymmetry drivers, including increased scale height (further discussed in Section \ref{subsubsec:comparison-with-theory}).
    
            More broadly, our absorption signature is qualitatively similar to WASP-76b's in that it blueshifts with phase and that signal strength significantly increases in phase bins after mid-transit, as evidenced in \citet{Ehrenreich2020, Kesseli2022}. A Ba II detection can also rudimentarily constrain the magnetic field strength when compared to a strong neutral species detection such as Fe I \citep{Savel2024}. For Ba II, we obtain a $2.9\sigma$ peak in the $K_p-\Delta V$ map at $-6.4$ km s$^{-1}$. However, there are several peaks of similar SNR across the map which violates our detection criteria, so we do not discuss this signal any further.

        \subsubsection{Comparison to theory}\label{subsubsec:comparison-with-theory}        
            
            In both Fe I and Fe II, the net redshift extends past the ingress bin. The absorption trace does not cross over to a net blueshift until after mid-transit in both species (See Figure \ref{fig:phase-binned-1d-ccfs-Fe/Fe+}). The Fe I absorption trace also jumps after mid-transit to a blueshift, followed by a "bottoming out" of the signal, similar to the \citet{Ehrenreich2020} WASP-76b trace. This behavior is consistent with the absorption signature of the \citet{Beltz2023} active drag GCM, which includes a $3$ G dipolar magnetic field aligned with the polar axis on WASP-76b. This detail is not present in the drag-free and uniform drag GCMs, making it a unique trait of the more computationally complex $3$ G active drag treatment of magnetic effects that localizes the magnetic drag timescale to regions of the atmosphere.

            Our phase-resolved morphology is also qualitatively similar to that predicted by \citet{Simonnin2024}'s GCM for TOI-1518b which incorporates drag, albeit uniform. Fe II is assumed to be probed at a lower pressure than Fe I, and in \citet{Simonnin2024}'s strong uniform drag model, wind speeds decrease in the upper atmosphere. This causes a continuous atmospheric trace of Fe II in the $T_{23}$ region, whereas the Fe I trace jumps to a blueshift after $T_C$. This result is consistent with our own, as can be seen in Figure \ref{fig:phase-binned-1d-ccfs-Fe/Fe+}. This again suggests magnetic drag effects are visible in our results, but it does require theoretical corroboration. 
            
            Distinct behavior between Fe I and Fe II suggests we are probing multidimensional wind structures. Fe II is most likely probed at a lower pressure region than Fe I since it has greater speed at ingress and egress and a stronger signal.
            
            \citet{Savel2023} models a 2200 K UHJ atmosphere and differentiates the impact of the scale height effect and thermal ionization on abundance asymmetries between leading and trailing limbs. Notably, \citet{Savel2023} finds that when isolating thermal ionization, most neutral species abundances are slightly greater in the leading limb. This behavior is consistent with a decrease in neutral abundances due to thermal ionization, as discussed in Section \ref{subsubsec:limb-asymmetry}. In \citet{Savel2023}'s model, the scale height effect is dominant, forcing nearly all species' signals to be stronger in the trailing limb. In Figure \ref{fig:phase-binned-1d-ccfs-Fe/Fe+}, Fe I has a considerably stronger signal in the latter half of transit, suggesting that the scale height effect is the dominant asymmetry driver in our observation. However, the S/N levels of the spectra, as shown in Figure \ref{fig:observation_quality}, are significantly higher in the second half of transit, which may also cause apparent signal strength differences. The S/N increase cannot exclusively be responsible for the signal strength asymmetry, given that the S/N increases $\sim20\%$ while the detection significance increases $\sim 250\%$.
            
            \citet{Chachan2025} obtained Hubble Space Telescope transmission spectra of KELT-20b and did not find significant limb asymmetry. They generated two GCMs: a drag-free model and an active $5$ G magnetic drag model, showing that the latter produces a nearly symmetric pressure-temperature structure, consistent with the observed limb symmetry. This result is certainly in tension with our own, suggesting that strong magnetic drag can suppress the limb asymmetry implied by our ground-based results. This apparent discrepancy comes with the caveat that more observations are required to make a strong claim for absent limb asymmetries, just as more observations are required to confidently claim asymmetry with our single ground-based observation.
            
        \section{Conclusions}\label{sec:Conclusions}
            This paper presented an analysis of the high-resolution cross-correlation transmission spectrum of KELT-20b using an observation obtained with the PEPSI spectrograph. KELT-20b has been studied extensively with transmission spectroscopy and cross-correlation transmission spectroscopy, revealing an abundance of absorbers. This study confirmed 3 prior detections and presented 1 new tentative detection from an inventory of 58 species. 
            
            Of the 4 species with tentative detections or detections, two (Fe I and Cr I) show clear net blueshfits, Fe II is consistent with a small blueshift or zero, and Na I shows a tentative redshift. Taken together with previous studies, the iron species continue to support the presence of day-to-night winds, although the Na I behavior remains puzzling. We presented stringent detection methodology to promote reproducibility and consistency in future HRCCTS studies and quantified discrepancies between systemic velocities of the discovery papers and observational works. When we correct for differences in the adopted systemic velocities between studies, the scatter in published Fe I and Fe II velocity offsets decreases and our Fe II offset becomes consistent with most previous measurements.

            We phase-resolved the CCFs into ingress, egress, and six intermediate bins, revealing limb asymmetries in Fe I and Fe II. We found that the SNR increases from $T_C$ to $T_4$. We found similarity between our absorption traces and those post-processed from \citet{Beltz2023, Simonnin2024} GCMs that incorporate atmospheric drag. We also showed that a simple linear blueshift with phase is a stronger fit to the Fe II absorption trace than a constant offset. Again, our results are largely inconsistent with the two other studies that presented phase-resolved traces for KELT-20b.

            Our phase-resolved results are broadly consistent with several dynamical mechanisms driving limb asymmetries, which motivates further theoretical treatment of KELT-20b---building on recent GCM work \citep{Chachan2025}---with non-ideal magnetohydrodynamics and explicit post-processing into phase-resolved absorption traces for comparison with this dataset and future observations. Likewise, our findings motivate an observational campaign of KELT-20b which combines time-series spectra from several high-resolution spectrographs across different wavelength ranges. With meticulous detection methodology imposed, a benchmark set of observations would resolve inconsistencies with previous studies and unite observation with theory.
            
\begin{acknowledgments}

    We thank the anonymous referee for a thorough and constructive report that increased the quality of this manuscript.
    
    CL thanks MCJ and the Summer Undergraduate Research Program in Astrophysics at The Ohio State University's Department of Astronomy and Center for Cosmology and AstroParticle Physics for the opportunity to take on this project.
    
    CL, MCJ, and JW are supported by NASA Grant 80NSSC23K0730.

    JW acknowledges support from the National Science Foundation under Grant No. 2143400.
    
    The LBT is an international collaboration among institutions in the United States, Italy, and Germany. LBT Corporation Members are: The Ohio State University, representing OSU, University of Notre Dame, University of Minnesota and University of Virginia; LBT Beteiligungsgesellschaft, Germany, representing the Max-Planck Society, The Leibniz Institute for Astrophysics Potsdam, and Heidelberg University; The University of Arizona on behalf of the Arizona Board of Regents; and the Istituto Nazionale di Astrofisica, Italy. Observations have benefited from the use of ALTA Center (\url{alta.arcetri.inaf.it}) forecasts performed with the Astro-Meso-Nh model. Initialization data of the ALTA automatic forecast system come from the General Circulation Model (HRES) of the European Centre for Medium Range Weather Forecasts.

    The PEPSI spectra used in this paper will be made publicly available through the NASA Exoplanet Archive\footnote{\url{https://exoplanetarchive.ipac.caltech.edu/docs/PEPSIMission.html}}.
\end{acknowledgments}

\begin{contribution}

CL contributed to conceptualization, data curation, formal analysis, methodology, software development, visualization, writing the original draft, and reviewing and editing the manuscript.

MCJ contributed to conceptualization, data curation, formal analysis, funding acquisition, investigation, methodology, software development, supervision, and reviewing the manuscript.

JW contributed to conceptualization, funding acquisition, methodology, supervision, and reviewing the manuscript.

APA contributed to software development and reviewing the manuscript.

SP contributed to methodology, software development, and reviewing the manuscript.

AD contributed to investigation.

KGS contributed to data curation, resources, and reviewing the manuscript.

II contributed to data curation and resources.
\end{contribution}

\facilities{LBT(PEPSI)}

    \software{astropy \citep{astropy}, 
            Jupyter \citep{jupyter},
            matplotlib \citep{Matplotlib},
            molecfit \citep{Kausch2015, Smette2015},
            numpy \citep{NumPy},
            petitRADTRANS \citep{petitRADTRANS},
            PyFastChem \citep{PyFastChem},
            uncertainties~\citet{uncertainties}
            }

\clearpage
\appendix

    \section{Detection criteria examples}
    We provide several examples of the detection criteria given in Section \ref{subsec:DetectionCriteria}. One template spectrum and 1D CCF combination meets the detection criteria and the other combination fails.
    
                \subsection{Example 2D CCF check}\label{sec:2d-ccf-check-example}
\begin{figure}[h]
                    \centering
                    \gridline{
                        \fig{KELT-20b.20190504.combined.Fe+.CCFs-shifted.pdf}{0.333\textwidth}{}
                        \fig{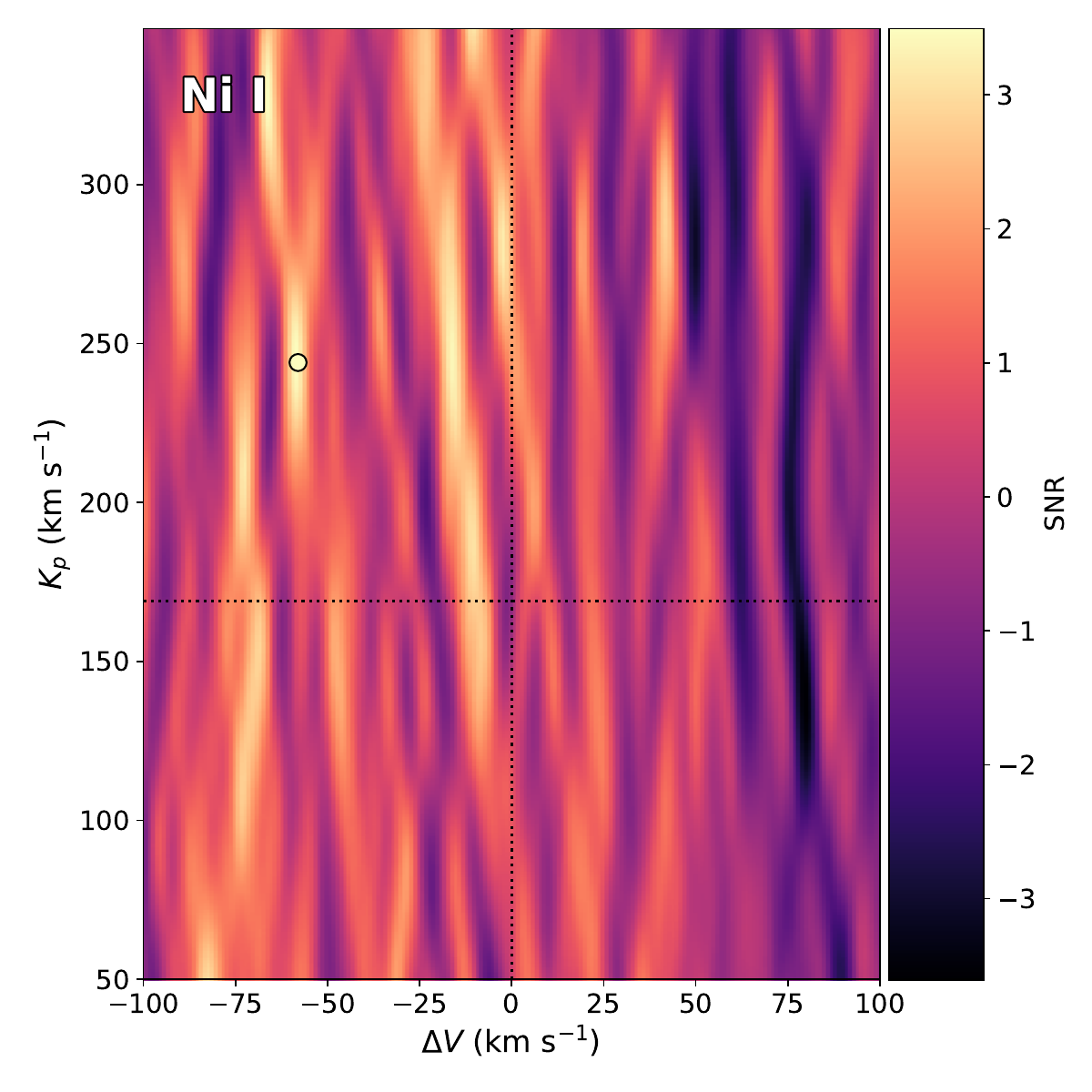}{0.333\textwidth}{}
                    }
                    \vspace{-10mm}
                    \caption{\textit{Left:} Example of a physical atmospheric signature within an Fe II 2D CCF shifted into the planetary rest frame, or $K_p-\Delta V$ space. 
                    \textit{Right:} Example of a likely unphysical 2D CCF map of Ni I. Note the peak signal occurs at a unphysical $\Delta V$ and $K_p$. Additionally, there are peaks throughout the map of a similar amplitude, indicating we are looking at noise. We do not consider this a detection.}
                    \label{fig:standard-shape}
                \end{figure}

                Figure \ref{fig:standard-shape} shows one physical and one unphysical 2D CCF.
            \clearpage

        \subsection{Example template spectra and 1D CCF check}\label{sec:template-spectra-and-1d-ccf-check-example}
\begin{figure*}[h]
                                \centering
                                \gridline{
                                    \fig{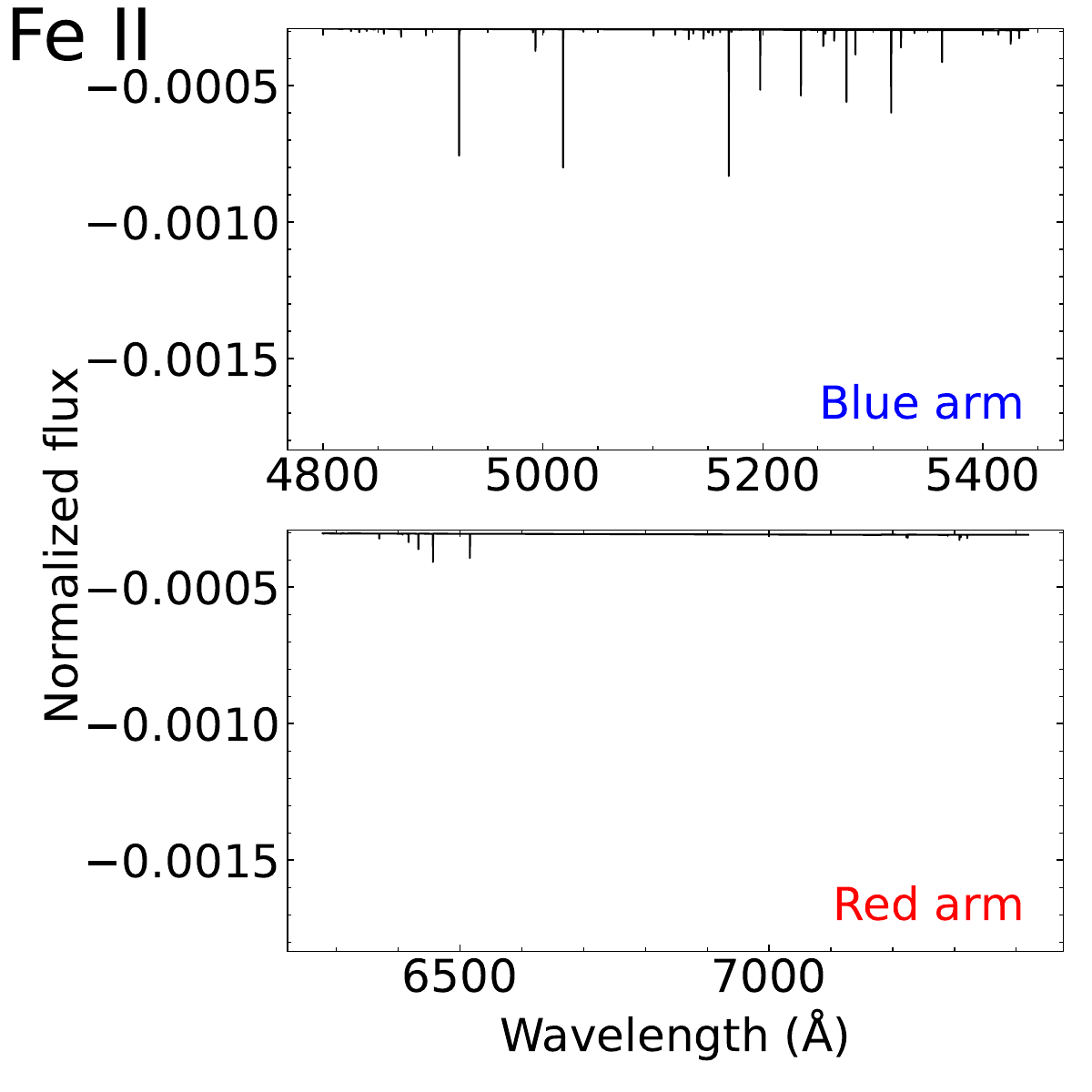}{0.345\textwidth}{}
                                    \fig{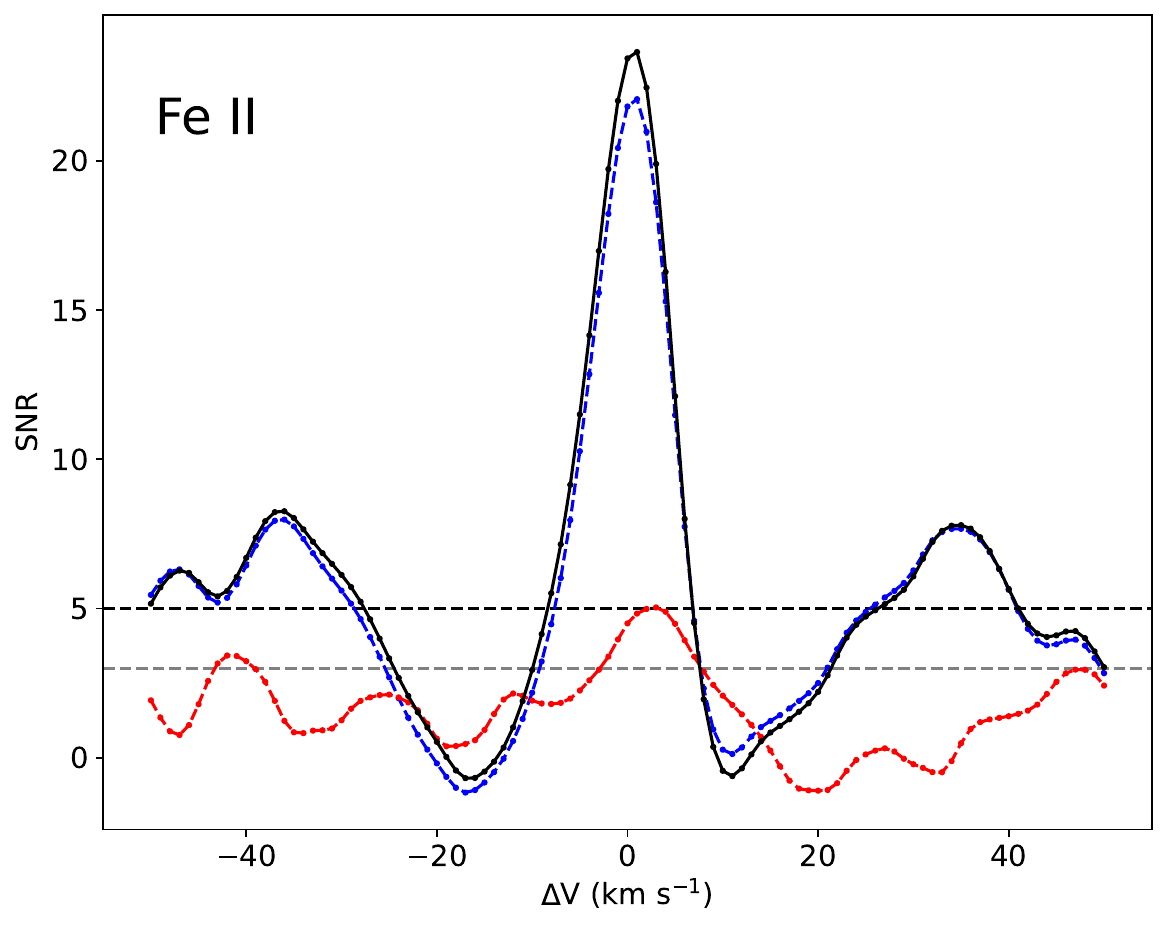}{0.4\textwidth}{}
                                }
                                \vspace{-30pt}
                                \gridline{
                                    \fig{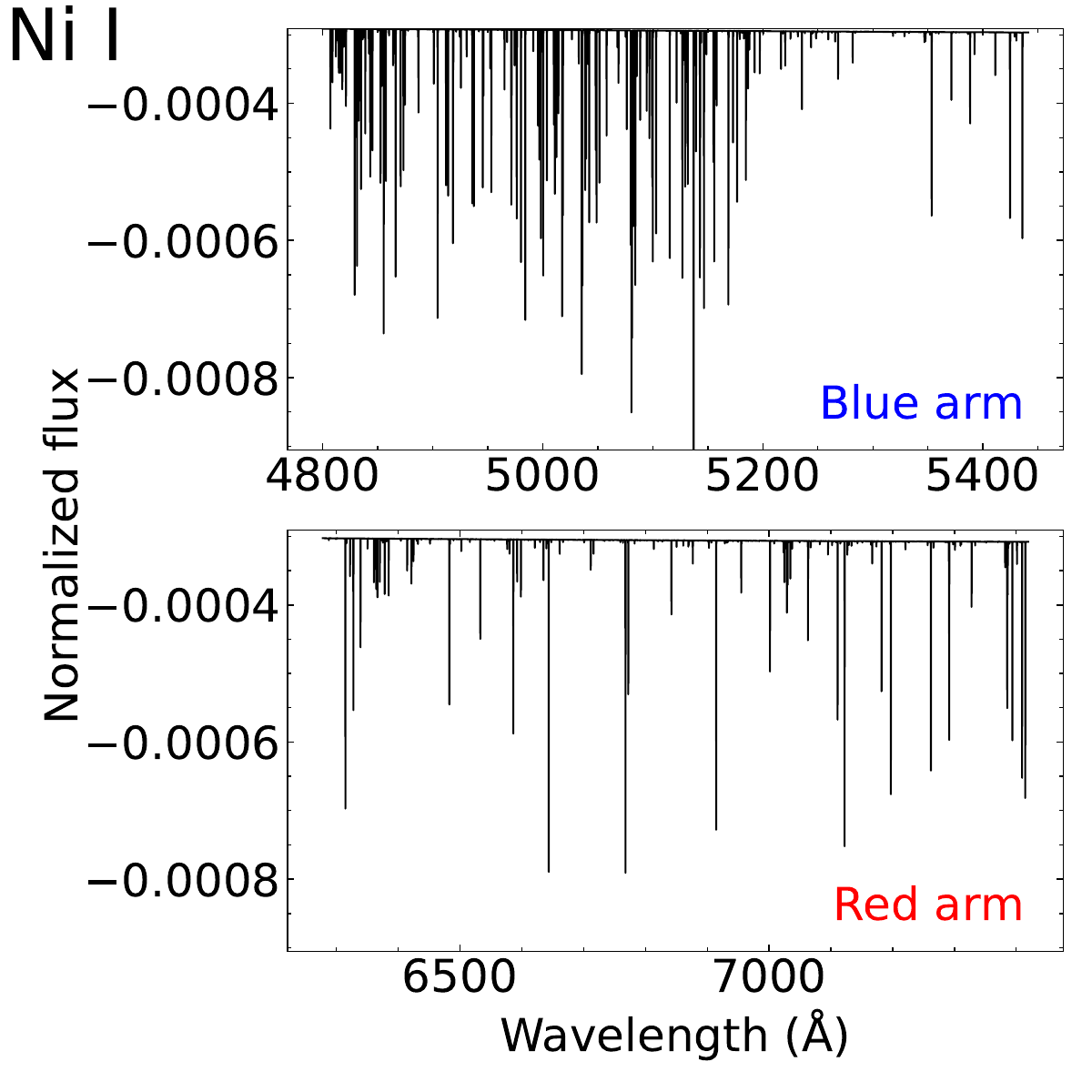}{0.345\textwidth}{}
                                    \fig{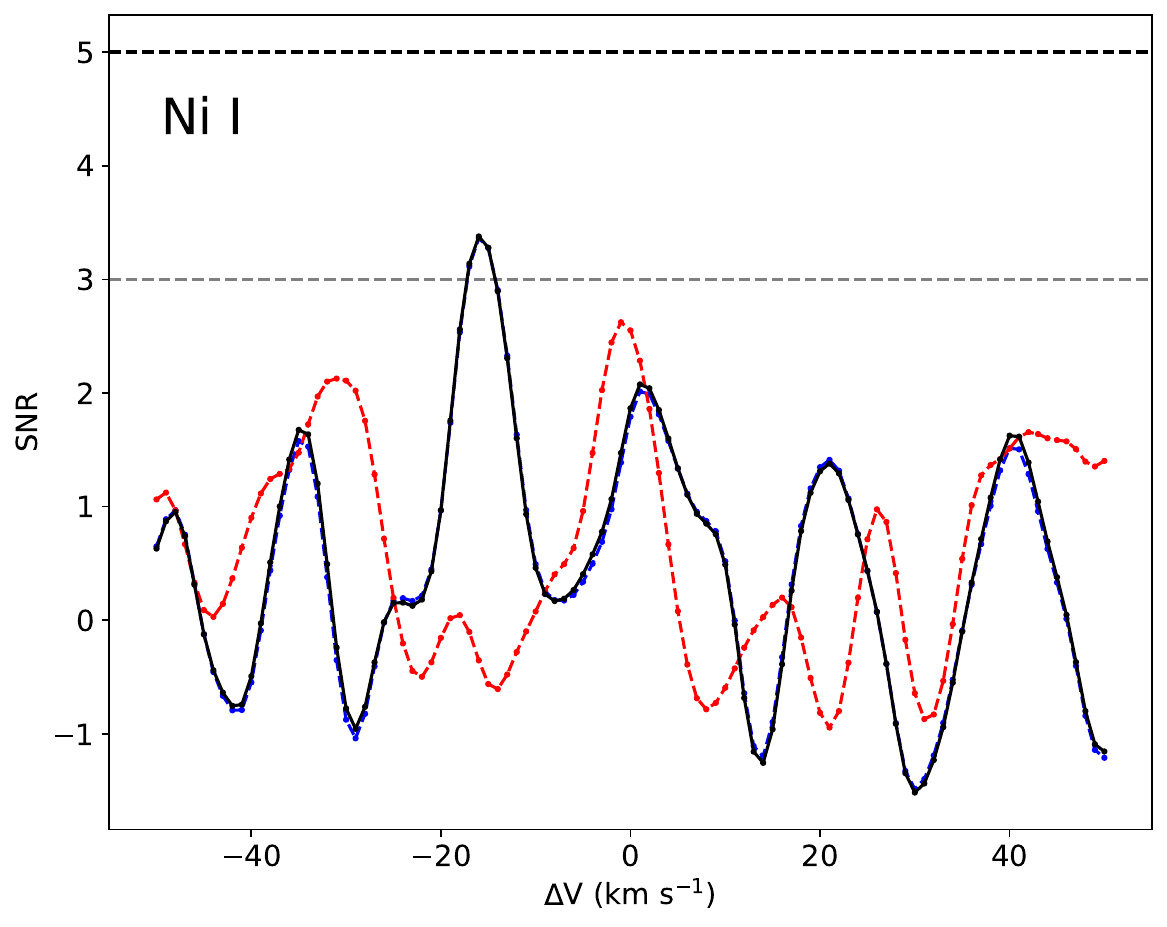}{0.4\textwidth}{}
                                }
                                \caption{\textit{Top:} On the left is the \code{petitRADTRANS}-generated transmission spectrum of Fe II, where we observe many deep lines in the blue arm, and few, weak lines in the red arm. On the right, the blue arm, red arm, and combined arm 1D CCFs are overlaid. The red arm 1D CCF is the red line, the blue arm 1D CCF is the blue line, and the combined arm 1D CCF is the black line. The gray dashed line is the tentative detection threshold and the black dashed line is the detection threshold. We observe a $\geq5\sigma$ signal from each arm, and both have a peak within $\pm5$ km s$^{-1}$. Thus, we fit a Gaussian to the combined arm 1D CCF, as described in Section \ref{subsec:Line Velocities}. ~\textit{Bottom:} The same plots for Ni I where we see an abundance of lines in both arms. Upon inspection of the 1D CCF of each arm, there are peaks in the blue and red arm that are $ >10$ km s$^{-1}$. Since these low-confidence signals do not agree, we discard this detection. The same methodology is followed for the remainder of detected species, found in Table \ref{tab:parameters_summary}.}
                                \label{fig:detection-methodology-example}
                            \end{figure*}

        Figure \ref{fig:detection-methodology-example} shows a template spectrum and 1D CCF for two species.

                               \clearpage
         
    \section{Examining the impact of Doppler shadow removal on the absorption traces}\label{app:doppler-shadow-impact}

    From Figure \ref{fig:doppler-shadow-removal}, it is clear that the Doppler shadow removal is not perfect: a residual Doppler shadow is still visible in the bottom panel. This is unsurprising since, as argued in \citet{Stephan2022}, the perturbation of the signal by the planet may not be sufficiently described by our Doppler shadow model.  We compared the iron absorption traces with and without a Doppler shadow removal to quantify its impact.
    
          \begin{figure}[h]
                \centering
                \vspace{-15pt}
                \gridline{
                    \fig{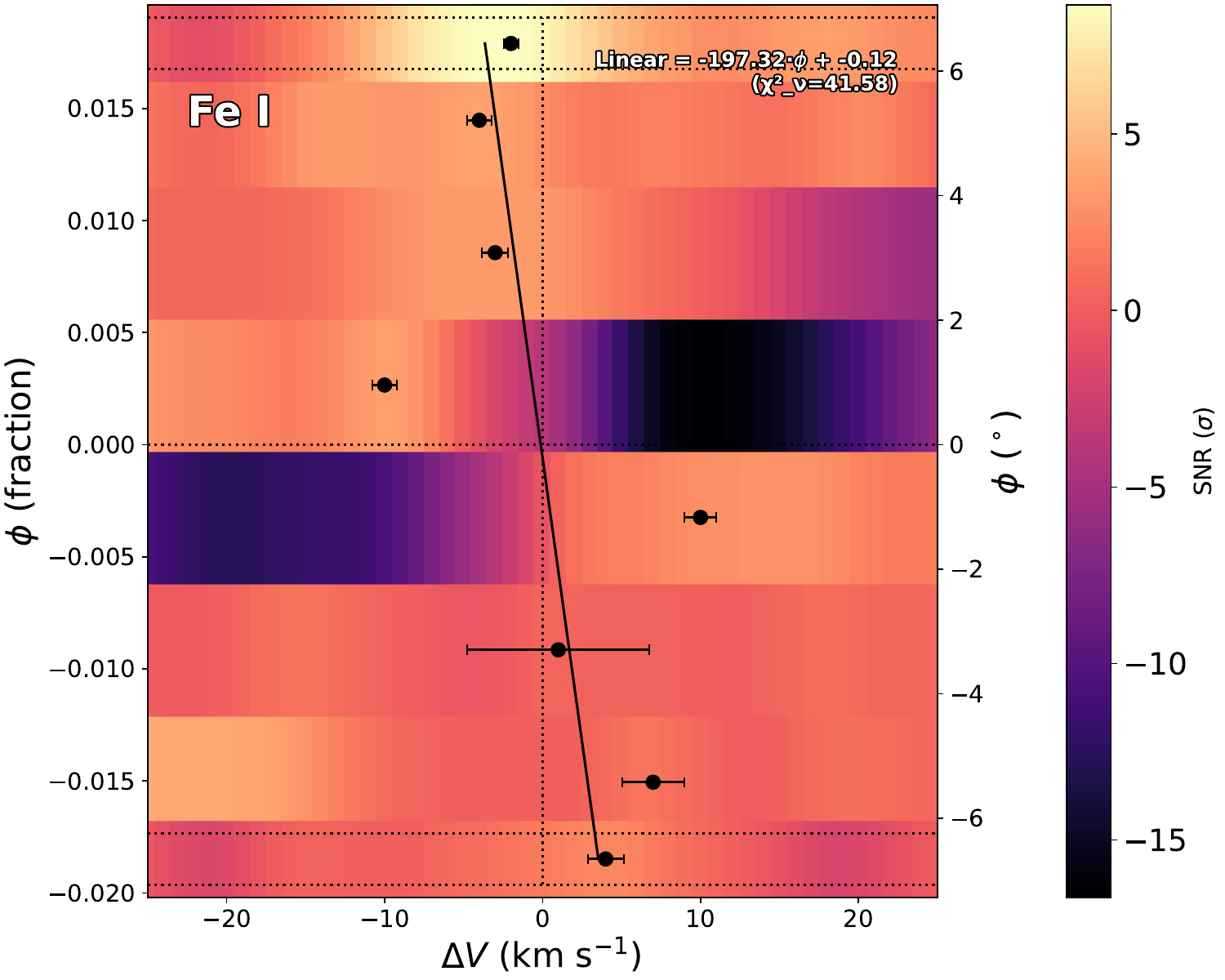}{0.4\textwidth}{}
                    \fig{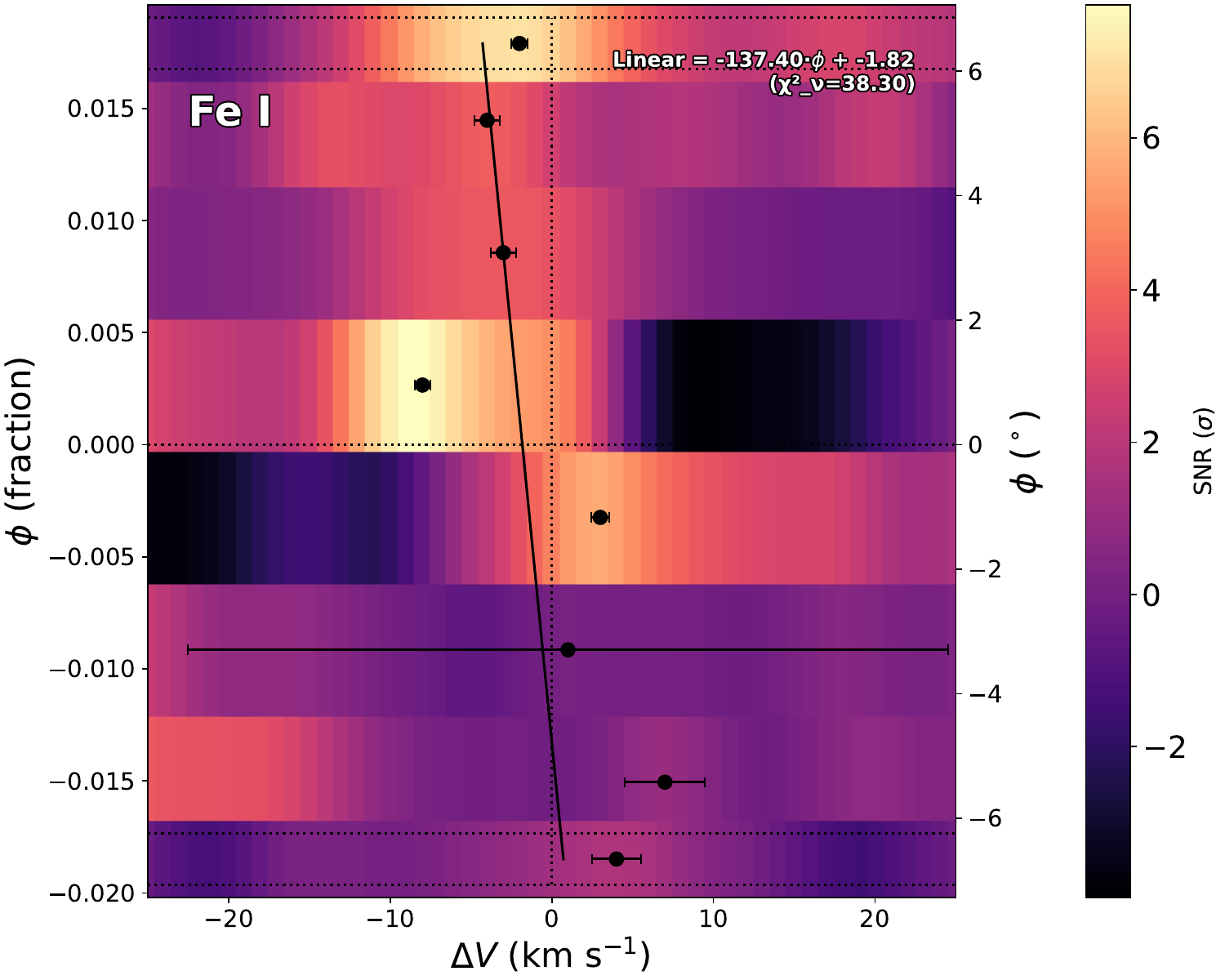}{0.4\textwidth}{}
                        }
                \vspace{-20pt}
                \caption{Phase-binned absorption traces of Fe I with the  parameters of a linear fit shown in the top right. The horizontal dashed lines at the bottom delineate the ingress phase bin and those at the top delineate the egress phase bin.
                \textit{Left:} Doppler shadow present.
                \textit{Right:} Doppler shadow removed. We calculated the reduced chi-squared statistic between the Doppler shifts of each absorption trace to be $\chi^2_\nu=0.50$.}
                \label{fig:Doppler-shadow-comparison-FeI}
            \end{figure}
            \begin{figure}[h]
                \centering
                \vspace{-15pt}
                \gridline{
                    \fig{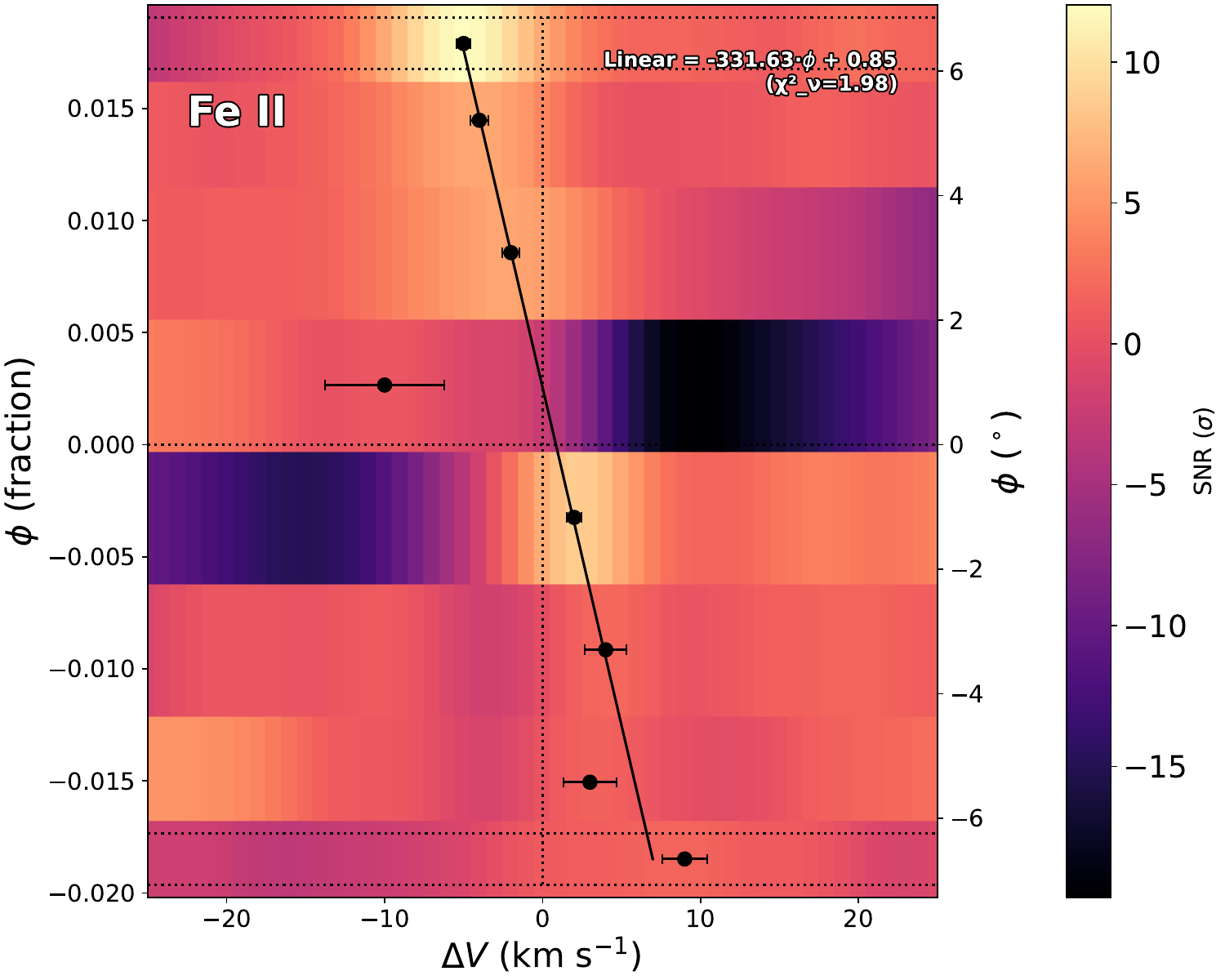}{0.4\textwidth}{}
                    \fig{KELT-20b.Fe+.combined.phase-binned+RVs.pdf}{0.4\textwidth}{}
                        }
                \vspace{-20pt}
                \caption{Same plots as Figure \ref{fig:Doppler-shadow-comparison-FeI}, but for Fe II. We calculated $\chi^2_\nu=0.58$. }
                \label{fig:Doppler-shadow-comparison-FeII}
            \end{figure}

          There is a discrepancy in mid-transit $\Delta V$ and SNR between the uncorrected and corrected absorption traces, but it does not result in a statistically significant difference in the time resolved velocity offsets. However, the Doppler shadow removal artificially increases the SNR at mid-transit near $\Delta V = 0$, effectively amplifying the $\Delta V$ peaks and shifting their centers closer to zero. This does not affect our conclusions, including the discussion in Section \ref{subsubsec:comparison-with-theory}, in part because much of the absorption trace analysis is qualitative, and all qualitative features used for comparison with theory are consistent between the uncorrected and corrected traces.

    \section{Detection profiles}\label{app:Detections}

    \begin{figure}[H]
        \centering
        \gridline{
            \fig{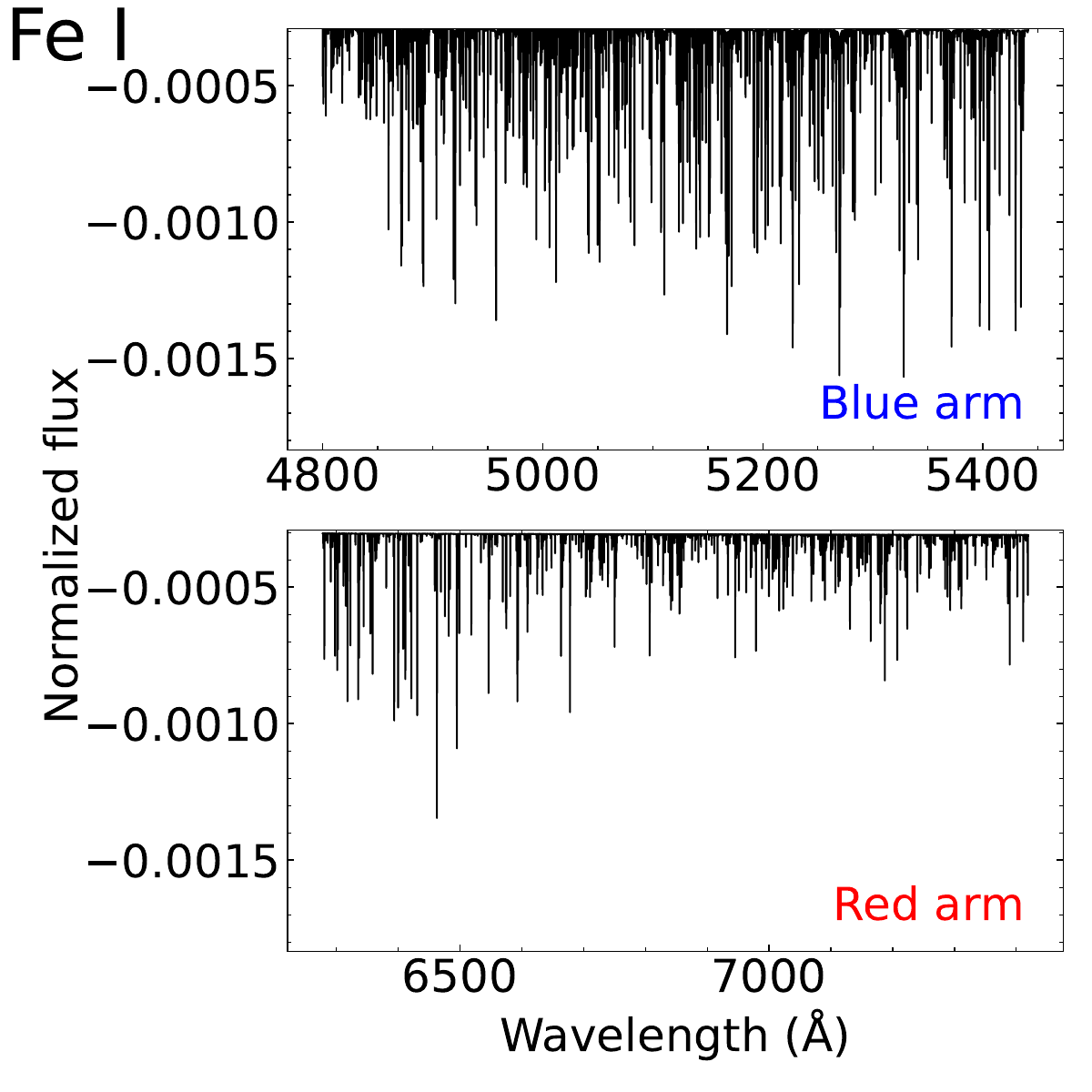}{0.333\textwidth}{}
            \fig{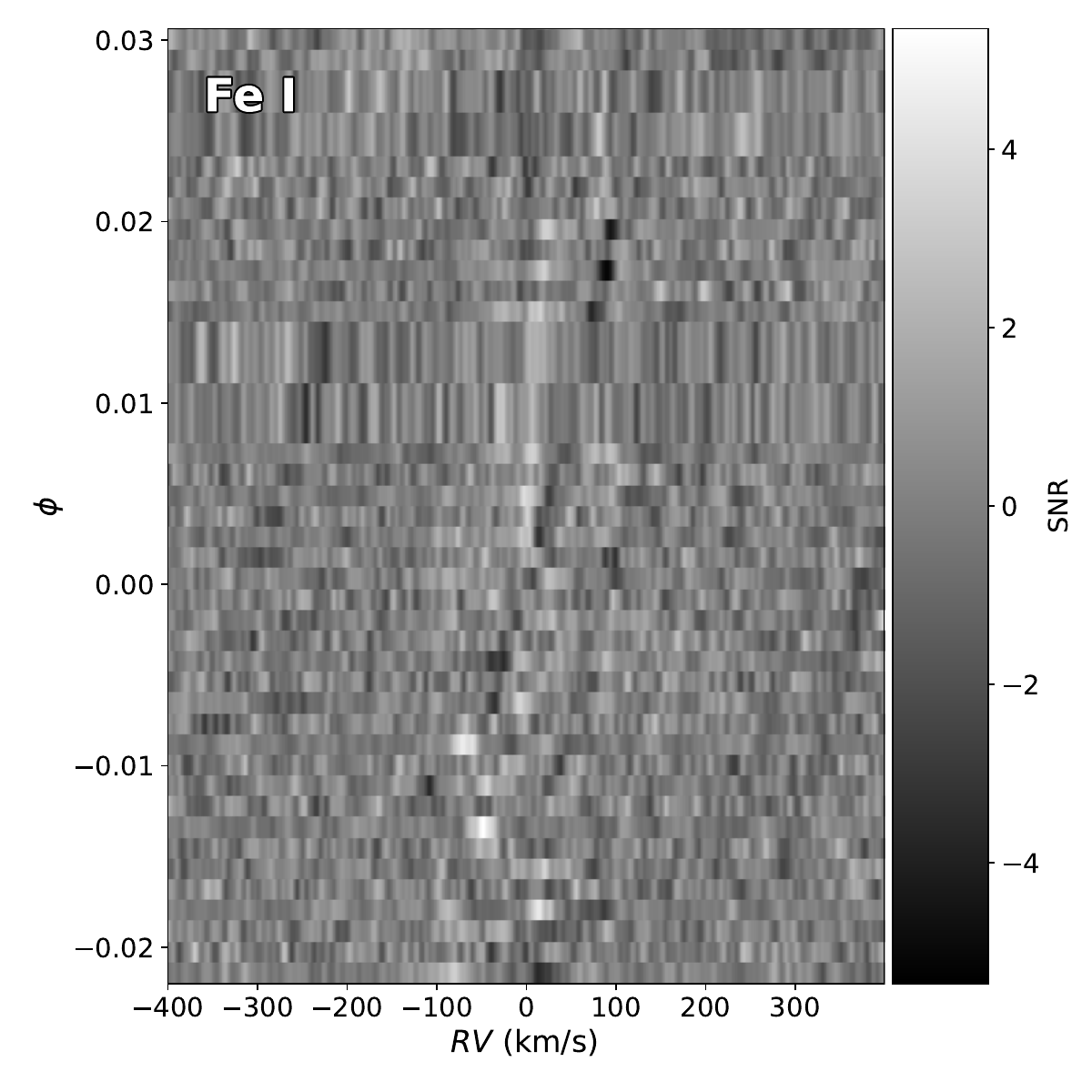}{0.333\textwidth}{}
            \fig{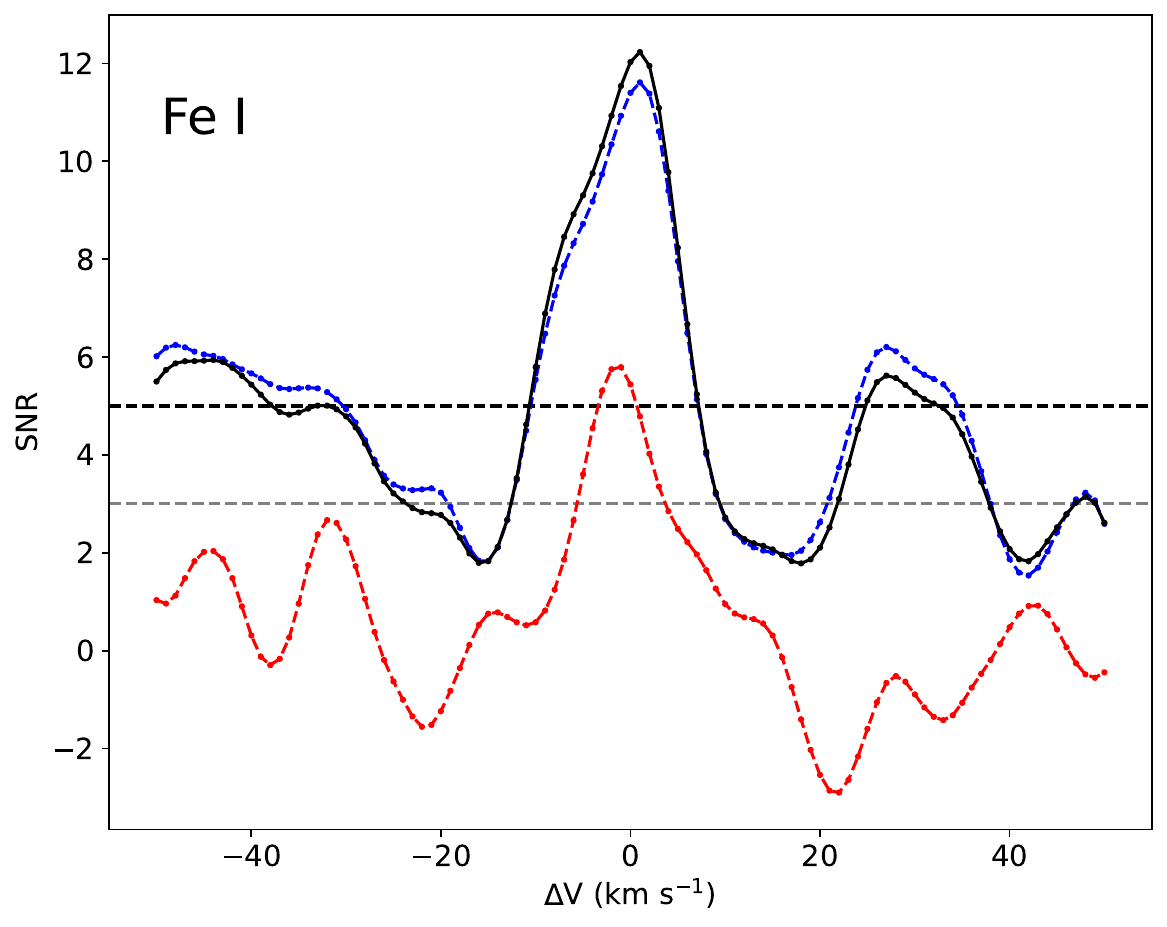}{0.333\textwidth}{}
            }
        \vspace{-30pt}
        \gridline{
            \fig{KELT-20b.20190504.combined.Fe.CCFs-shifted.pdf}{0.333\textwidth}{}   
            \fig{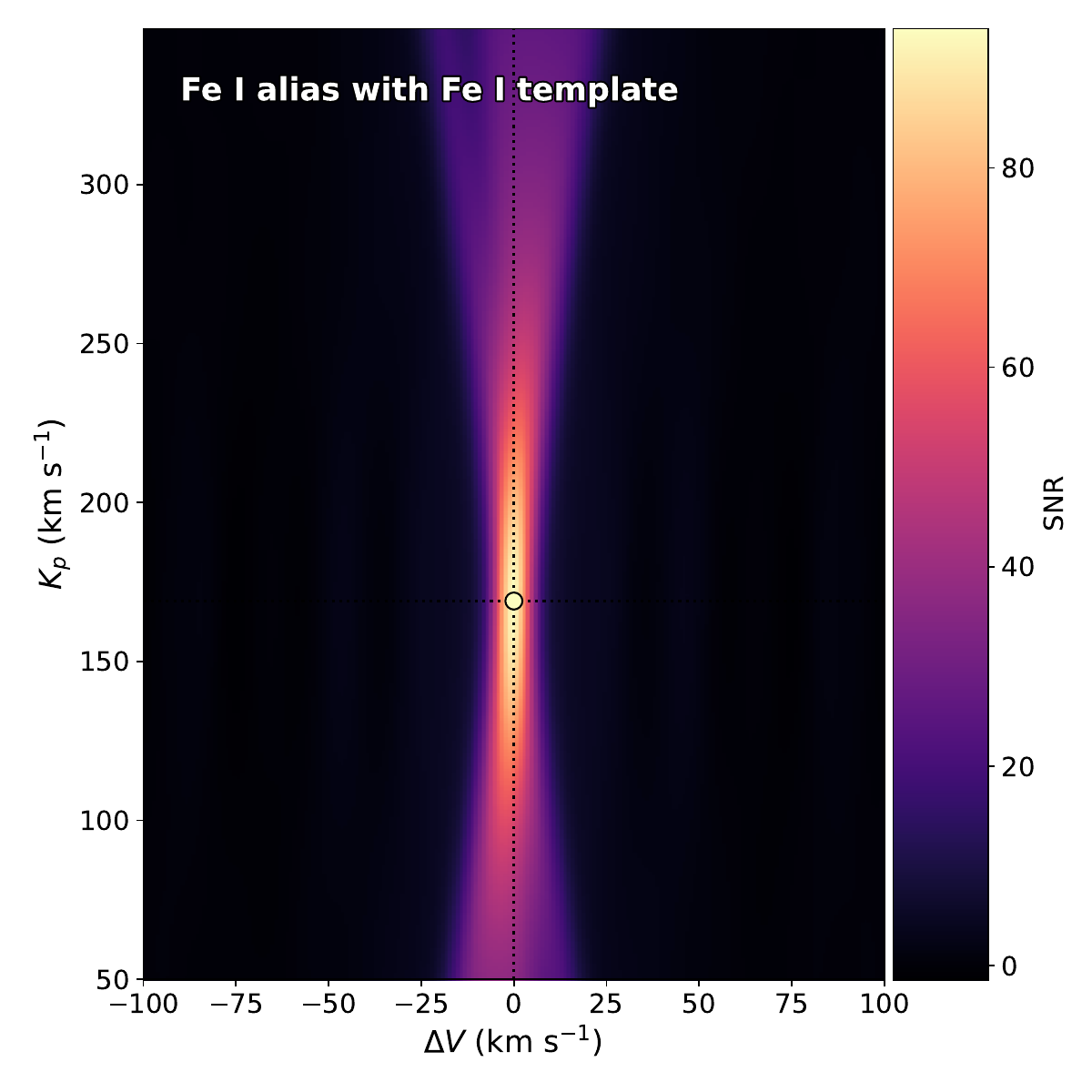}{0.333\textwidth}{}
            \fig{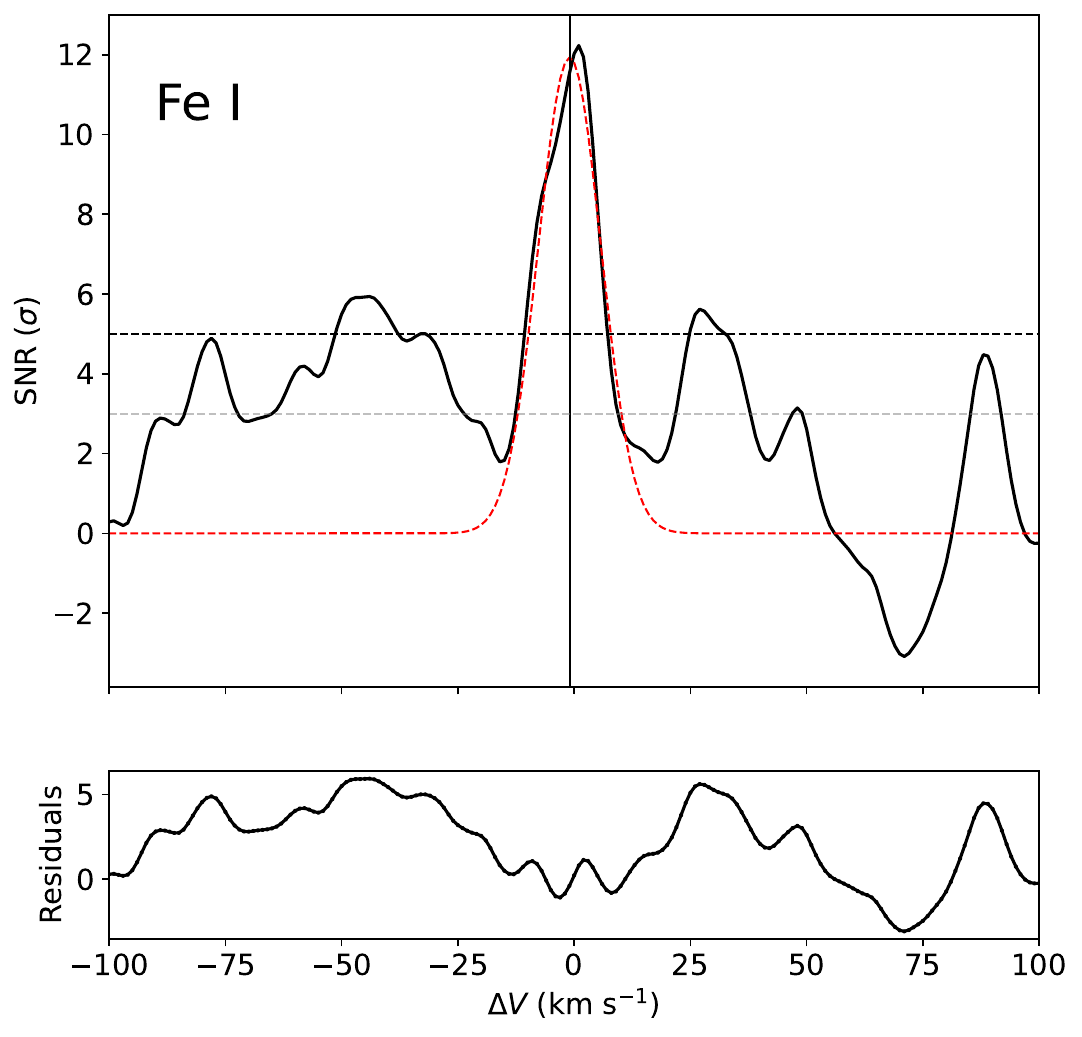}{0.333\textwidth}{}
            }           
        \gridline{
            \fig{KELT-20b.Fe.combined.phase-binned+RVs.pdf}{0.4\textwidth}{}
            \fig{KELT-20b.20190504.combined.Fe..halves.SNR-Gaussian-Asymmetry.pdf}{0.4\textwidth}{}
             }
             \caption{Detection profile for Fe I. \textit{First row, left}: \code{petitRADTRANS}-generated transmission spectrum of Fe I containing many lines across the bandpass of both arms. \textit{First row, middle}: 2D CCF with Doppler shadow correction applied, shifted into stellar rest frame. \textit{First row, right}: Blue arm, red arm, and combined arm 1D CCFs overlaid. The red arm 1D CCF is the red line, the blue arm 1D CCF is the blue line, and the combined arm 1D CCF is the black line. The gray dashed line is the tentative detection threshold and the black dashed line is the detection threshold. We observe a $\geq5\sigma$ signal from each arm, and both have a peak within $\pm5$ km s$^{-1}$. \textit{Second row, left}: 2D $K_p-\Delta V$ map shifted into the planetary rest frame. \textit{Second row, middle}: 2D CCF with alias between the species and Fe I template. Note that in this case, since we are searching for Fe I and aliasing with the Fe I template, the alias is an autocorrelation. \textit{Second row, right}: 1D full-transit CCF at peak $K_p$. \textit{Third row, left}: Phase-binned velocity offsets, i.e. centers of a Gaussian fit to the 1D CCF slice at $K_{p,\text{expected}}$ within each phase bin. The SNR is plotted to show the line strengths in each bin relative to each other. \textit{Third row, right:} Coadded Fe I 1D CCFs from first half of transit in green, second half of transit in blue, and full-transit in black. The gray horizontal dashed line is the tentative detection threshold and the black horizontal dashed line is the detection threshold. }
             \label{fig:CCFs-appendix-FeI}
    \end{figure}
    
    \begin{figure}[H]
        \centering
        \gridline{
            \fig{spectraIndArms.Fe_II.KELT-20b.inverted-transmission-better.pdf}{0.333\textwidth}{}
            \fig{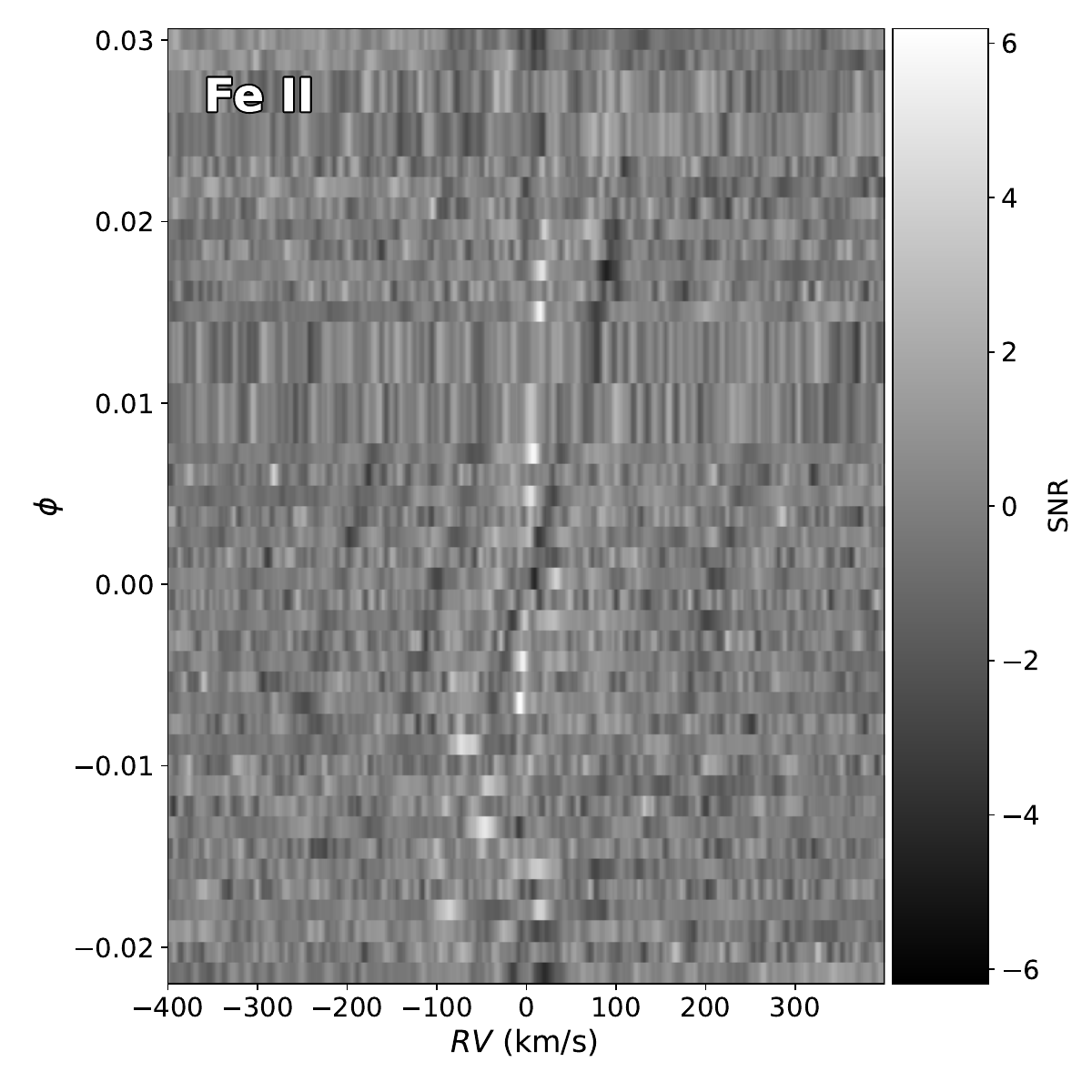}{0.333\textwidth}{}
            \fig{KELT-20b.20190504.combined.Fe_II.line-profiles-overlaidarms.pdf}{0.333\textwidth}{}
            }
        \vspace{-30pt}
        \gridline{
            \fig{KELT-20b.20190504.combined.Fe+.CCFs-shifted.pdf}{0.333\textwidth}{}   
            \fig{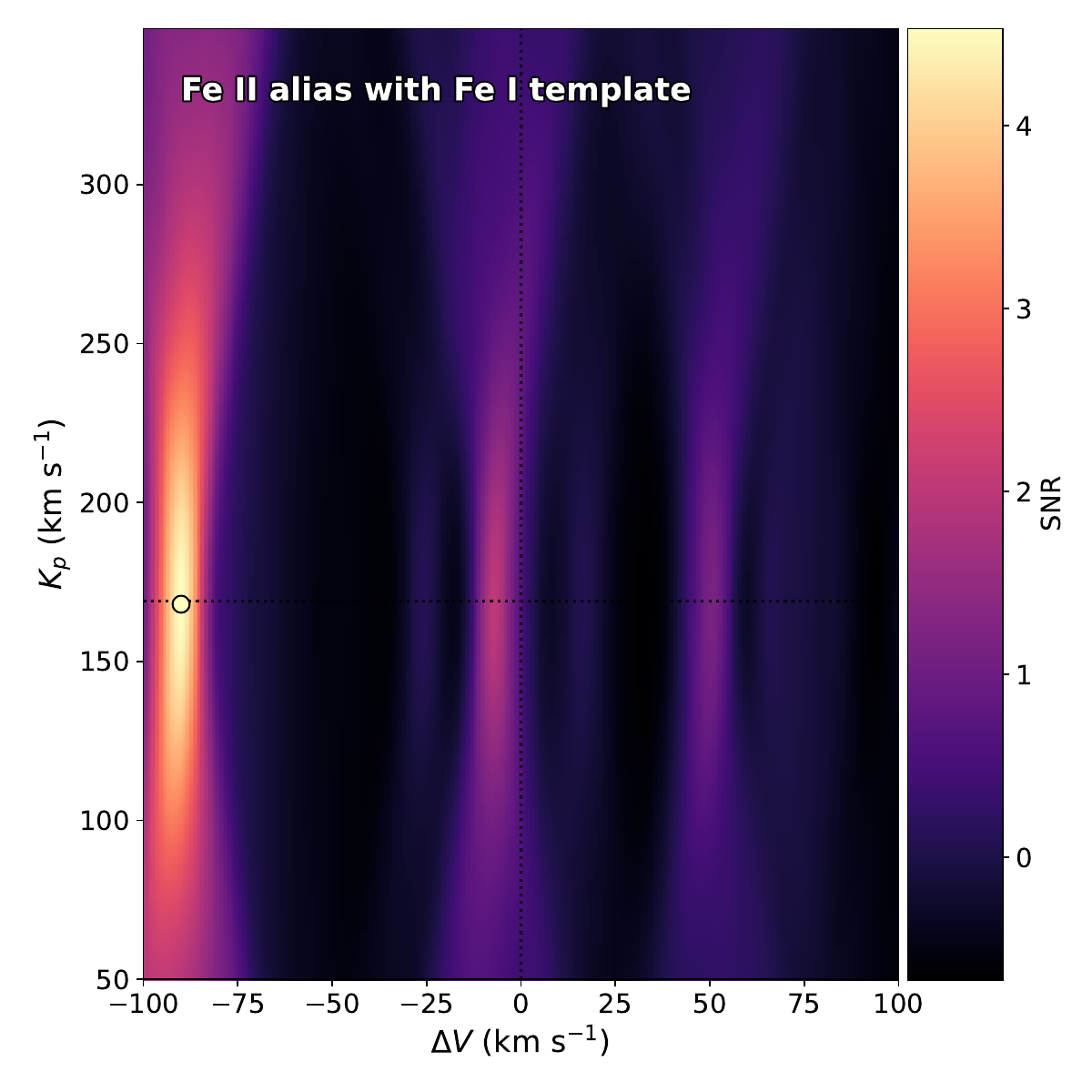}{0.333\textwidth}{}
            \fig{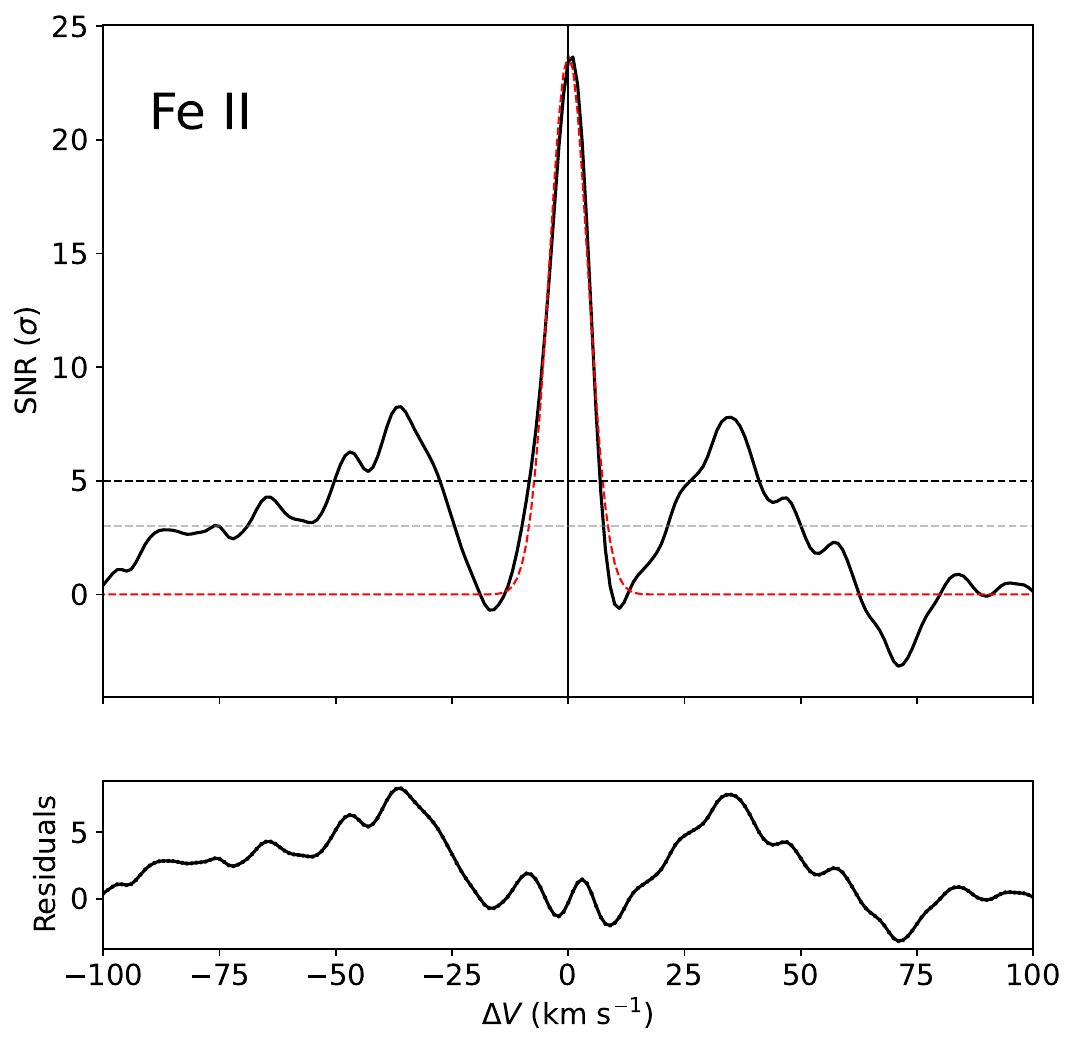}{0.333\textwidth}{}
            }      
        \gridline{
            \fig{KELT-20b.Fe+.combined.phase-binned+RVs.pdf}{0.4\textwidth}{}
            \fig{KELT-20b.20190504.combined.Fe+..halves.SNR-Gaussian-Asymmetry.pdf}{0.4\textwidth}{}
             }
             \caption{Same as Figure \ref{fig:CCFs-appendix-FeI}, but for Fe II.}
    \end{figure}
    
    \begin{figure}[H]
        \centering
        \gridline{
            \fig{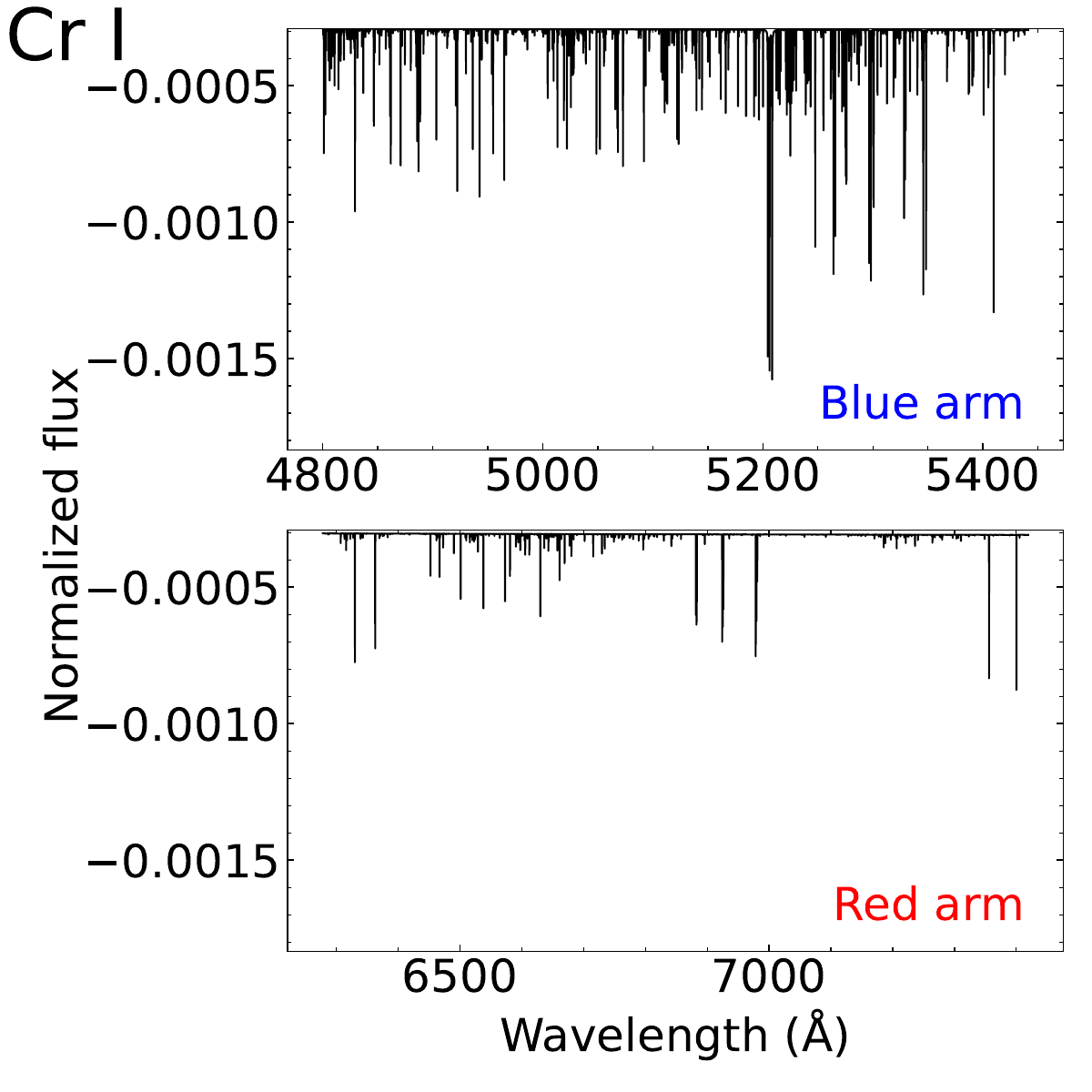}{0.333\textwidth}{}
            \fig{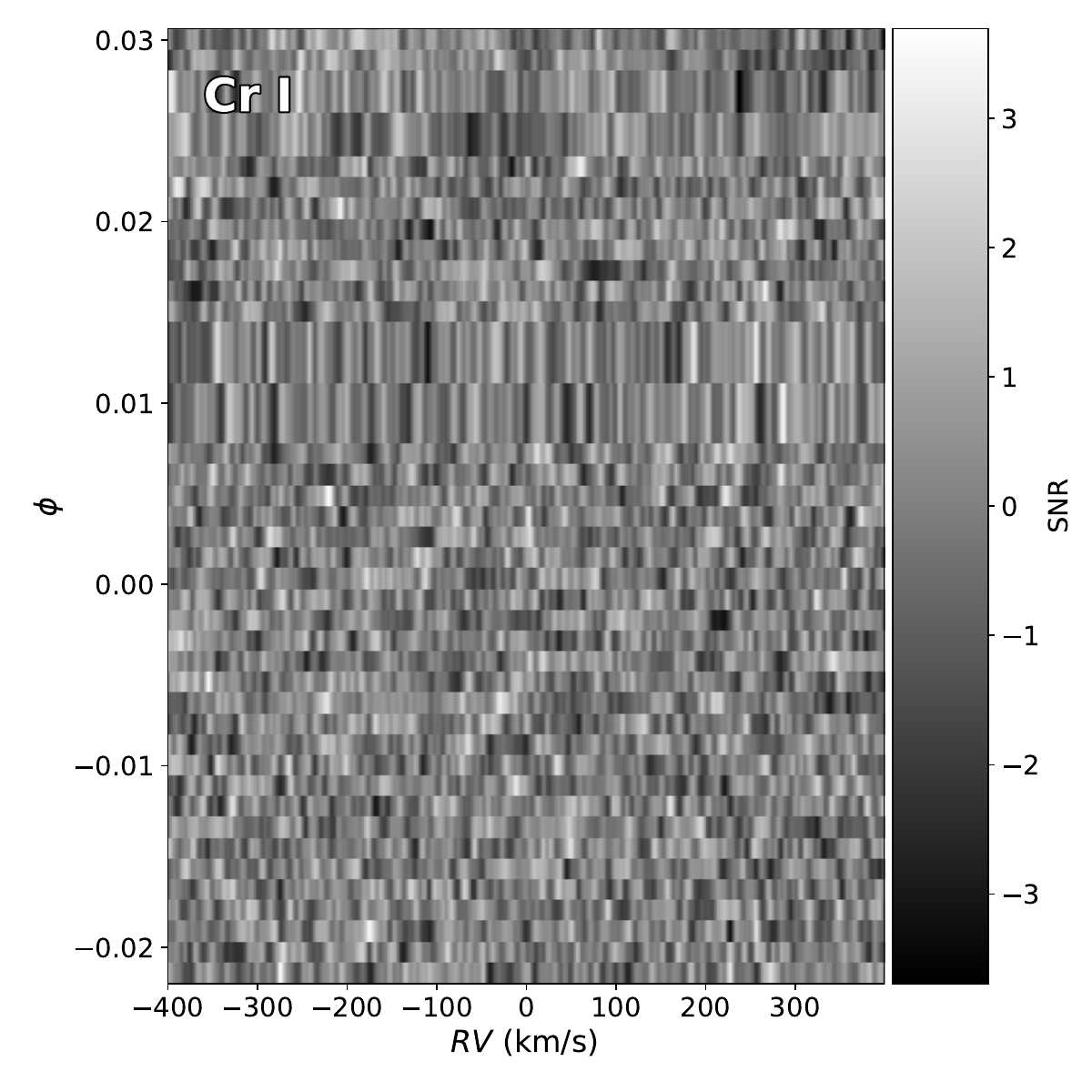}{0.333\textwidth}{}
            \fig{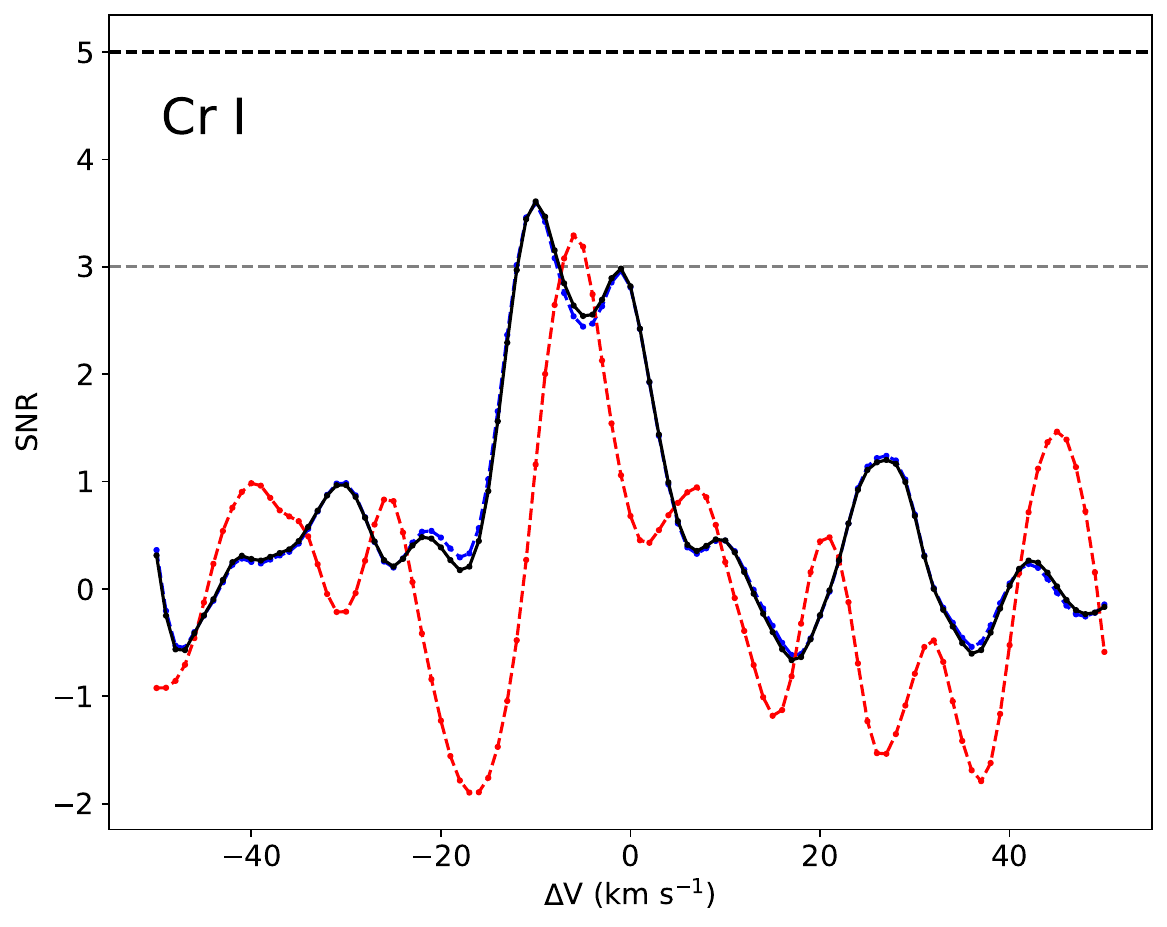}{0.333\textwidth}{}
            }
        \vspace{-30pt}
        \gridline{
            \fig{KELT-20b.20190504.combined.Cr.CCFs-shifted.pdf}{0.333\textwidth}{}            
            \fig{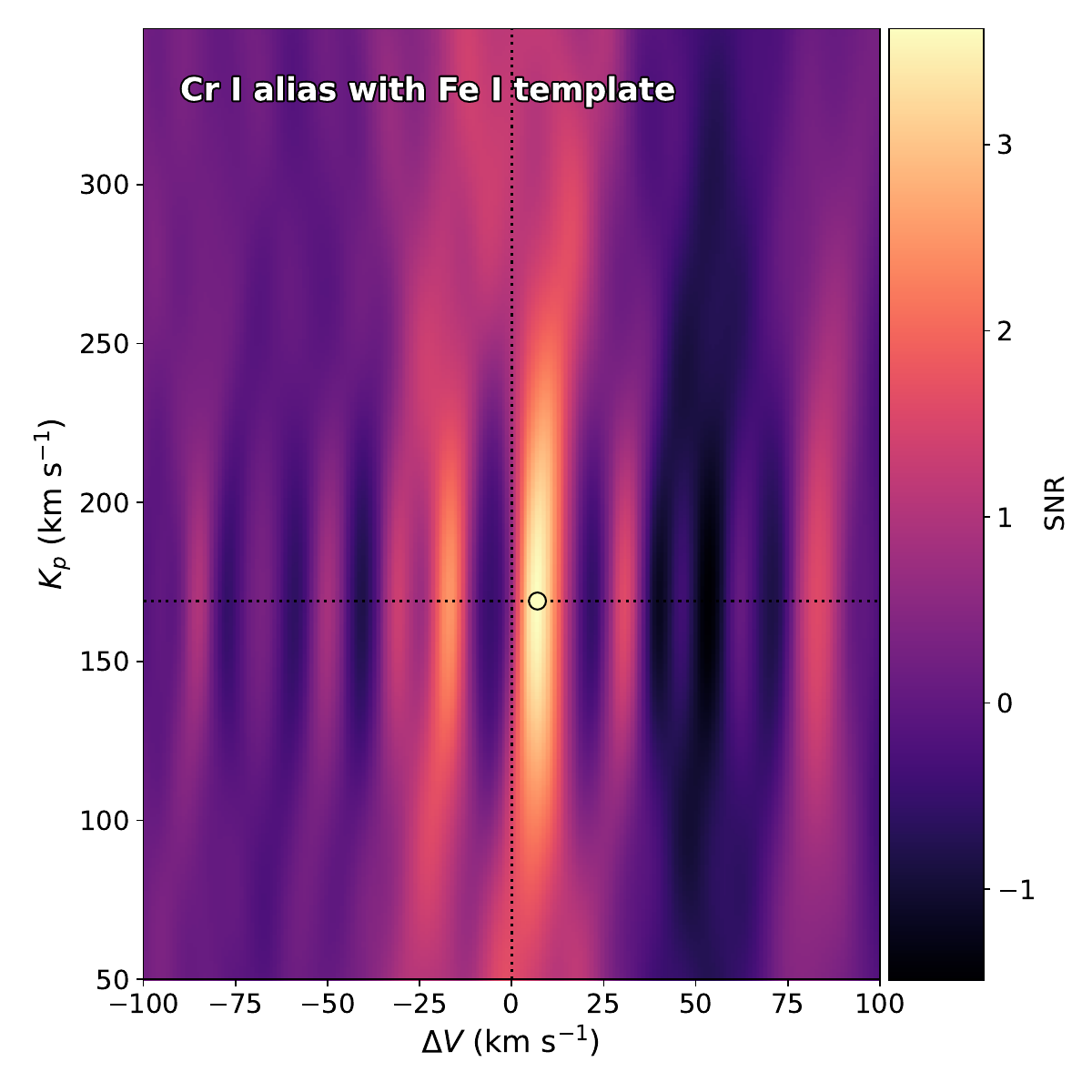}{0.333\textwidth}{}
            \fig{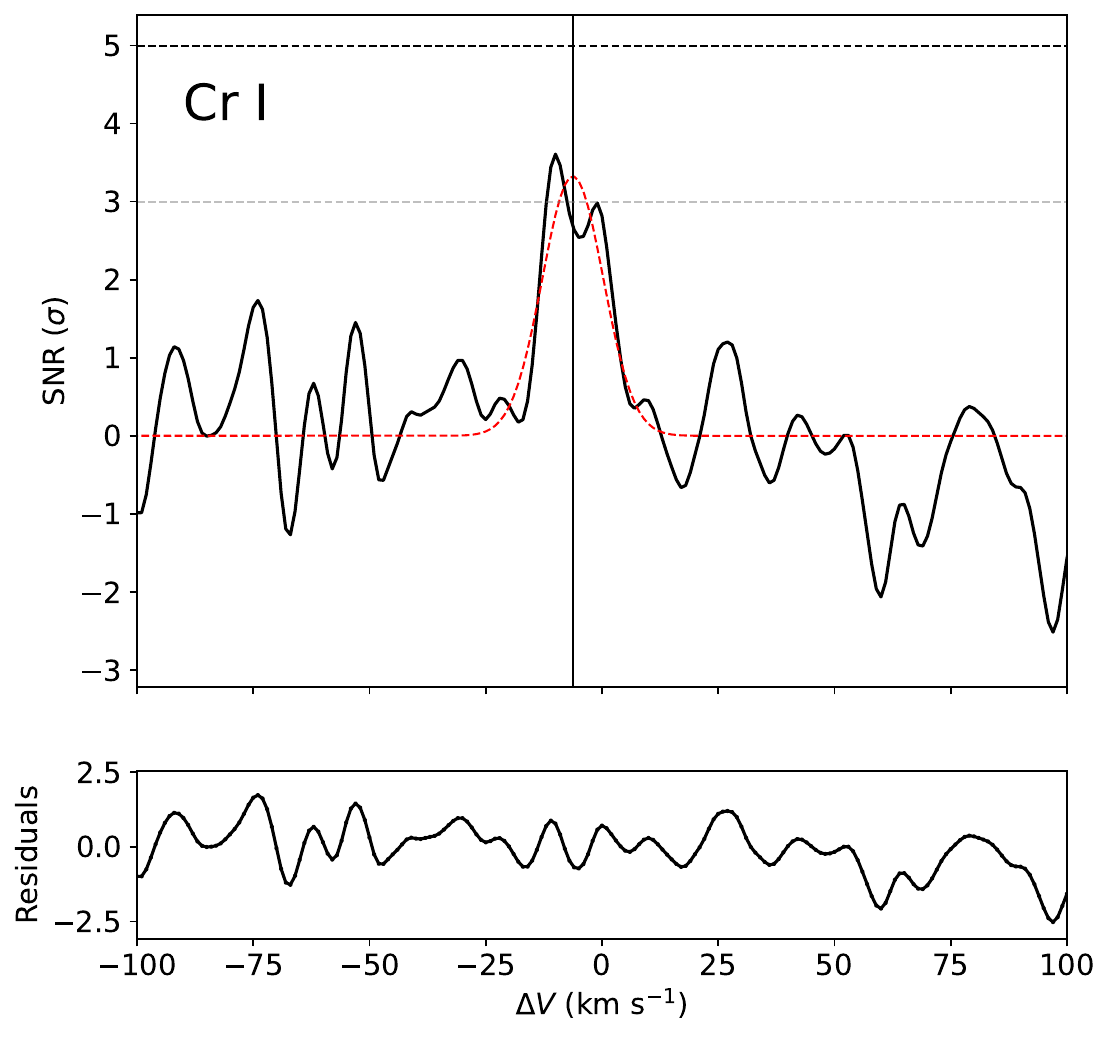}{0.333\textwidth}{}
            }     
        \gridline{  
            \fig{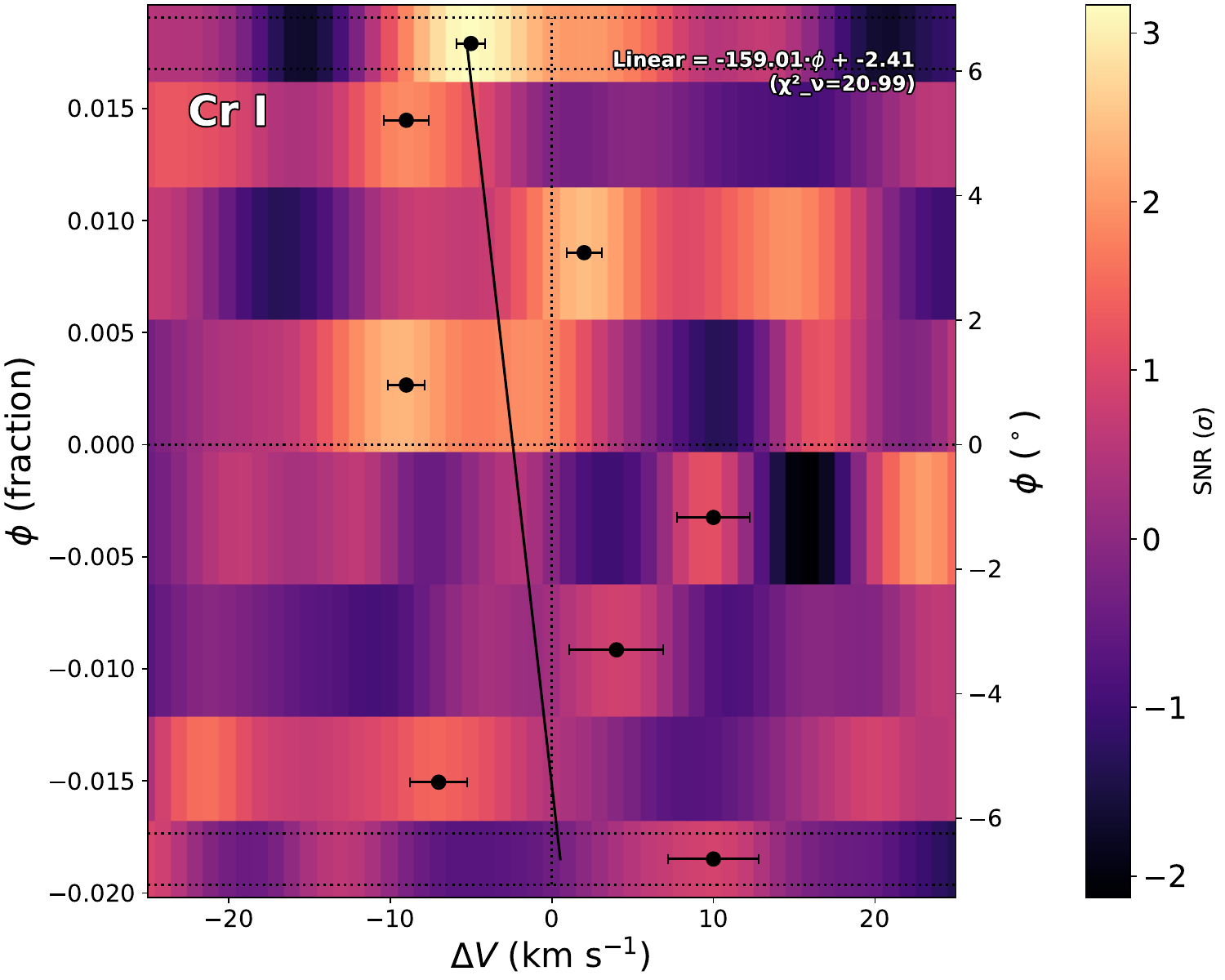}{0.4\textwidth}{}
            \fig{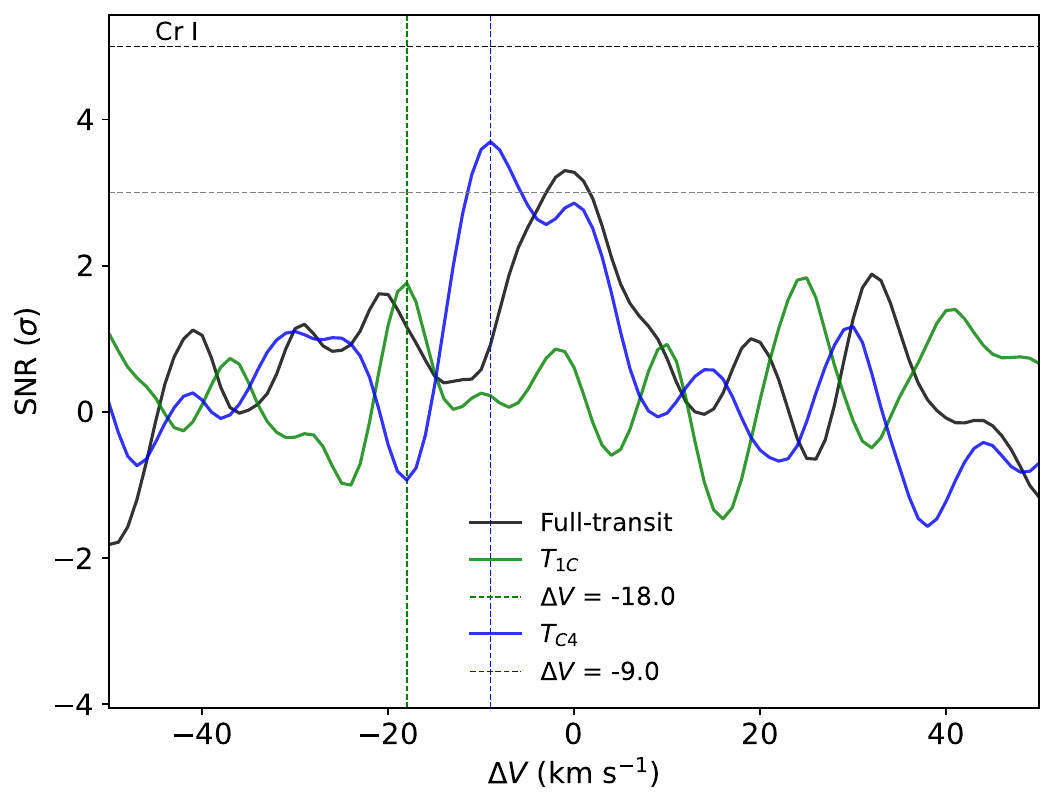}{0.4\textwidth}{}
             }
             \caption{Same as Figure \ref{fig:CCFs-appendix-FeI}, but for Cr I.}
    \end{figure}

    \begin{figure}[H]
        \centering
        \gridline{
            \fig{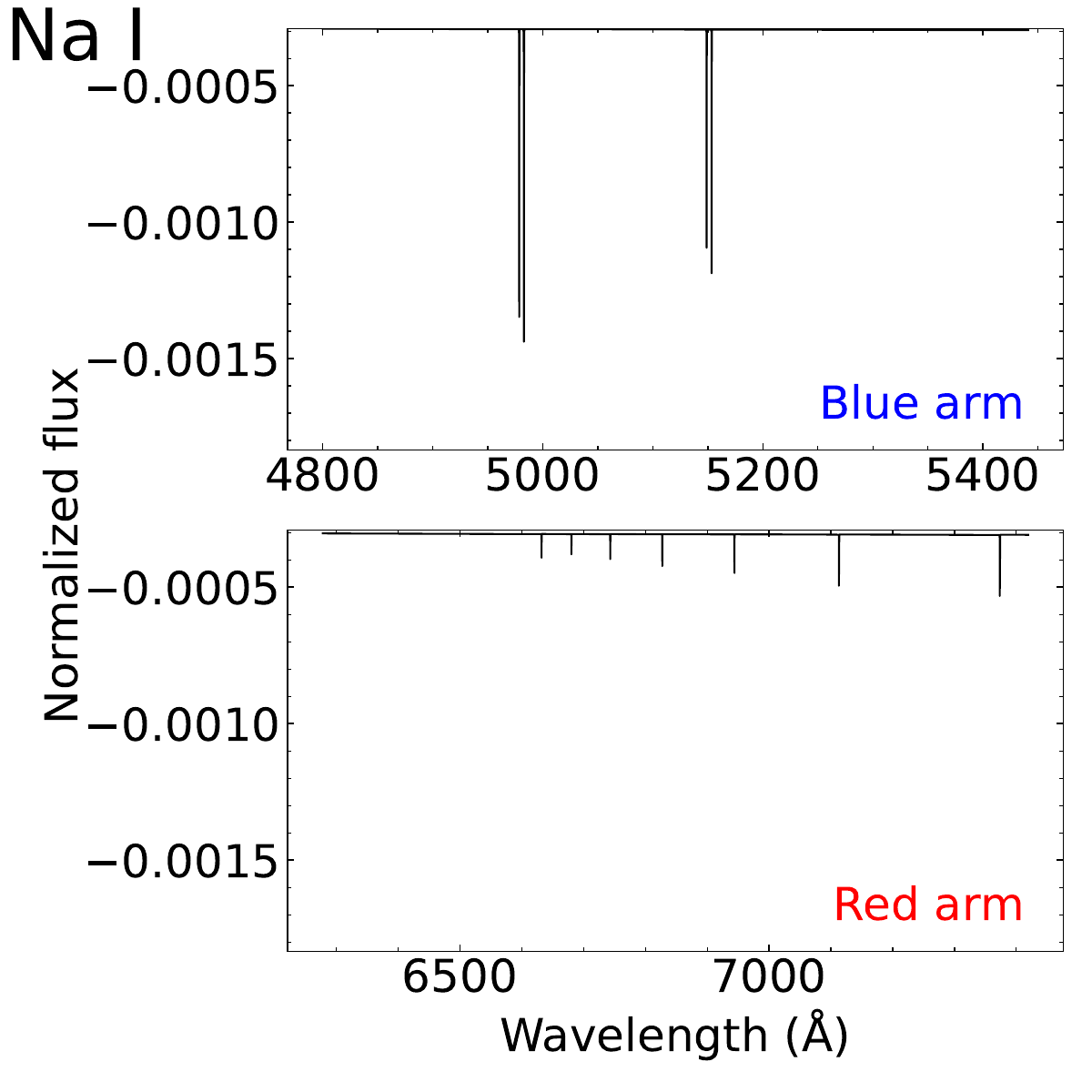}{0.333\textwidth}{}
            \fig{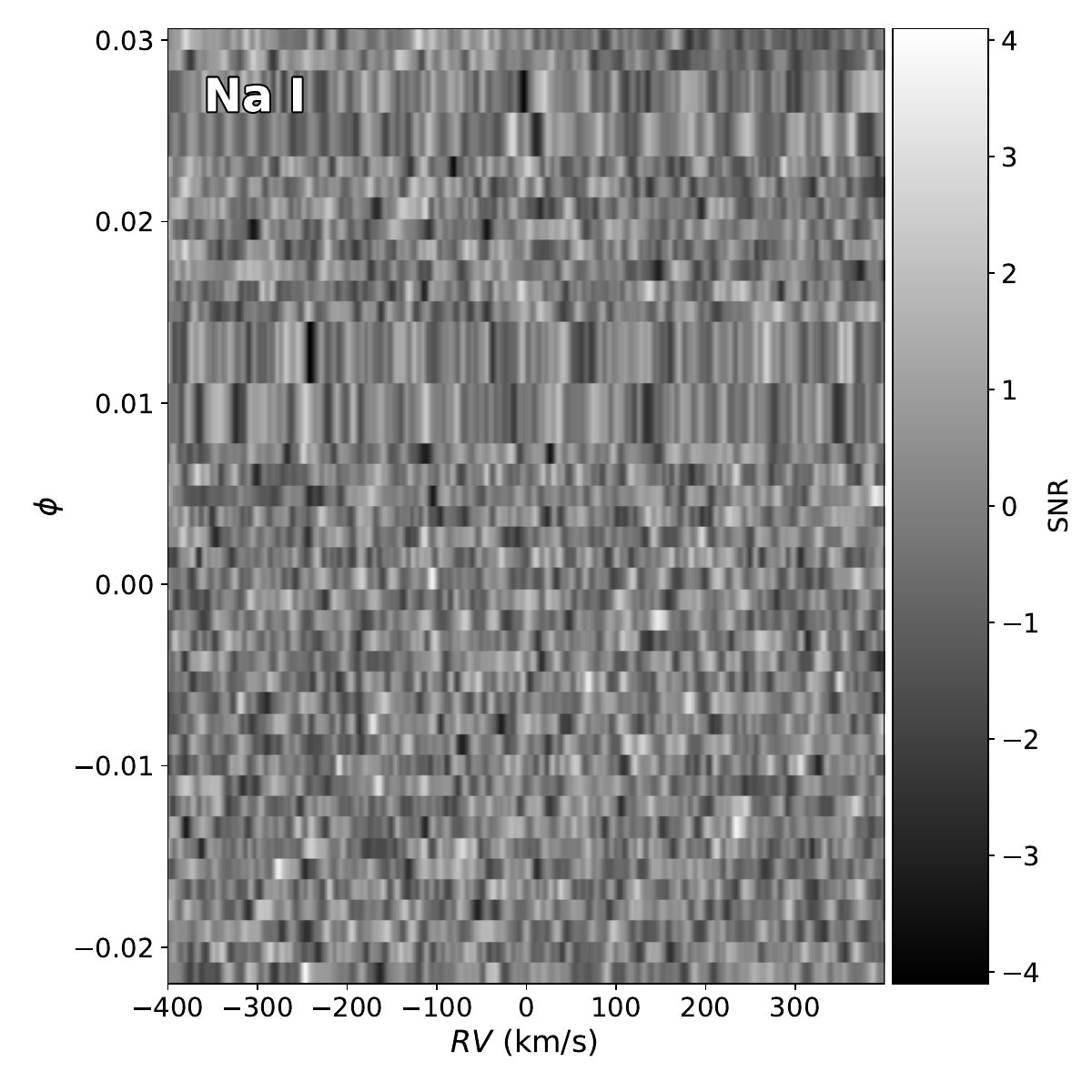}{0.333\textwidth}{}
            \fig{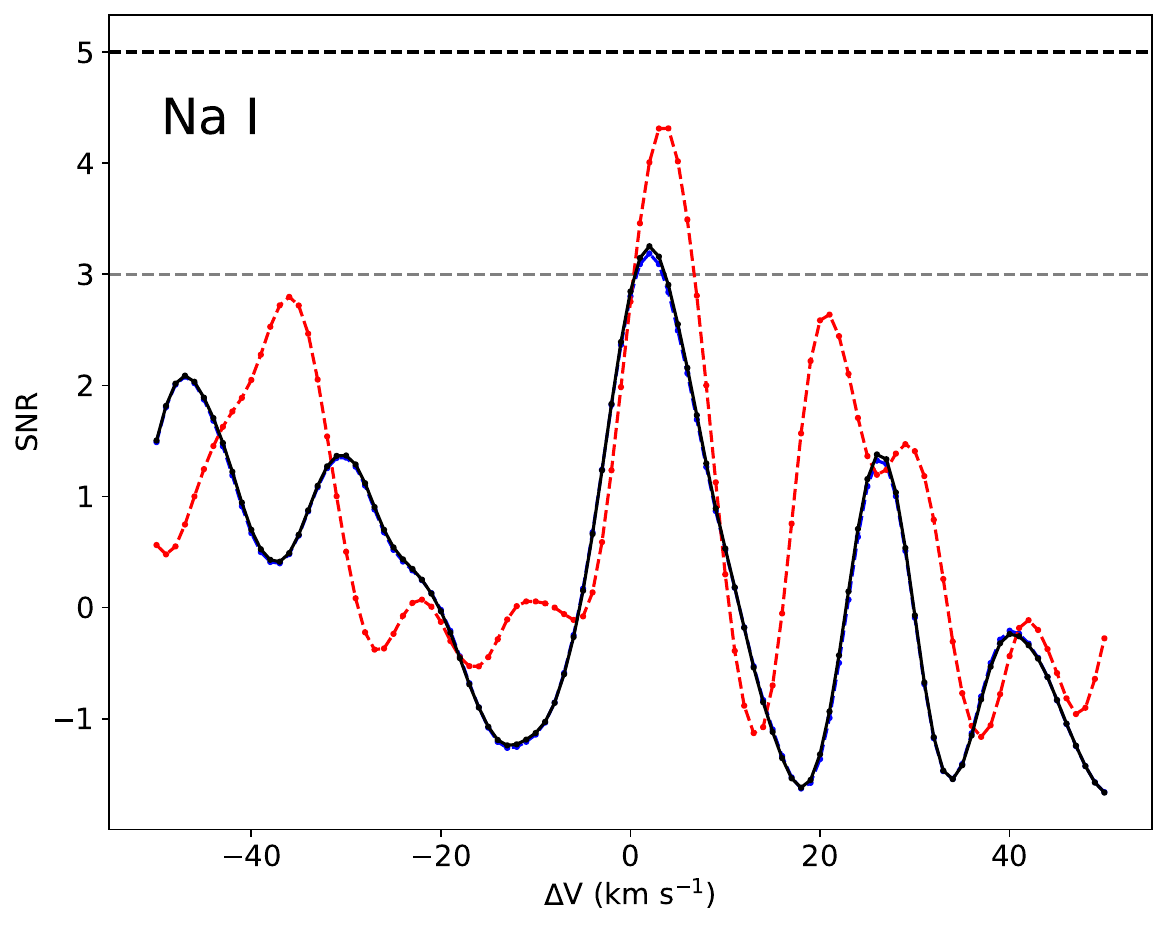}{0.333\textwidth}{}}
        \vspace{-30pt}
        \gridline{
            \fig{KELT-20b.20190504.combined.Na.CCFs-shifted.pdf}{0.333\textwidth}{}          
            \fig{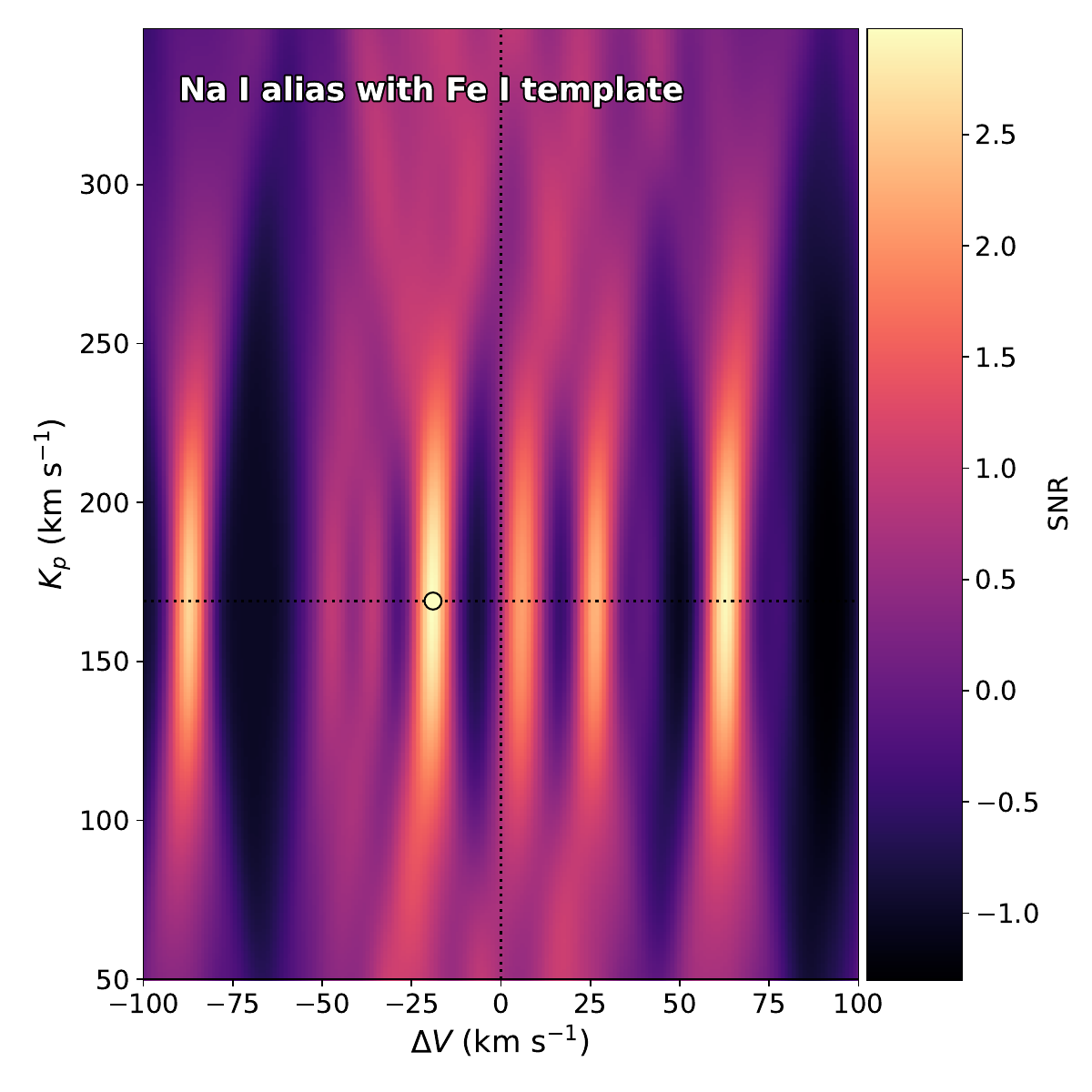}{0.333\textwidth}{}
            \fig{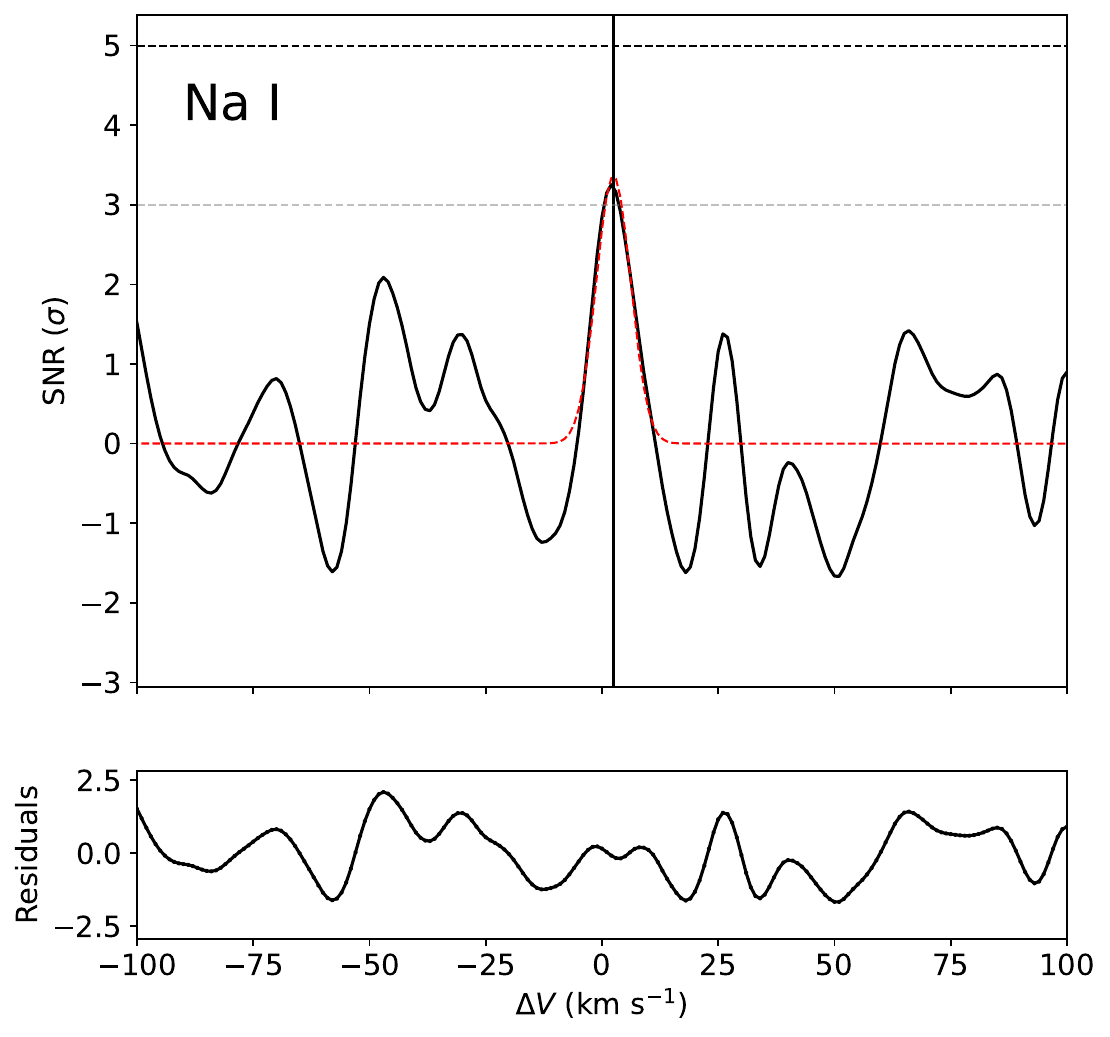}{0.333\textwidth}{}
            }      
        \gridline{
            \fig{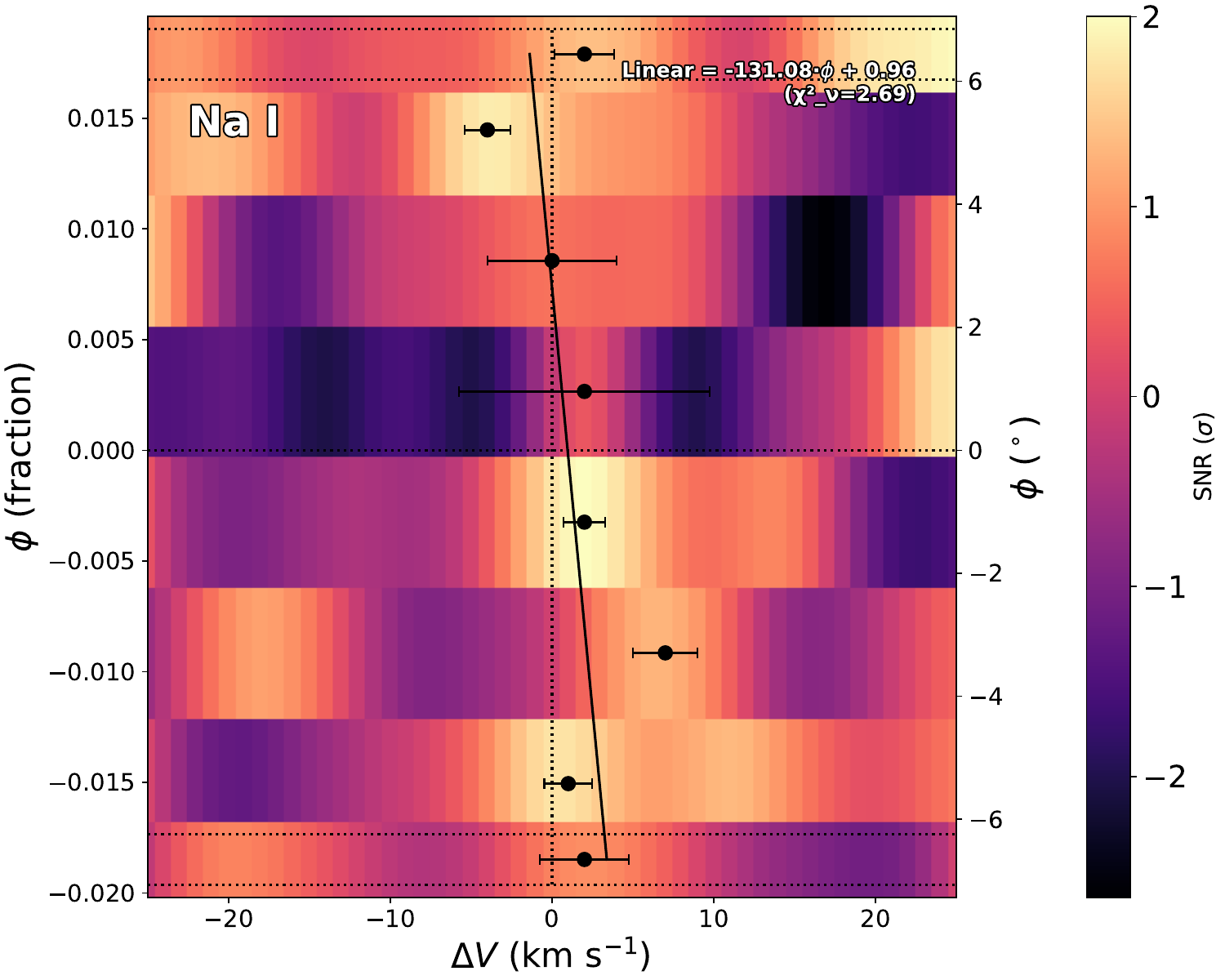}{0.4\textwidth}{}
            \fig{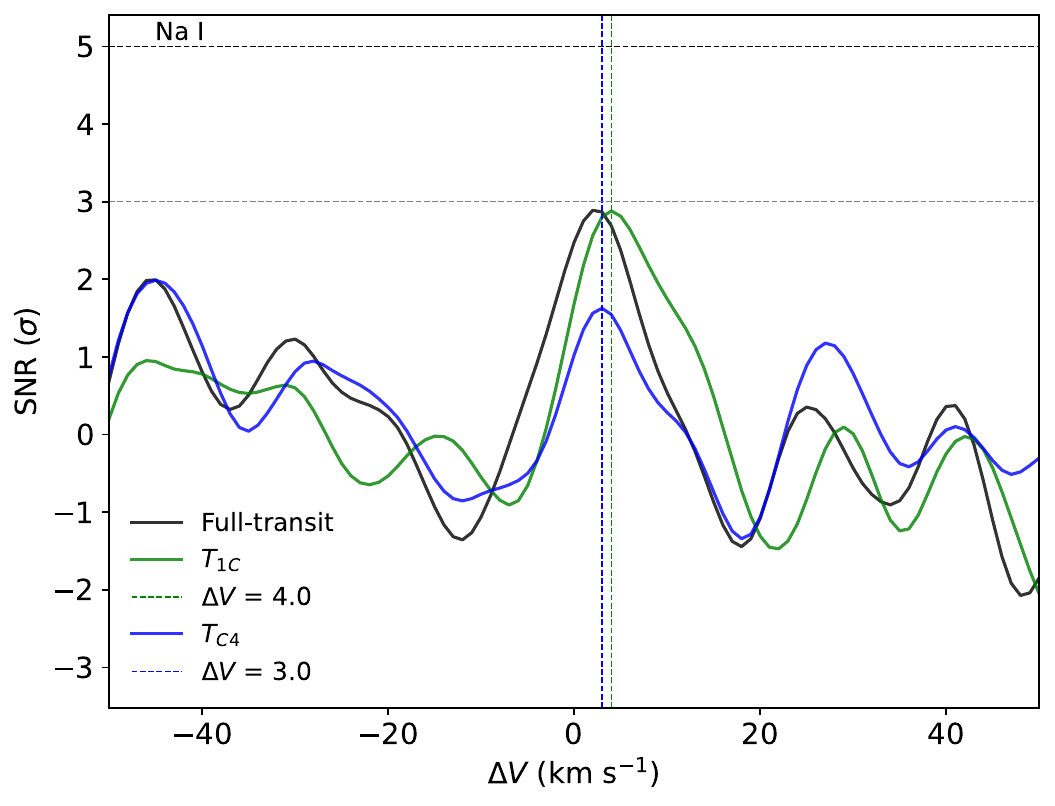}{0.4\textwidth}{}
             }
             \caption{Same as Figure \ref{fig:CCFs-appendix-FeI}, but for Na I.}
    \end{figure}
                \clearpage

\bibliography{sample7}{}
\bibliographystyle{aasjournal}

\end{CJK*}
\end{document}